\documentclass[12pt,a4paper]{article}
\usepackage{epsf,epsfig,rotating}
\def\newpic#1{}

\newcommand{\beq}{\begin{equation}}
\newcommand{\eeq}{\end{equation}}
\newcommand{\ff}{\frac}

\begin{document}

\begin{center}

{\Large \bf Higgs Bosons in the Two-Doublet Model
with {\itshape CP} Violation\\
} \vspace{4 mm}

E. Akhmetzyanova$^{\$}$, M. Dolgopolov$^{\$}$, M. Dubinin$^\#$

\vspace{4 mm}

$^{\$}$Samara State University, Physics Department, Russia;\\
$^\#$Institute of Nuclear Physics, Moscow State University,
Russia.\\

\begin{abstract}
We consider the effective two-Higgs-doublet potential
with complex parameters, when the $CP$ invariance is broken
both explicitly and spontaneously. 
Diagonal mass term in the local
minimum of the potential is constructed
for the physical basis of Higgs fields, 
keeping explicitly the limiting case of $CP$-conservation,
if the parameters are taken real.
For special case of the two-doublet Higgs sector of the
minimal supersymmetric model, when $CP$ invariance is violated by the 
Higgs bosons interaction with scalar quarks of the third generation, 
we calculate by means of the effective potential method 
the Higgs boson masses and evaluate
the two-fermion Higgs boson decay widths and the widths of rare 
one-loop mediated decays $H\rightarrow \gamma \gamma$, $H\rightarrow gg$. 
\end{abstract}

\end{center}

\section{Introduction}

It is well-known that the Cabibbo-Kobayashi-Maskawa (CKM) 
mixing matrix 
originates from
the Standard Model (SM) Lagrangian terms, describing the Higgs boson
interaction with quarks (the Yukawa terms)
\begin{eqnarray}
L&=&- g^u_{ij} \,{\bar{\psi}}^{i\prime}_L \, H \,u^{j\prime}_R 
    - g^d_{ij} \,{\bar{\psi}}^{i\prime}_L 
\, \tilde{H} \, d^{j\prime}_R \, +\mbox{h.c.},
\label{yukawaterms}
\end{eqnarray}
where $\bar{\psi}^{1\prime}_L=(\bar{u'}, \bar{d'})_L$,
      $\bar{\psi}^{2\prime}_L=(\bar{c'}, \bar{s'})_L$,
      $\bar{\psi}^{3\prime}_L=(\bar{t'}, \bar{b'})_L$,
$u^{1\prime}_R=u_R'$, $u^{2\prime}_R=c_R'$, $u^{3\prime}_R=t_R'$,
$d^{1\prime}_R=d_R'$, $d^{2\prime}_R=s_R'$, $d^{3\prime}_R=b_R'$,
and $H$ denotes the scalar complex field doublet,
$\tilde{H}_k=\epsilon_{kl}H^*_l$ and $g^u_{ij}$, $g^d_{ij}$ are the
3$\times$3 matrices with matrix elements that are generally
speaking complex and defined with an uncertainty coming from
the phases of $CP$ transformation
\footnote{Let us remind, for example, that from the definition
of the $P$ transformation $P a^+_{\sigma} (\vec{p}) P^+= \eta_{\sigma}
a^+_{\sigma} (-\vec{p})$, where the complex factor $|\eta_{\sigma}|=$1
contains the $P$ transformation phase, and $\sigma=$0 or 1/2,
it follows that
$P \phi(x) P^+ = \stackrel{*} \eta_0 \phi(x^{'})$, $P
\psi(x) P^+ = \stackrel{*} \eta_{1/2} \gamma_0 \psi(x^{'})$, where
$x^{'}=Px$.} 
for the quark spinor fields and the Higgs boson scalar field.
In order to diagonalize the quark mass term after spontaneous
symmetry breaking $H\to(0,v/\sqrt{2})$, the unitary transformations
of the $u^{i\prime}$ and $d^{i\prime}$ quark fields 
$u^i_{L,R}=U_{L,R} \,u^{i\prime}_{L,R}$, $d^i_{L,R}=D_{L,R}\,
d^{i\prime}_{L,R}$ are needed. After the diagonalization
of the quark mass term the unitary matrices $U_L$ and $D_L$
do not appear neither in the Yukawa Lagrangian terms (\ref{yukawaterms})
nor in the quark neutral current interactions, but arise
in the quark $u^{i\prime}$, $d^{i\prime}$ charged current interaction 
terms $g 
\bar{u}_L' \gamma_{\mu} d_L' W^{\mu}= g \bar{u}_L \gamma_{\mu} U_L
D^\dagger_L d_L W^{\mu}$. The product $V_{CKM}=U_L D^\dagger_L$ defines
the complex CKM matrix, which decribes $CP$ violation in the
quark charged currents sector.     
In the framework of the SM the $CP$ violation takes place since
it is generally speaking not possible to get the mixing matrix
with real matrix elements using $CP$ transformations for six
up- and down- quarks. In other words, $CP$~violation takes place in the SM
because the number of quark generations is exactly three.

There are other sources of $CP$ violation besides the CKM mechanism. It is 
possible to introduce explicitly $CP$ noninvariant
hermitian Lagrangians \cite{[2]} for the system of several scalar fields.
For example, if we have three complex scalar fields 
$\varphi_1, \, \varphi_2, \, \varphi_3$
\begin{eqnarray*}
L= \lambda \varphi_1 \varphi^*_2 \varphi^*_3+
   \lambda^* \varphi^*_1 \varphi_2 \varphi_3 ,  &&
CP\, L\, P^+ C^+=L^{CP}= \lambda e^{i\alpha} \varphi^*_1 \varphi_2
\varphi_3
     + \lambda^* e^{-i\alpha} \varphi_1 \varphi^*_2 \varphi^*_3 ,
\end{eqnarray*}
where $\lambda$ is complex parameter and $\alpha$ is the $CP$
transformation phase, not essential in this case. It can be rotated
away by the phase transformation of the fields, related to charge 
conservation. One can see that $L$ and $L^{CP}$ have different
signs of the imaginary part of $\lambda$. In this simple example  
the difference in the sign does not lead to any observable consequences,
because the phase of  $\lambda$ can be also rotated away by the
$U(1)_Q$ transformation. However for the system with trilinear interactions
of the four complex scalar fields it is generally speaking not possible
to rotate away all phase factors. It is easy to show that the Lagrangian
of such a system will be $CP$ invariant only if the phases of the
four parameters $\lambda_i$ respect certain conditions, which ensure
the possibility to remove them by $U(1)$ rotations of the fields
$\varphi_{i}$. From this point of view the models with extended
Higgs sector, where $CP$ invariance of the Higgs potential
with complex parameters is explicitly broken, are of particular interest.
The simplest example is represented by the two-doublet Higgs potential
of the MSSM, including (if the possibility of spontaneous
$CP$ violation \cite{[3]} is not considered) ten parameters,
four of them can be complex. In the framework of MSSM 
the dominant loop-mediated contributions from the third 
generation scalar quarks could lead to substantial violation of
$CP$ invariance of the two-doublet effective Higgs potential 
\cite{PilaftsisWagner}. Various models with radiatively induced
$CP$ violation in the two-doublet Higgs sector have been studied
\cite{overall,Dubinin02}.

In this paper we develop further on our approach to the Higgs
boson phenomenology in the scenario with $CP$ violation considered
in \cite{Dubinin02}. In Section 2, after brief introductory remarks, we 
calculate the effective $\lambda_i$ parameters of the two-doublet
MSSM Higgs potential at the $m_{top}$ scale. In section 3 we consider in 
details the 
diagonalization
of the mass term for the two-doublet Higgs potential with $CP$ 
invariance broken both explicitly and spontaneously. In the Appendix
some numerical results for the Higgs boson masses and the two-particle 
Higgs decay widths are presented. Our numerical results are compared
with the output of other approaches.

\section{The effective two-doublet Higgs potential with {\itshape CP} 
violation}

In the general two-Higgs-doublet model (THDM)
two $SU(2)$ doublets of complex scalar fields are introduced:
\beq
\Phi_1 = \left( \begin{array}{cc} \phi_{1}^+(x)\\[2mm]
 \phi_{1}^0(x) \end{array}\right)= \left( \begin{array}{cc}
 - \, i\omega_1^+\\[2mm]
 \frac{1}{\sqrt 2}(v_1 + \eta_1 + i\chi_1) \end{array}\right) \, ,
\label{eq:Phi1} \eeq \beq \Phi_2 = e^{\,i\,\xi} \, \left(
\begin{array}{cc}
\phi_{2}^+(x)\\[2mm]
 \phi_{2}^0(x) \end{array}\right) = e^{\,i\,\xi} \, \left( \begin{array}{cc}
 - \, i\omega_2^+\\[2mm]
 \frac{1}{\sqrt 2}(v_2 e^{\,i\,\zeta} + \eta_2 + i\chi_2) \end{array}\right)
\label{eq:Phi2}
\eeq
Their vacuum expectation values (VEV's) \beq
\langle \Phi_1 \rangle = \frac{1}{\sqrt 2} \left( \begin{array}{cc} 0\\
     v_1 \end{array}\right) , \qquad
\langle \Phi_2 \rangle = \frac{e^{\,i\,\xi}}{\sqrt 2}\left( \begin{array}{cc} 0\\
 v_2 \, e^{\,i\,\zeta} \end{array}\right) \equiv
\frac{1}{\sqrt 2}\left( \begin{array}{cc} 0\\
v_2 \, e^{\,i\,\theta} \end{array}\right) . \label{eq:vev} \eeq
where $v_1$ and $v_2$ are real. 
The phases $\zeta$, relative phase of the
VEV's, and $\xi$, relative phase of the $SU(2)$ doublets, are 
introduced to consider the general case, their sum $\theta$
will be used for convenience of notations (section 3.3). 
For special case $\xi=$0
the analysis of Yukawa 
sector with the two fermion generations can be found in \cite{W87},
where somewhat simpler form without the dimension 
2 terms $\Phi^\dagger_1 \Phi_2+\Phi^\dagger_2 \Phi_1$ and 
real $\mu^2_{12}$, $\lambda_{5,6,7}$
parameters of the THDM potential with spontaneous violation of $CP$ invariance
($\zeta=\theta \neq 0$) has been considered in the context of superweak
(i.e. flavor-changing Higgs boson exchage mediated) $CP$ violation in 
meson decays.

The most general renormalizable hermitian $SU(2)\times U(1)$
invariant Lagrangian for the system of scalar fields 
(\ref{eq:Phi1}),~(\ref{eq:Phi2})
can be written as
\beq {\cal L}_H=({\cal D}_\nu \Phi_1)^\dagger {\cal
D}^{\,\nu} \Phi_1 + ({\cal D}_\nu \Phi_2)^\dagger {\cal D}^{\,\nu}
\Phi_2 + \kappa\,({\cal D}_\nu \Phi_1)^\dagger {\cal D}^{\,\nu}
\Phi_2 + \stackrel{*}{\kappa}({\cal D}_\nu \Phi_2)^\dagger {\cal
D}^{\,\nu} \Phi_1 - U(\Phi_1,\Phi_2) , \label{eq:kinet} \eeq где
where
\beq U(\Phi_1,\Phi_2) =
- \, \mu_1^2 (\Phi_1^\dagger\Phi_1) - \, \mu_2^2 (\Phi_2^\dagger
\Phi_2) - \mu_{12}^2 (\Phi_1^\dagger \Phi_2) -
\stackrel{*}{\mu_{12}^2} (\Phi_2^\dagger \Phi_1) + \label{eq:genU}
\eeq
$$ + \lambda_1
(\Phi_1^\dagger \Phi_1)^2
      +\lambda_2 (\Phi_2^\dagger \Phi_2)^2
+ \lambda_3 (\Phi_1^\dagger \Phi_1)(\Phi_2^\dagger \Phi_2) +
\lambda_4 (\Phi_1^\dagger \Phi_2)(\Phi_2^\dagger \Phi_1) + $$
$$ + \frac{\lambda_5}{2}
       (\Phi_1^\dagger \Phi_2)(\Phi_1^\dagger\Phi_2)
 +\frac{\stackrel{*}{\lambda}_5}{2}
(\Phi_2^\dagger \Phi_1)(\Phi_2^\dagger \Phi_1) + $$
$$
 + \lambda_6
(\Phi^\dagger_1 \Phi_1)(\Phi^\dagger_1 \Phi_2)+
\stackrel{*}{\lambda}_6(\Phi^\dagger_1 \Phi_1)(\Phi^\dagger_2
\Phi_1) + \lambda_7 (\Phi^\dagger_2 \Phi_2)(\Phi^\dagger_1 \Phi_2)
+\stackrel{*}{\lambda}_7(\Phi^\dagger_2 \Phi_2)(\Phi^\dagger_2
\Phi_1) \label{eq:genU} $$
The parameters $\mu_{12}^2$,
$\lambda_{\,5}$, $\lambda_{\,6}$ и $\lambda_{\,7}$ are complex.
Complex parameter $\kappa$ could be introduced to describe
an interesting possibility of a mixing in the kinetic term 
\cite{ginzkraw}. However, strong  
restrictions on the real part of $\kappa$ are imposed by
precise experimental data on the gauge boson masses $m_{W,Z}$.
Moreover, mixing in the kinetic term does not allow to construct
the diagonal 4$\times$4 matrix of the Higgs boson kinetic terms 
consistently with the diagonal matrix for their mass 
terms\footnote{We analysed these conditions written in the form of ten 
linear
equations, having the solution practically only in the case $\kappa=0$.
The mixed term is not obligatory to ensure the renormalizability.
It is shown below that the contributions of 
self-energy diagrams absorbed by the Higgs boson wave-function 
renormalization to the effective
parameters $\lambda_{5,6,7}$ are zero, see also \cite{HH1993}.  }.
In the following we consider the case
$\kappa=0$.

Special case of the two-Higgs-doublet potential is the potential
of the MSSM Higgs sector. At the energy scale $M_{SUSY}$ (i.e. at the
energy of the order of the sparticle masses) the tree level parameters
$\lambda_{1,...,7}$ are real and can be expressed through the $SU(2)\times U(1)$
gauge couplings $g_1$ and $g_2$ \cite{Inoue}
\begin{eqnarray}
\label{eq:lMSUSY} &\lambda_1(M_{SUSY}) = \lambda_2(M_{SUSY}) =
\ff{1}{8}(g_2^2(M_{SUSY})+ g_1^2(M_{SUSY})) ,&  \\ \nonumber
&\lambda_3(M_{SUSY}) = \ff{1}{4}(g_2^2(M_{SUSY})-g_1^2(M_{SUSY}))
, \qquad \lambda_4(M_{SUSY}) = - \, \ff{1}{2}g_2^2(M_{SUSY}) ,& \\
\nonumber &\lambda_5(M_{SUSY}) = \lambda_6(M_{SUSY})=
\lambda_7(M_{SUSY}) = 0.&
\end{eqnarray}
At the scale $M_{SUSY}$ the potential is $CP$ invariant.
However, the potential parameters of any model depend, generally speaking,
on the energy scale where they are fixed or measured. The dependence
is described by the renormalization group equations (RGE). The conditions
(\ref{eq:lMSUSY}) are the boundary conditions for the RGE. At the energies
smaller than $M_{SUSY}$ they are affected by large quantum corrections
\cite{RGE91} where the main contribution is coming from 
the Higgs bosons - third generation quarks and scalar quarks interaction
(the interactions with the first and second generations are suppressed).
The potential of the Higgs bosons - scalar quarks interaction
can be written in the form \cite{HH1993}
\beq {\cal V}^{\,0} =
{\cal V}_M + {\cal V}_\Gamma + {\cal V}_\Lambda + {\cal
V}_{\widetilde Q}\,, \label{eq:HHpot}
\eeq
where
\beq {\cal V}_M =
(-1)^{i+j}m_{ij}^2\Phi_i^{\dag}\Phi_j+ M_{\widetilde
Q}^{\,2}\left(\widetilde
  Q^{\,\dag}\widetilde Q\right)
+M_{\widetilde U}^{\,2}\widetilde U^*\widetilde U +M_{\widetilde
D}^{\,2}\widetilde D^*\widetilde D\,, \eeq \beq {\cal V}_\Gamma =
\Gamma_i^D\left(\Phi^{\dag}_i\widetilde Q\right)\widetilde D
+\Gamma_i^U\left(i\Phi_i^T\sigma_2\widetilde Q\right)\widetilde U
+\stackrel{*}{\Gamma_i^D}\left({\widetilde
Q}^{\,\dag}\Phi_i\right)\widetilde D^*
-\stackrel{*}{\Gamma_i^U}\left(i\widetilde
Q^{\,\dag}\sigma_2\Phi_i^*\right) \widetilde U^* \, ,
\label{eq:VGamma}
\eeq
\beq
{\cal V}_\Lambda =
\Lambda_{ik}^{jl}\left(\Phi^{\dag}_i\Phi_j\right)
\left(\Phi^{\dag}_k\Phi_l\right) +\left(\Phi^{\dag}_i\Phi_j\right)
\left[\Lambda_{ij}^Q\left(\widetilde Q^{\,\dag}\widetilde Q\right)
+\Lambda_{ij}^U \widetilde U^*\widetilde U +\Lambda_{ij}^D
\widetilde D^*\widetilde D\,\right] + \label{eq:VL} \eeq
$$\quad+
\overline\Lambda_{\,ij}^{\,Q}\left(\Phi^{\dag}_i\widetilde
Q\right) \left(\widetilde Q^{\,\dag}\Phi_j\right)
+\ff{1}{2}\left[\Lambda\epsilon_{ij}
\left(i\Phi_i^T\sigma_2\Phi_j\right)\widetilde D^*\widetilde
U+h.c.\right]\,, \quad i,j,\,k,l=1,2 \,, $$ ${\cal V}_{\widetilde
Q}$ 
denotes the four scalar quarks interaction terms, Pauli matrix
$\sigma_2\equiv\left(\begin{array}{cc} 0 & i\\ - i &
0\end{array}\right)$. 
The Yukawa coupligs for the third generation of scalar quarks are
defined in a standard way
$h_{\,t} = \frac{\sqrt{2}\, m_{\,t}}{v \sin\beta }$, $h_{\,b} =
\frac{\sqrt{2}\, m_b }{v \cos\beta }$. Following
\cite{CEPW0003180}
\footnote{For the case of $CP$ conservation, considered in
\cite{HH1993}, the trilinear parameters in (\ref{eq:VGamma})
are real. Then 
$\Gamma_{\{1;\, 2\}}^{\,U} \equiv h_U\, \{-\mu; A_U\}, \,\,
\Gamma_{\{1;\, 2\}}^D \equiv h_D \,\{A_D\,; -\mu\}$.}:
\beq \Gamma_{\{1; \,2\}}^{\,U} = h_U\, \{-\mu^*; A_U\}, \qquad
\Gamma_{\{1; \,2\}}^{\,D} = h_D \,\{A_D\,; -\mu^*\} ,
\label{eq:Gammi} \eeq
they are complex in the case under consideration.
One can observe $CP$ violating terms of the structure similar
to (\ref{yukawaterms}) in the sector of Higgs-scalar quark interactions,
so complex mixing matrices are expected to appear there.
The trilinear parameters $A_t$, $A_b$ and the Higgs 
mass parameter $\mu$ should be taken complex,
the
imaginary parts of the mixing matrix elements could be large.

In the framework of the effective field theory approach \cite{HH1993}
the MSSM potential (\ref{eq:HHpot}) which explicitly describes sparticle 
interactions at the energy 
scale above $M_{SUSY}$ is matched to an effective Standard Model-like
Lagrangian at the energy scale below $M_{SUSY}$, where the sparticles
decouple.
So the MSSM effective Higgs potential at the energy scale
$m_{top}$, much smaller than $M_{SUSY}$,
is represented by the general two-Higgs-doublet model potential
(6), the parameters of the latter are expressed
by means of the Higgs bosons - scalar quarks interaction parameters
(\ref{eq:Gammi}) and the scalar quark masses, playing the role of 
ultraviolet 
Pauli-Villars regulators.
The RGE boundary conditions (\ref{eq:lMSUSY}) modified
by the interactions of the third generation squarks with the Higgs
bosons (these modifications are sometimes called the "threshold" effects, 
since the stops decouple at the $M_{SUSY}$ scale), 
are imposed at the energy scale $M_{SUSY}$. 
They affect the evolution of $\lambda_i$ parameters, the Yukawa
couplings $h_{t,b}$ and the gauge couplings $g_{1,2}$. We 
calculated radiative
corrections to the boundary conditions (\ref{eq:lMSUSY}) for
$\lambda_i$ parameters at the scale $m_{top}$ 
using the effective potential method \cite{Quiros97}. The squark
mass matrices 
$({\cal M}_X^2)_{ab} \equiv
\frac{\partial^2 {\cal V}_X}{\partial \tilde Q_a\, \partial \,
 \tilde Q_b^* }$
defined by (\ref{eq:HHpot}) were calculated and then
substituted to the one-loop effective potential
\begin{eqnarray*}
{\cal V} ={\cal V}^0 + \frac{N_C}{32\pi^2} \mbox{tr}{\cal M}^4
\left[\ln\left(\frac{{\cal M}^2}{\sigma^2}\right) - \, \frac{3}{2}\right]
\, ,
\end{eqnarray*}
decomposed in the inverse powers of $M_{SUSY}$.
Taking into account 
the one-loop wave-function renormalization terms (i.e. terms
introduced to absorb the contributions of self-energy diagrams to the 
Higgs bosons kinetic term, which are 
beyond the
calculation by means of the effective potential method), 
the effective parameters can be evaluated as 
follows:
\begin{eqnarray}
\label{eq:lambda1} \lambda_1 &=& \, \frac{g_2^2+g_1^2}{8} +
\frac{3}{32\pi^2} \Big[ h^4_b \frac{|A_b|^2}{M^2_{\rm
SUSY}}\left(2-\frac{|A_b|^2}{6 M^{\,2}_{\rm SUSY}}\right)
-h^4_t\frac{|\,\mu|^4}{6 M^{\,4}_{\rm SUSY}}\,+\,2 h_b^4 l + \\
\nonumber && +\,\frac{g_2^2+g_1^2}{4\,M^{\,2}_{\rm
SUSY}}(h^2_t|\,\mu|^2-h^2_b|A_b|^2) \Big] \, +\\
\nonumber && + \,
\Delta\,\lambda_{\,1}^{field}\,+\ff{1}{768\pi^2}\,
\left(11 g_1^4 + 9g_2^4 - 36 \,(g_1^2+g_2^2)\,h_b^2\right) l , \\
\nonumber \lambda_2 &=& \lambda_1 \,\,(\,t \longleftrightarrow b)
,
\label{eq:lambda2} \\
 \lambda_3 &=&
   \, \frac{g_2^2-g_1^2}{4}\left[1-\,\ff{3}{16\pi^2}\,(h_t^2+h_b^2)\, l
\right] +
   \ff{3}{8\pi^2}\,h_t^2h_b^2\left[l+\ff{1}{2} X_{tb}\right]\, + \\
\nonumber
\label{eq:lambda3}
&&   +\, \frac{3}{96\pi^2}\, \ff{|\mu|^2}{M^2_{\rm
   SUSY}} \left[ h^4_t\,  \left(\, 3\, -\, \frac{|A_t|^2}{M^2_{\rm
SUSY}}\,
\right)\,
   +\,  h^4_b\, \left(\, 3\, -\, \frac{|A_b|^2}{M^2_{\rm SUSY}}\,
   \right)\right] \, +
\\ \nonumber
&& + \,
\ff{3(g_2^2-g_1^2)\left[h_b^2(|\mu|^2-|A_b|^2)+h_t^2(|\mu|^2-|A_t|^2)\right]}
{128\, \pi^2 M^{\,2}_{\rm SUSY}} +\, \Delta\,\lambda_{\,3}^{field}
+ \,\ff{9g_2^4 - 11g_1^4}{384 \,\pi^2} \,l , \\
\lambda_4 &=& - \, \frac{g_2^2}{2}
\left[1-\,\ff{3}{16\pi^2}\,(h_t^2+h_b^2)\, l \right] -
   \ff{3}{8\pi^2}\,h_t^2h_b^2\left[l+\ff{1}{2} X_{tb}\right]\, +
  \label{eq:lambda4}
\\ \nonumber
&&  +\, \frac{3}{96\pi^2}\, \ff{|\mu|^2}{M^2_{\rm
   SUSY}} \left[ h^4_t\,  \left(\, 3\, -\, \frac{|A_t|^2}{M^2_{\rm
SUSY}}\,
\right)\,
   +\,  h^4_b\, \left(\, 3\, -\, \frac{|A_b|^2}{M^2_{\rm SUSY}}\,
   \right)\right] \, -
\\ \nonumber
&& - \,\, \ff{3
g_2^2\left[h_b^2(|\mu|^2-|A_b|^2)+h_t^2(|\mu|^2-|A_t|^2)\right]}{64
\pi^2 M^{\,2}_{\rm SUSY}}+\, \Delta\,\lambda_{\,4}^{field}\, - \,
\ff{3 g_2^4}{64 \pi^2} \,\,l \,, \label{eq:add4}
\end{eqnarray}
where
\beq
X_{tb}\equiv\ff{|A_t|^2+|A_b|^2+2{\tt Re}(A_b^*A_t)}{2M^{\,2}_{\rm
SUSY}}-\,\ff{|\mu|^2}{M^{\,2}_{\rm
SUSY}}-\,\ff{||\mu|^2-A_b^*A_t|^2}{6M^{\,4}_{\rm SUSY}}\, . 
\eeq
The effective complex parameters $\lambda_{5,6,7}$
\begin{eqnarray}
\label{eq:lambda5} \lambda_5= \, -
\,\Delta\lambda_{\,5} &=& - \,
\frac{3}{96\,\pi^2}\, \left(h^4_t\,
   \left(\frac{\mu A_t}{M^{\,2}_{\rm SUSY}}\right)^2\,
+\, h^4_b\, \left(\frac{\mu A_b}{M^{\,2}_{\rm
SUSY}}\right)^2\right) ,
\end{eqnarray}
\begin{eqnarray}
\label{eq:l6g}
\lambda_6= \, - \,\Delta\lambda_{\,6} &=& \,
\frac{3}{96\,\pi^2}\, \left[h^4_t\,
   \frac{|\mu|^2 \mu A_t}{M^{\,4}_{\rm SUSY}}\,
   -\, h^4_b\, \frac{\mu A_b}{M_{\rm SUSY}^2}\,
   \Big(\, 6\, -\, \frac{|A_b|^2}{M^{\,2}_{\rm
   SUSY}}\, \Big)\,\, +   \right.\\ \nonumber
&& \left. + \,\, (h^2_b A_b - h^2_t A_t)\, \frac{ 3\, \mu
}{M^{\,2}_{\rm SUSY}}\, \frac{g_2^2\, +\, g_1^2}{4} \right],
\end{eqnarray}
\begin{eqnarray}
\label{eq:lambda7}
\lambda_7= \, - \,\Delta\lambda_7 &=& \,
\frac{3}{96\,\pi^2}\, \left[h^4_b\,
   \frac{|\mu|^2 \mu A_b}{M^{\,4}_{\rm SUSY}}\,
   -\, h^4_t\, \frac{\mu A_t}{M_{\rm SUSY}^{\,2}}\,
   \Big(\, 6\, -\,
   \frac{|A_t|^2}{M^{\,2}_{\rm SUSY}}\,\Big)\,\, +
\right.\\ \nonumber && \left. + \,\, (h^2_t A_t - h^2_b A_b)\,
\frac{ 3\, \mu }{M^{\,2}_{\rm SUSY}}\, \frac{g_2^2\, +\, g_1^2}{4}
\right] .
\end{eqnarray}
Some details of the calculation can be found in \cite{qfthep1}.
The one-loop wave-function renormalization terms in
(\ref{eq:lambda1})-(\ref{eq:lambda4}) are
\beq
\Delta\,\lambda_{\,1}^{field} = \ff{1}{2} (g_1^2+g_2^2) A'_{11} ,
\qquad \Delta\,\lambda_{\,2}^{field} = \ff{1}{2} (g_1^2+g_2^2)
A'_{22} , \label{eq:frc}
\eeq
$$ \Delta\,\lambda_{\,3}^{field} = -\,\ff{1}{4}
(g_1^2-g_2^2) (A'_{11}+A'_{22}) , \qquad
\Delta\,\lambda_{\,4}^{field} = -\,\ff{1}{2} g_2^2
(A'_{11}+A'_{22}) , $$
$$ \Delta\,\lambda_{\,5}^{field} =0, $$
$$ \Delta\,\lambda_{\,6}^{field}  = \ff{1}{8} (g_1^2+g_2^2)
(A'_{12}-{A'_{21}}^*) = 0, \qquad \Delta\,\lambda_{\,7}^{field} =
\ff{1}{8} (g_1^2+g_2^2) (A'_{21}-{A'_{12}}^*) = 0 . $$
They are similar to the case of $CP$ conservation \cite{HH1993}
containing the logarithmic contributions and imaginary parameters as a 
consequence of  
(\ref{eq:Gammi}), and can be written as
\beq A'_{\,ij} = - \, \ff{3}{96\,\pi^2
\,M_{\,SUSY}^{\,2}} \left[h_t^2 \left[\begin{array}{cc} |\mu|^2 &
- \mu^* A_t^*\\ - \mu A_t & |A_t|^2 \end{array}\right] \,+\, h_b^2
\left[\begin{array}{cc} |A_b|^2 & - \mu^* A_b^*\\ - \mu A_b &
|\mu|^2
\end{array}\right]\right] \, \times \eeq
$$ \times \, \left(1-\,\ff{1}{2}\, l\,\right) \,.
$$
Here and in the formulas given below
$l\equiv\ln\left(\ff{ M^{\,2}_{\rm SUSY}}{\sigma^2}\right) $, where
$\sigma=m_{top}$ is the renormalization scale. 
The one-loop wave-function renor\-ma\-lization does not yield a
$CP$ violating contribution to $\lambda_i$.
For convenience we introduce the notation for the deviations of  
effective
parameters $\lambda_i$ from $\lambda^{SUSY}_i=\lambda_i (M_{SUSY})$
following \cite{Dubinin02}:
\beq \lambda_{1, 2} \equiv
\lambda_{1, 2}^{SUSY}-\Delta\lambda_{1, 2}/2 , \quad \lambda_{3,
4} \equiv \lambda_{3, 4}^{SUSY}-\Delta\lambda_{3, 4} , \quad
\lambda_{5, 6, 7} \equiv - \,\Delta\lambda_{5, 6, 7} ,
\label{eq:effsc} \eeq \beq \mbox{where}\, 
\qquad \Delta\lambda_{\,i} \equiv
\Delta\lambda_{\,i}^{eff. pot.} - \Delta\lambda_{\,i}^{field} \,,
\qquad \Delta\,\lambda_{\,i}^{\{eff. pot.; \,field\}} \equiv
\Delta\lambda_{\,i}^{log} +
\Delta\lambda_{\,i}^{finite}\,,\label{eq:wheredelta}\eeq \beq
\mbox{причем} \qquad \Delta\,\lambda_{\,5, 6, 7}^{log} = 0 \,,
\qquad \Delta\,\lambda_{\,5, 6, 7}^{field} =0\,.
\eeq

In the end of this section we would like to make some general
comments as well as some comments in connection with results obtained
by other authors. Like in the existing effective field theory approach
\cite{HH1993}
we are using the standard scheme of leading logarithmic
terms resummation by means of RGE, additionally taking into account
in the boundary conditions at the scale $M_{SUSY}$ the effects
of Higgs bosons - third generation of scalar quarks interaction.
The one-loop effective parameters (\ref{eq:lambda1}) - (\ref{eq:lambda7})
satisfy the boundary conditions defined by (\ref{eq:lMSUSY}) and 
modified by the soft supersymmetry breaking potential terms (\ref{eq:HHpot})
("threshold effects").
The terms with the logarithmic factor $l$ describe the 
parameters evolution 
from the energy scale $M_{SUSY}$ down to the scale $\sigma=m_{top}$.
Finite power term threshold corrections to $\lambda_{1,...,7}$ appear from 
the
so-called $F$-terms (the trilinear interaction terms in (\ref{eq:VGamma}))
and $D$-terms (contained in (\ref{eq:VL})). The corrections to 
$\lambda_{5}$
come from the $F$-terms only. Radiative corrections to the
parameters $\lambda_{1...,7}$ of the effective two-Higgs-doublet
potential have been considered earlier in \cite{PilaftsisWagner}
for the case of broken $CP$ invariance and in \cite{HH1993}, \cite{CEQW}
for the case of $CP$ conservation. Phenomenological consequences
of the two-doublet system are usually analysed assuming for simplicity
$A_t=A_b$ and introducing the universal phase ${\tt arg}\mu A_{t,b}$,
so that $\lambda_{\,5} = |\,\lambda_{\,5}| \, \exp\,[ i \,2 \,  
{\tt arg} (\mu A) ]$, $\lambda_{\,6} = |\,\lambda_{\,6}| \,
\exp\,[ i\, {\tt arg} (\mu A) ]$, $\lambda_{\,7} =
|\,\lambda_{\,7}| \, \exp\,[ i\, {\tt arg} (\mu A) ]$.

\begin{table}[t]
\begin{tabular}{|c||c|c|c|c|c|c|c|}
\hline
$i$ & $1$ & $2$ & $3$ & $4$ & $5$ & $6$ & $7$ \\
\hline\hline
only ${\cal O}(h_t^4)$ terms \cite{Quiros97} & 0.907 & -0.203 & 0.057 &
0.057 & 0.227 &
-0.453 & 0.057 \\
\hline
$\Delta\lambda_{\,i}$ & 0.860 & -0.182 & 0.054 & 0.072 & 0.227 & -0.442 &
0.046 \\
\hline
1-loop~\cite{PilaftsisWagner} & 0.907 & -0.191 & 0.064 & 0.043 & 0.227 &
-0.453 & 0.057 \\
\hline 1-loop + 2-loop \cite{PilaftsisWagner} & 0.761 & -0.152
& 0.052 & 0.032 & 0.135 & -0.371 & 0.044 \\ \hline
2-loop~\cite{PilaftsisWagner} & -0.146 & 0.039 & -0.012 & -0.011
& -0.092 & 0.082 & -0.013 \\ 
\hline 
1-loop(D+wfr) & -0.047 & 0.009 & -0.010 & 0.028 & 0 & 0.011 & -0.011 \\ 
\hline
$\ff{\Delta\lambda(\mbox{D+wfr})}{\Delta\lambda(\mbox{2-loop})}$
&0.32  & 0.23 & 0.83 & -2.55 & 0 &
0.13 & 0.85 \\ \hline 
1-loop+2-loop
+ 1-loop(D+wfr) & 0.715 & -0.143 & 0.042 & 0.061 & 0.135 & -0.360 &
0.033 \\ \hline
\end{tabular}
\label{tab:lambdas} 
\caption{Numerical comparison of various corrections to the
$\lambda_{\,i}$ parameters at the scale $m_{top}$. For convenience of
the following Higgs boson masses comparison, the same parameter values
as in the package CPsuperH \cite{CPsuperH} are chosen here: $m_Z
=91.19$\,GeV, $m_b=3$\,GeV, $m_t=175$\,GeV, $m_W=79.96$\,GeV, 
$g_2=0.6517$, $g_1=0.3573$,
$v=245.4$\,GeV, $G_F=1.174\cdot10^{-5}$\,GeV$^{-2}$,
$\alpha_S(m_t)=0.1072$, $\tan\beta=5$, $M_{SUSY}=500$\,GeV,
$\sigma=m_t$, $m_{H^{\pm}}=300$\,ГэВ, $|A_t|=|A_b|=A=1000$\,GeV,
$|\mu|=2000$\,GeV, $\varphi\equiv{\tt arg}(\mu A_{t, b})=0$. The
abbreviation 'wfr' stands for the 'wave-function renormalization'.}
\end{table}
  
Only the leading $D$-term contributions were 
calculated in \cite{PilaftsisWagner},
\cite{CEQW}. In our expressions for the effective parameters 
(\ref{eq:lambda1})-(\ref{eq:lambda7}) the nonleading $D$-term 
contributions are
represented by the power terms containing gauge couplings $g^2_1$, $g^2_2$. 
The one-loop contributions of the wave-function renor\-ma\-lization 
$\Delta \lambda^{field}_{1,...,4}$
are neglected in \cite{PilaftsisWagner}, \cite{CEQW}.
However, the QCD and weak corrections to Yukawa couplings up to two loops, 
not calculated in our case, have been included there.
The expressions for 
$\lambda_{1,2,3,4}$ (\ref{eq:lambda1})-(\ref{eq:lambda4}) do not
contain imaginary parts up to the two-loop approximation  
and coincide with the results of \cite{PilaftsisWagner}, \cite{CEQW}
if we omit the contributions of nonleading $D$-terms and 
$\Delta \lambda^{field}_{1,...,4}$ terms. If $\mu$ and $A$ are real, the 
expressions (\ref{eq:lambda1})-(\ref{eq:lambda7})
are consistent with the results of \cite{HH1993}, where the $D$-terms
contribution was calculated\footnote{In (\ref{eq:lambda1})-(\ref{eq:lambda4}) 
we kept the terms of the order of $g^4_{1,2}$.}. 
Let us note that it is not possible
to generalize the expressions for real $\lambda_{5,6,7}$ in the case of
$CP$ violating potential by the straightforward replacement of
the real $\mu$, $A$ parameters to the complex ones.

If we neglect the contributions of $D$-terms, the wave-function
renormalization terms $\Delta \lambda^{field}_{1,...,4}$ and the terms
of the order of $h^2_b$ for the $b$-quark couplings, only the one-loop
corrections of the order of ${\cal O}(h_t^4)$ remain.
This approximation was discussed in \cite{CEQW,Quiros97}. For example,
$\lambda_2$ is given by
\begin{equation}
\label{eq:lam2approx} \lambda_2 \approx \, \frac{g_2^2+g_1^2}{8} +
\frac{3}{32\,\pi^2} \left[h^4_t \frac{|A|^2}{M^{\,2}_{\rm
SUSY}}\left(2-\frac{|A|^2}{6 M^{\,2}_{\rm SUSY}}\right) \,+\,2
h_t^4 l \right]\,,\eeq
The beta-function for $\lambda_2$
contains large negative contribution $-6h^4_t$ \cite{HH1993}, or 
equivalently, $\lambda_2$
(13) contains the large logarithmic term $6 h^4_t \, l \,/(32 \pi^2)$
which was observed in the first calculations \cite{RGE91}. In
the following the negative $\Delta \lambda_2$ defined by (22) gives large 
positive contribution to the light Higgs boson mass in (38).

Numerical comparison of the $\lambda_i$ parameters evaluated
using different approximations is presented in the Table 1,
where for our case in the second line of the Table
$$ \Delta\lambda_{\,i} = \{\mbox{one-loop contribution}\} +
\{\mbox{one-loop}(D
-\mbox{terms}+\mbox{wave-func.renormalization})\} .
$$
One can conclude that the one-loop corrections from $D$-terms
and wave-function renormalization can be of the order of the leading
two-loop corrections.
Difference of the effective $\lambda_i$ of the order of 10$^{-1}$
may result in the deviation of Higgs boson masses around 5 GeV and 
even more.

\section{Diagonalization of the effective potential mass term
in the local minimum}

\subsection{Complex $\mu^2_{12}$, $\lambda_{5,6,7}$ parameters,
$\theta=$0}
\vskip 3mm
The components $\omega_i$, $\eta_i$, $\chi_i$ of the $SU(2)$ doublets
(\ref{eq:Phi1}), (\ref{eq:Phi2}) are not a physical Higgs fields
(mass eigenstates). In order to extract the Higgs boson masses
and the self-interaction of the physical fields from the potential
(\ref{eq:genU}) it is necessary to diagonalize the mass term of
the latter in the local minimum. This problem has been considered
in \cite{Dubinin02} for the case of complex $\mu^2_{12}$, $\lambda_{5,6,7}$
parameters and the zero phase of the $\Phi_2$ VEV $\theta=$0.
The diagonalization of the mass term is performed in two stages.
First the $CP$-even fields $h$,$H$, the $CP$-odd field $A$ ('pseudoscalar')
\footnote{The fields  $h$,$H$,$A$ are the physical fields at
$\varphi={\tt arg}(\mu A_{t,b})=0,\, n\pi$.}
and the Goldstone field $G^0$ are defined by the linear transformation
\beq h = - \, \eta_1\sin\alpha +
\eta_2\cos\alpha , \label{eq:MSSMh} \eeq \beq H = \eta_1\cos\alpha
+ \eta_2\sin\alpha , \label{eq:MSSMH}\eeq \beq A = - \,
\chi_1\sin\beta + \chi_2\cos\beta , \label{eq:MSSMA}\eeq \beq G^0
= \chi_1\cos\beta + \chi_2\sin\beta ,\label{eq:MSSMG} 
\eeq 
where
${\tt tg}\beta=v_2/v_1$ and
\begin{eqnarray}
{\tt tg} 2\alpha &\hspace{-3mm} =& \hspace{-5mm}
\frac { s_{2\beta} (m^2_A + m^2_Z)  +v^2 ((\Delta
\lambda_3 + \Delta \lambda_4) s_{2\beta}+2c^2_{\beta}
{\tt Re}\Delta \lambda_6
+ 2s^2_{\beta}  {\tt Re}\Delta \lambda_7) }
                        {c_{2\beta}(m^2_A - m^2_Z) +v^2 (\Delta
\lambda_1 c^2_{\beta} - \Delta \lambda_2 s^2_{\beta} - {\tt Re}
\Delta \lambda_5 c_{2\beta}+({\tt Re}\Delta \lambda_6- {\tt Re}
\Delta \lambda_7)s_{2\beta} )}\,. \label{t2b}
\end{eqnarray}
Here the relations $g^2_1+g^2_2=g^2m^2_Z/m^2_W$,
$g^2_2-g^2_1=g^2_1 \, (2-m^2_Z/m^2_W)$ are used.
Then we substitute to the effective potential 
the real parameters $\mu_{1,2}$, $\lambda_{1,2,3,4}$
and the real parts ${\tt Re}\mu^2_{12}$, ${\tt Re}\lambda_{5,6,7}$,
which are related by linear transformation 
\cite{Dubinin02,Dubinin01,GunionHaber}:
\begin{eqnarray}
\label{diaglambda1}
 \lambda_1& =&
\frac{1}{2v^2}
         [(\frac{s_{\alpha}}{c_{\beta}})^2 m_h^2
        + (\frac{c_{\alpha}}{c_{\beta}})^2 m_H^2
     -  \frac{s_{\beta}}{c^3_{\beta}}{\tt Re}\mu_{12}^2  ]
+\frac{1}{4}( {\tt Re} \lambda_7 {\tt tg}^3 {\beta}
-3 {\tt Re} \lambda_6{\tt tg} {\beta}) , \\
 \lambda_2& =&
          \frac{1}{2v^2}
         [(\frac{c_{\alpha}}{s_{\beta}})^2 m_h^2
        + (\frac{s_{\alpha}}{s_{\beta}})^2 m_H^2
     -  \frac{c_{\beta}}{s^3_{\beta}}{\tt Re}\mu_{12}^2  ]
+\frac{1}{4}({\tt Re} \lambda_6 {\tt ctg}^3 {\beta}
-3 {\tt Re} \lambda_7 {\tt ctg} {\beta}) , \\
\label{diaglambda3}
 \lambda_3& =&
\frac{1}{v^2}[2m^2_{H^\pm}
                - \frac{{\tt Re}\mu_{12}^2}{s_{\beta} c_{\beta}}
              +\frac{s_{2\alpha}}{s_{2\beta}} (m_H^2-m_h^2)]
-\frac{{\tt Re} \lambda_6}{2} {\tt ctg}\beta
- \frac{{\tt Re} \lambda_7}{2} {\tt tg}\beta , \\
 \lambda_4& =&
\frac{1}{v^2}(\frac{{\tt Re}\mu_{12}^2}{s_{\beta} c_{\beta}}
                   +m^2_A- 2 m^2_{H^\pm} )
-\frac{{\tt Re} \lambda_6}{2} {\tt ctg}\beta  -
\frac{{\tt Re} \lambda_7}{2} {\tt tg}\beta , \\
\label{diaglambda5}
{\tt Re} \lambda_5& = &
\frac{1}{v^2} (\frac{{\tt Re}\mu_{12}^2}{s_{\beta} c_{\beta}}
                    -m^2_{A} )
-\frac{{\tt Re} \lambda_6}{2} {\tt ctg}\beta - \frac{{\tt Re}
\lambda_7}{2} {\tt
tg}\beta , \\
\label{diagmu1}
\mu^2_{1}& = &
 \lambda_1 v^2_1+( \lambda_3+
\lambda_4+{\tt Re} \lambda_5)\frac{v^2_2}{2}
- {\tt Re}\mu_{12}^2 {\tt tg}\beta
+\frac{v^2 s^2_{\beta}}{2}
(3 {\tt Re} \lambda_6 {\tt ctg} \beta + {\tt Re} \lambda_7 {\tt tg}
\beta) , \\
\label{diagmu2}
\mu^2_{2}& = &  {\hskip -2mm}
 \lambda_2 v^2_2+( \lambda_3+
\lambda_4+{\tt Re} \lambda_5)\frac{v^2_1}{2} - {\tt Re}\mu_{12}^2
{\tt ctg}\beta +\frac{v^2 c^2_{\beta}}{2} ( {\tt Re} \lambda_6
{\tt ctg}\beta + 3  {\tt Re} \lambda_7 {\tt tg} \beta ) .
\end{eqnarray}
At the purely real parameters (in the following we shall name
this case of $\varphi=0$ as the $CP$-conserving limit,
${\tt Re} \lambda_i=|\lambda_i|$, ${\tt Re}
\Delta \lambda_i=|\Delta \lambda_i|$)
the relations (\ref{diagmu1}), (\ref{diagmu2}) 
set to zero the potential terms which are
linear in the fields, so they are the minimization conditions.
It follows from the equations (\ref{diaglambda1})-(\ref{diaglambda5}) that
in the $CP$ conserving limit the $CP$-even Higgs boson masses
and the real part of the $\mu^2_{12}$ parameter can be expressed as
\begin{eqnarray}
m^2_h&=&s^2_{\alpha+\beta} m^2_Z + c^2_{\alpha-\beta} m^2_A - \\
\nonumber &&\hspace{-16mm} - v^2 (\Delta  \lambda_1 s^2_{\alpha}
c^2_{\beta} +\Delta
 \lambda_2
c^2_{\alpha} s^2_{\beta}- 2(\Delta  \lambda_3+\Delta
 \lambda_4)c_{\alpha}
c_{\beta} s_{\alpha} s_{\beta}+ {\tt Re}\Delta \lambda_5
(s^2_{\alpha} s^2_{\beta} +c^2_{\alpha} c^2_{\beta}) - \\
\nonumber && -2 c_{\alpha+\beta} ({\tt Re}\Delta \lambda_6
s_{\alpha} c_{\beta}
                    -{\tt Re}\Delta \lambda_7 c_{\alpha} s_{\beta})) , \\
m^2_H&=&c^2_{\alpha+\beta} m^2_Z + s^2_{\alpha-\beta} m^2_A - \\
\nonumber &&\hspace{-16mm}
 -v^2 (\Delta  \lambda_1 c^2_{\alpha} c^2_{\beta} +\Delta
 \lambda_2
s^2_{\alpha} s^2_{\beta}+ 2(\Delta  \lambda_3+\Delta
 \lambda_4)c_{\alpha}
c_{\beta} s_{\alpha} s_{\beta}+ {\tt Re}\Delta \lambda_5
(c^2_{\alpha} s^2_{\beta} +s^2_{\alpha} c^2_{\beta}) + \\
\nonumber && +2 s_{\alpha+\beta} ({\tt Re}\Delta \lambda_6
c_{\alpha} c_{\beta}
                    +{\tt Re}\Delta \lambda_7 s_{\alpha} s_{\beta})) , \\
m^2_{H^\pm}&=&m^2_W+m^2_A-\frac{v^2}{2}( {\tt Re} \Delta \lambda_5-
 \Delta
\lambda_4) , \\
\nonumber {\tt Re}\mu^2_{12}&=&s_{\beta}
c_{\beta}[m^2_A-\frac{v^2}{2}(2 {\tt Re}\Delta \lambda_5+{\tt Re}
\Delta \lambda_6 {\tt ctg} \beta+{\tt Re}\Delta \lambda_7 {\tt tg}
\beta)] . \nonumber
\end{eqnarray}
After the substitution of (\ref{diaglambda1})-(\ref{diaglambda5}),
(\ref{diagmu1}), (\ref{diagmu2}) to (\ref{eq:genU})
we find the mass term of the effective potential
\begin{eqnarray}
\label{massterm}
 U_{mass}(h,H,A)& =& c_0 A + c_1 hA + c_2 HA + \\ \nonumber
&&+ \frac{m^2_h}{2} h^2 + \frac{m^2_H}{2} H^2 +\frac{m^2_A}{2} A^2
+ m^2_{H^\pm} {H^+ H^-} .
\end{eqnarray}
The minimization condition $c_0=$0 fixes the imaginary part of the
$\mu^2_{12}$ parameter
\begin{eqnarray}
\label{immu12}
{\tt Im} \mu^2_{12}&=&
\frac{v^2}{2} (  s_{\beta} c_{\beta} {\tt Im}  \lambda_5 +c^2_{\beta}
  \; {\tt Im}  \lambda_6
+s^2_{\beta} \; {\tt Im}  \lambda_7) ,
\end{eqnarray}
and the factors in front of the nondiagonal terms $hA$ and $HA$
in the local minimum $c_0=0$ have the form
\begin{eqnarray}
\label{c1c2}
c_1&=&\frac{v^2}{2}( s_{\alpha} s_{\beta}-c_{\alpha} c_{\beta})
{\tt Im} { \lambda_5}
+v^2 \, (s_{\alpha} c_{\beta} {\tt Im} { \lambda_6} -
              c_{\alpha} s_{\beta} {\tt Im} { \lambda_7}) ,
\\ \nonumber
c_2&=&-\frac{v^2}{2}( s_{\alpha} c_{\beta}+c_{\alpha} s_{\beta})
{\tt Im} { \lambda_5}
-v^2 \, (c_{\alpha} c_{\beta} {\tt Im} { \lambda_6} +
              s_{\alpha} s_{\beta} {\tt Im} { \lambda_7}) .
\nonumber
\end{eqnarray}
They include only the imaginary parts of the parameters
${\tt Im}\mu^2_{12}$, ${\tt Im}\lambda_{5,6,7}$. The nondiagonal
term $hH$ does not appear in (\ref{massterm}), so in the mixing matrix
(\ref{matrix}) $M_{12}=M_{21}=0$.

At the second stage in order to remove the nondiagonal terms $hA$ and $HA$
we perform the orthogonal transformation in the $h$, $H$, $A$ sector
\begin{eqnarray}
(h,H,A) \; M^2 \;  \left( \begin{array}{c} h\\ H\\ A
\end{array} \right)
&=& (h_1, h_2, h_3) \; a^T_{ik} \; M^2_{kl} \; a_{lj} \;
\left( \begin{array}{c} h_1 \\ h_2\\ h_3
\end{array} \right) \, ,
\end{eqnarray}
where the mass matrix is
\begin{eqnarray}
\label{matrix}
M^2 & = & \frac{1}{2} \left( \begin{array}{ccc}
m^2_h         &      0        & c_1 \\
0             &     m^2_H     & c_2 \\
 c_1          &     c_2       &     m^2_A
\end{array} \right) \, ,
\end{eqnarray}
and get the physical Higgs bosons $h_1$, $h_2$, $h_3$
without a definite $CP$ parity\footnote{Note that this picture is
different from the well-known description of 
weak $CP$ violation in meson decays, when the mass splitting $\Delta m$ of 
the states is given by 2${\tt Re}M_{12}$, $M_{12}$ the off-diagonal 
elements of the complex 2$\times$2 mass matrix, and the meson mixing 
$\epsilon$ parameter is ${\tt Im}M_{12}/(\sqrt{2} \Delta m)$.
The meson decay formalism uses 
the non-hermitian effective Hamiltonian and not precisely orthogonal mass 
'eigenstates'.}.
The eigenvalues of the $M^2$ 
matrix
define their masses squared and the components of normalized eigenvectors
are the matrix elements in the rows of the mixing matrix $a_{ij}$.
The squared masses of Higgs bosons are
($m^2_{h_1}\leq m^2_{h_2}\leq m^2_{h_3}$)
\begin{eqnarray}
m^2_{h_1} &=& 2 \sqrt{(-q)} \cos \left(\frac{ \Theta+ 2
\pi}{3}\right) - \, \frac{a_2}{3} \,\, , \label{eq:massi123} \\
\nonumber m^2_{h_2} &=& 2 \sqrt{(-q)} \cos \left(\frac{ \Theta + 4
\pi}{3}\right) - \,
 \frac{a_2}{3} \,\, ,  \\
\nonumber m^2_{h_3} &=& 2 \sqrt{(-q)} \cos \left(\frac{
\Theta}{3}\right) - \, \frac{a_2}{3} \,\, , \nonumber
\end{eqnarray}
where
\begin{eqnarray*}
\Theta = \arccos \frac{r}{\sqrt{(-q^3)}} \,\, ,&& \\
r=\frac{1}{54}(9 a_1 a_2 -27 a_0 - 2 a^3_2) \,\, ,&& \,\,\,
q=\frac{1}{9}(3 a_1-a^2_2) \,\, ,
\\
a_1=m^2_{ h} m^2_{ H}+m^2_{ h}
m^2_{ A}+m^2_{ H} m^2_{ A}-
{ c}^2_1-{ c}^2_2 \,\, , && \,\,\,
a_2=-m^2_{ h}-m^2_{ H}-m^2_{ A} \,\, ,\\
a_0={ c}^2_1 m^2_{ H}+{ c}^2_2
m^2_{ h}-m^2_{ h} m^2_{ H}
m^2_{ A} \,\, .&&
\end{eqnarray*}
The normalized eigenvector components $(h,H,A)=a_{ij} h_j$,
$a_{ij}=a^{'}_{ij}/n_j$ are given by
\begin{eqnarray*}
a^{'}_{11}=((m^2_H-m^2_{h_1})(m^2_A-m^2_{h_1})-c^2_2), \;
a^{'}_{21}=c_1 c_2, \;
a^{'}_{31}=-c_1 (m^2_H-m^2_{h_1}) \\
a^{'}_{12}=-c_1 c_2, \;
a^{'}_{22}=-((m^2_h-m^2_{h_2})(m^2_A-m^2_{h_2})-c^2_1), \;
a^{'}_{32}=c_2 (m^2_h-m^2_{h_2}), \\
a^{'}_{13}=-c_1 (m^2_H-m^2_{h_3}), \; a^{'}_{23}=-c_2
(m^2_h-m^2_{h_3}),  \;
a^{'}_{33}=(m^2_h-m^2_{h_3})(m^2_H-m^2_{h_3}),
\end{eqnarray*}
$n_i=\sqrt{(a^{'2}_{1i}+a^{'2}_{2i}+a^{'2}_{3i})}$. 
The Higgs boson masses $m_{h_1}$, $m_{h_2}$, $m_{h_3}$ and
the mixing matrix elements $a_{ij}$, which describe the mixed states,
are shown in Fig.2-4 as a function of the $A_{t,b}$, $\mu$ parameters 
and/or 
the universal phase $\varphi={\tt arg} (\mu A_{t,b})$.
Different to the figures in \cite{Dubinin02}, the $m_{H^\pm}$, ${\tt tg}\beta$
parametrization is used for the convenience of comparison with \cite{CPsuperH}
and \cite{feynh}.
The parameters $c_1$ and  $c_2$ can change a sign with the variation of the phase 
$\varphi$, the ranges of positively or negatively defined $c_1$ and  $c_2$
depend on the primary choice of the $m_{H^\pm}$, ${\tt tg}\beta$, $A$, $\mu$ and $M_{SUSY}$  
in the $CP$ conserving limit. When we pass the zeroes of $c_1$ and  $c_2$,
the matrix elements $a_{ij}$ are expected to change their signs respecting the requirement
of the left orthonormal basis for the eigenvectors. It is essential that
$m_{h_1}$, $m_{h_2}$ and $m_{h_3}$ are positioned in the mass matrix along
the diagonal from the upper left to the lower right corner, satisfying in the
limiting case $c_1=c_2=0$ the correspondences $m_{h_1} \to min(m_h,m_H,m_A)$, 
$m_{h_3} \to max(m_h,m_H,m_A) $ ("the mass ordering"). Note also that as
$\Delta \lambda_i$ increases, the denominator of (\ref{t2b}) can change sign,
so for the mass ordering one must define the angle $\alpha(\varphi)$
consistently with the boundary condition at the scale $M_{SUSY}$,
which has the known form $m^2_A+m^2_Z=-{\tt sin}2\alpha/ {\tt sin}2\beta(m^2_H-m^2_h)$,
following from (\ref{diaglambda1})-(\ref{diaglambda5}) and (\ref{eq:lMSUSY}).

Some numerical values for the Higgs boson masses $m_{h_1}$, $m_{h_2}$, 
$m_{h_3}$ as a function of the phase
$\varphi$ in our approach,
and masses of the states $H_1$, $H_2$, $H_3$ evaluated by means of 
CPsuperH  \cite{CPsuperH} and FeynHiggs \cite{feynh} packages are shown in 
the Table 2. These packages are using the 
renormalization group improved diagrammatic calculation that uncludes 
radiative corrections to Yukawa couplings up to two-loops. Detailed 
general
discussion on the conciliation of results obtained in the frameworks
of the diagrammatic and the effective field theory approaches can be found 
in \cite{conciliate}. Different renormalization schemes in which 
calculations in the two approaches are performed, may lead to the
deviations of results evaluated with parameters taken at different
renormalization scales, so the untrivial reevaluation of parameters is 
needed for consistency. 
Besides this it is important to notice that 
in the CPsuperH and FeynHiggs packages
the $SU(2)$ eigenstates $\eta_{1,2}$, $\xi_{1,2}$ 
are directly transformed 
to the Higgs boson mass eigenstates, which is different from 
our procedure, when we first transform to the states of the 
$CP$-conserving limit and then rotate to $h_{1,2,3}$. 
The 'intermediate'
Higgs boson states $(h,H,A)$ of the $CP$ conserving limit are not 
used, so the $\eta_1$, $\eta_2$ mixing angle $\alpha$ is 
not introduced there. 
For this reason at
$\varphi=$0 the analogue of the mixing matrix $a_{ij}$, see (44),
has nonzero off-diagonal matrix elements $a_{12}=a_{21} \neq$0 and
in the analogue of the mass matrix (47) $m_{12}$ and $m_{21}$ (the $hH$
mixing terms in our notation) are also nonzero. In the framework
of the 'direct' diagonalization procedure the matrix elements of
(45) have the form
\begin{eqnarray*}
m_{11} &=& m^2_A \, s^2_{\beta} + v^2 {\tt Re} \lambda_5 \, s^2_{\beta}
          + v^2 {\tt Re} \lambda_6 \, s_{2\beta} + 2 v^2 
\lambda_1 c^2_{\beta}, \\
m_{22} &=& m^2_A \, c^2_{\beta} + v^2 {\tt Re} \lambda_5 \, c^2_{\beta}
          + v^2 {\tt Re} \lambda_7 \, s_{2\beta} + 2 v^2
\lambda_2 s^2_{\beta}, \\
m_{12} &=& v^2 {\tt Re} \lambda_6 \, c^2_{\beta} +
           s_{\beta}\, (v^2 {\tt Re} \lambda_7 s_{\beta} + c_{\beta}
           \, (-m^2_A + v^2 \lambda_3 + v^2 \lambda_4)), \\
m_{13} &=& -\frac{1}{2} v^2 (2 \, {\tt Im} \lambda_6 c_{\beta}
                + {\tt Im} \lambda_5 s_{\beta}), \\
m_{23} &=& -\frac{1}{2} v^2 ( {\tt Im} \lambda_5 c_{\beta}
                + 2 \, {\tt Im} \lambda_7 s_{\beta}), \\
m_{33} &=& m^2_A
\end{eqnarray*}
and the parameters $a_0$, $a_1$, $a_2$ in (46) should be redefined as 
follows
\begin{eqnarray*}
a_0 &=& m^2_{12} \, m_{33} + m^2_{23} \, m_{11} + m^2_{13} m_{22}
         - 2 \, m_{12} \, m_{23} \, m_{13} 
         - m_{11} \, m_{22} \, m_{33}, \\
a_1 &=& m_{11} \, m_{22} + m_{11} \, m_{33} + m_{22} \, m_{33}
         - m^2_{12} - m^2_{13} - m^2_{23}, \\
a_2 &=& - \, m_{11} - \, m_{22} - \, m_{33}
\end{eqnarray*}
We checked that both the 'two-step' and the 'direct' diagonalization
methods lead within our procedure, as expected, to the same
masses of Higgs states $m_{h_1}$, $m_{h_2}$, $m_{h_3}$ (see Table 2).
For the parameter values in the comparison, Table 2, the benchmark point 
of the maximal CP violation 'CPX scenario' \cite{cpx} at $M_{SUSY}=$500 
GeV was used.
Extended list of numbers (Table 5) including also the rare one-loop 
mediated decay 
widths 
$h_1\rightarrow \gamma \gamma$, $h_1\rightarrow g g$
and the tree-level two-particle decays $h_1\rightarrow f \bar f$ 
can be found in the Appendix. Good qualitative 
agreement of results is observed, but diversity of
approaches to the calculation of radiative corrections makes precise 
numerical comparisons difficult.  
\begin{table}[t]
\begin{center}
\begin{tabular}{|c||c|c|c|c|c|c|c|}
\hline
   & $\varphi=0$ & $\pi/6$ & $\pi/3$ & $\pi/2$ & $2\pi/3$ &
$5\pi/6$ & $\pi$ \\
\hline\hline
$m_{h_1}$ & 115.4 & 118.7 & 125.9 & 131.4 & 130.7 & 125.2 & 122.0 \\
$m_{H_1}$ \cite{CPsuperH} & 106.8 & 109.0 & 113.9 & 117.4 & 114.9 & 105.7
& 99.4 \\
$m_{H_1}$ \cite{feynh} & 115.8 & 118.8 & 125.5 & 130.2 & 123.2 & 98.2
& 78.0 \\
\hline
$m_{h_2}$ & 295.5 & 289.6 & 279.7 & 269.3 & 262.2 & 259.8 & 259.6 \\
$m_{H_2}$ \cite{CPsuperH} & 302.2 & 297.8 & 290.9 & 282.2 & 273.9 & 268.3
& 264.4 \\
$m_{H_2}$ \cite{feynh} & 295.6 & 290.0 & 279.1 & 264.3 & 249.2 & 239.7
& 236.9 \\
\hline
$m_{h_3}$ & 297.1 & 299.5 & 300.4 & 299.9 & 298.8 & 297.6 & 297.1 \\
$m_{H_3}$ \cite{CPsuperH} & 302.3 & 304.4 & 305.0 & 304.5 & 303.5 & 302.4
& 302.0 \\
$m_{H_3}$ \cite{feynh} & 297.6 & 300.0 & 301.1 & 301.3 & 300.9 & 300.4
& 300.2 \\
\hline \hline
\end{tabular}
\end{center}
\label{tab:MandD} \caption{The Higgs boson masses (GeV) 
in our case and calculated by the packages CPsuperH \cite{CPsuperH}
and FeynHiggs \cite{feynh} (in the one-loop regime)
at the same parameter values $\alpha_{EM}(m_Z)=0.7812\cdot$10$^{-2}$,
$\alpha_S(m_Z)=0.1172$, $G_F=1.174\cdot10^{-5}$\,GeV$^{-2}$,
$\tan\beta=5$, $M_{SUSY}=500$\,GeV, $|A_t|=|A_b|=A$,
$|\mu|=2000$\,GeV, $A=1000$\,GeV, $m_{H^{\pm}}=300$\,GeV.}
\end{table}

\subsection{Real $\mu^2_{12}$, $\lambda_{5,6,7}$ parameters,
$\theta\neq$0}

If the parameters $\mu_{12}^2$, $\lambda_{5,6,7}$ of the effective
potential (\ref{eq:genU}) are real, the latter is $CP$ invariant.
It is easy to show 
\cite{PilaftsisWagner, Dubinin02, GunionHaber},
that the phases of complex parameters $\mu_{12}^2$, $\lambda_{5,6,7}$
can be rotated away by the $U(1)_Y$ hypercharge transformation if
the conditions
\begin{eqnarray}
\label{conditions}
&{\tt Im}({\mu}^4_{12} \stackrel{*}{ \lambda_5})=0, \quad
{\tt Im}({\mu}^2_{12} \stackrel{*}{ \lambda_6})=0, \quad
{\tt Im}({\mu}^2_{12} \stackrel{*}{ \lambda_7})=0.&
\end{eqnarray}
are satisfied. Insofar as the physical motivation of these 'fine tuning'
conditions is not available, the case of real parameters and nonzero
phase $\theta$ of the VEV, when $CP$ is broken spontaneously, looks rather artificial.
The local minimum of the effective potential (\ref{eq:genU}) occurs
at $\lambda_{\,5}> 0$ (i.e. purely imaginary $\mu A$, see (\ref{eq:lambda5}))
and 
\beq \cos\theta = \ff{\mu_{12}^2-\ff{v_1^2}{2}\lambda_6-
\ff{v_2^2}{2}\lambda_7}{\lambda_{\,5}v_1v_2} \, . \label{eq:Recos}
\eeq 
Combining this equation with the diagonalization condition
(\ref{diaglambda5}) we get
\beq
 \cos\theta = \ff{m_{A}^2}{\lambda_{\,5}v^2} + 1 \, ,
\label{eq:Reco2}
\eeq
so there is no minimum if $m_A^2>0$. In the case $\lambda_{\,5}< 0$
(\ref{eq:Recos}) corresponds to the maximum, the absolute minimum
is achieved at the endpoints $\cos\theta=\pm 1$. For example,
the absolute minimum at $\theta=$0 (taking into account again the
diagonalization condition (\ref{diaglambda5})) is absent if
\beq
m_A^2>2|\lambda_{\,5}|v^2 .
\eeq
and it follows that for the case of real $\mu_{12}^2$, $\lambda_{5,6,7}$
and $CP$ broken spontaneously there are no mass eigenstates in the 
framework of our diagonalization procedure, at least if $m_A$ is not
extremely small.

\subsection{Complex $\mu^2_{12}$, $\lambda_{5,6,7}$ parameters,
$\theta\neq$0}

In the case of complex parameters and the nonzero phase of 
$\Phi_2$ vacuum expectation value
\footnote{The upper component of $\langle \Phi_2 \rangle$ in (\ref{eq:vev}) is taken to 
be zero.
Otherwise additional constraint for the VEV components should be imposed to ensure
the existence of the massless gauge field (photon) \cite{Veltman}}
, the $CP$
invariance of the potential is broken both explicitly and spontaneously.
The condition to set to zero the derivative $\partial U/ \partial \theta$
includes both the real and the imaginary parts of $\mu_{12}^2$ and
$\lambda_{5,6,7}$:
\beq \cos\theta ( 2 {\tt Im}\mu_{12}^2 - v_1^2
{\tt Im}\lambda_6 - v_2^2 {\tt Im}\lambda_7 )
  - v_1v_2 {\tt Im}\lambda_{\,5} \cos2\theta + \label{eq:comcos} \eeq
$$ + \sin\theta ( 2 {\tt Re}\mu_{12}^2 - v_1^2
{\tt Re}\lambda_6 - v_2^2 {\tt Re}\lambda_7 ) - v_1v_2
{\tt Re}\lambda_{\,5} \sin2\theta = 0 . $$ 
The condition of the extremum for ${\tt Im}\mu_{12}^2$
depends on the phase between the VEV's $\theta$, while
the diagonalization condition for ${\tt Re}\mu_{12}^2$
depends also on the relative phase $\xi$ (see (\ref{eq:Phi2}), (\ref{eq:vev}))
of the $SU(2)$ doublets. At the real $\mu_{12}^2$, $\lambda_{5,6,7}$
and $\theta\not=0$ the equation (\ref{eq:comcos}) is reduced to
(\ref{eq:Recos}).

For convenience we present the extremum conditions 
$\partial U/ \partial \eta=0$,
$\partial U/ \partial \xi=0$ in the cases of zero and nonzero $\theta$
in the form of Tables 3 and 4, where the factors in front of the potential
parameters are shown. Bulky condition for the real part of $\mu^2_{12}$ to define the 
pseudoscalar mass $m_A$
for the general case of nonsero phases can explicitly be evaluated as 
follows:
\begin{table}[ht!]
\begin{center}
\begin{tabular}{|c||c|c||c|c|}
\hline
& \multicolumn{2}{c||}{ } & \multicolumn{2}{c|}{ } \\[-3mm]
 & \multicolumn{2}{c||}{$\mu_1^2$} & \multicolumn{2}{c|}{$\mu_2^2$}
\\ \cline{2-5}
& & & & \\[-3mm]
 & $\theta \not= 0$ & $\theta = 0$ & $\theta \not= 0$ & $\theta = 0$
 \\ \hline\hline
 & & & & \\[-3mm]
$\lambda_1$ &  $v_1^2$   &   $v_1^2$     & 0         & 0    \\[1mm] \hline
 & & & & \\[-3mm]
$\lambda_2$ &    0       &    0   & $v_2^2$   &    $v_2^2$     \\[1mm] \hline
 & & & & \\[-3mm]
$\lambda_3$ & $\frac{1}{2}v_2^2$  &  $\frac{1}{2}v_2^2$
& $\ff{1}{2}v_1^2$ & $\ff{1}{2}v_1^2$  \\[1mm] \hline
 & & & & \\[-3mm]
$\lambda_4$ &  $\ff{1}{2}v_2^2$ &  $\ff{1}{2}v_2^2$      & $\ff{1}{2}v_1^2$
 & $\ff{1}{2}v_1^2$  \\[1mm] \hline
 & & & & \\[-3mm]
${\tt Re} \lambda_{\,5}$ & $\ff{1}{2}v_2^2$ & $\ff{1}{2}v_2^2$
& $\ff{1}{2}v_1^2$   & $\ff{1}{2}v_1^2$  \\[1mm] \hline
 & & & & \\[-3mm]
${\tt Im} \lambda_{\,5}$ & $-\ff{1}{2}v_2^2 \mbox{tg}\theta$ &  0
&   $-\ff{1}{2}v_1^2 \mbox{tg}\theta$ & 0 \\[1mm] \hline
 & & & & \\[-3mm]
${\tt Re} \lambda_6$ & $\ff{1}{2}v_1 v_2 (2+\cos 2\theta)\sec\theta$ &
$\ff{3}{2}v_1 v_2$ & $\ff{1}{2}v_1^2\sec\theta \mbox{ctg}\beta$
 & $\ff{1}{2}v_1^2 \mbox{ctg}\beta$ \\[1mm] \hline
${\tt Im} \lambda_6$ & $-v_1 v_2 \sin\theta$  & 0 & 0 & 0 \\ \hline
 & & & & \\[-3mm]
${\tt Re} \lambda_7$ & $\ff{1}{2}v_2^2\sec\theta \mbox{tg}\beta$ &
$\ff{1}{2}v_2^2\mbox{tg}\beta$
& $\ff{1}{2}v_1 v_2 (2+\cos 2\theta)\sec\theta$  &
 $\ff{3}{2} v_1 v_2$ \\[1mm] \hline
${\tt Im} \lambda_7$ & 0 & 0 & $-v_1 v_2 \sin\theta$ & 0 \\ \hline
 & & & & \\[-3mm]
${\tt Re} \mu_{12}^2$ & $- \mbox{tg}\beta \sec\theta$ & $- \mbox{tg}\beta$ &
 $- \mbox{ctg}\beta \sec\theta$ & $- \mbox{ctg}\beta$ \\[1mm] \hline
\end{tabular}
\end{center}
\caption{The factors of the extremum conditions for $\mu_1^2$ и $\mu_2^2$ at
zero and nonzero $\theta$.}
\end{table}
\begin{table}[ht!]
\begin{center}
\begin{tabular}{|c||c||c|c|}
\hline
& & \multicolumn{2}{c|}{ } \\[-3mm]
 & ${\tt Re} \mu_{12}^2$ &
  \multicolumn{2}{c|}{${\tt Im} \mu_{12}^2$}\\  \cline{2-4}
& & & \\[-3mm]
& $\theta=0$ и $\xi = 0$
& $\theta \not= 0$ & $\theta = 0$   \\ \hline\hline $\lambda_1$ &
0    &  0  & 0    \\ \hline $\lambda_2$ &  0    & 0   & 0    \\
\hline $\lambda_3$ & 0 & 0 & 0 \\ \hline $\lambda_4$ & 0 & 0 & 0
\\ \hline ${\tt Re} \lambda_{\,5}$ & $v_1 v_2$ & $v_1 v_2 \sin\theta$
& 0 \\ \hline
 & & & \\[-3mm]
${\tt Im} \lambda_{\,5}$ & 0 & $\ff{1}{2}v_1 v_2 \cos 2\theta
\sec\theta$ &
$\ff{1}{2}v_1 v_2$ \\[1mm] \hline
 & & & \\[-3mm]
${\tt Re} \lambda_6$ & $\ff{1}{2}v_1^2$ & $\ff{1}{2}v_1^2 \mbox{tg}\theta$ &
0 \\[1mm] \hline
 & & & \\[-3mm]
${\tt Im} \lambda_6$ & 0  & $\ff{1}{2}v_1^2$ & $\ff{1}{2}v_1^2$\\[1mm] \hline
 & & & \\[-3mm]
${\tt Re} \lambda_7$ & $\ff{1}{2}v_2^2$  & $\ff{1}{2}v_2^2 \mbox{tg}\theta$ &
0 \\[1mm] \hline
 & & & \\[-3mm]
${\tt Im} \lambda_7$ & 0  & $\ff{1}{2}v_2^2$   &
 $\ff{1}{2}v_2^2$  \\[1mm] \hline
 & & & \\[-3mm]
$m_A^2$ & $\sin\beta\cos\beta$ & 0 & 0 \\ \hline
 & & & \\[-3mm]
${\tt Re} \mu_{12}^2$ & -- & $ - \mbox{tg}\theta$ & 0 \\ \hline
\end{tabular}
\end{center}
\caption{The factors of the extremum condition for ${\tt Re} \mu_{12}^2$ at
$\theta=0$ and for ${\tt Im} \mu_{12}^2$ for zero and nonzero $\theta$.
}
\end{table}
\beq
{\tt Re} \mu_{12}^2 = \label{eq:Remu12} \eeq
$$ = - \lambda_2 \, \ff{v^2\cos\theta\sin^3(2\beta)\sin^2(\theta+\xi))}
{3+(1-\cos\theta\cos\xi)(\cos^4\beta-\ff{3}{2}\sin^2(2\beta))+\sin^4\beta+
\cos\theta\cos\xi(1-\sin^4\beta)}\,+ $$
$$ + \, {\tt Re}\lambda_{\,5} \, \ff{v^2(\cos^4\beta\cos^2\xi+\cos^2\theta\sin^4\beta
+\cos\beta\cos(\theta-\xi)\sin\beta\sin(2\beta)}
{\cos^2\beta\mbox{ctg}\beta\sec\theta+\cos\xi\sin(2\beta)+
\sec\theta\sin^2\beta\mbox{tg}\beta} \, - $$
$$ - \, {\tt Im}\lambda_{\,5} \, \ff{v^2(\sin^2(2\beta)\sin(\theta-\xi)+
\sin^4\beta(\sin(2\theta)+\mbox{tg}\theta)+
\cos^4\beta(\mbox{tg}\theta-\sin(2\xi))}
{2(\cos^2\beta\mbox{ctg}\beta\sec\theta+\cos\xi\sin(2\beta)+
\sec\theta\sin^2\beta\mbox{tg}\beta)} + $$
$$ + \,{\tt Re}\lambda_6 \, \ff{1}{2} v^2 \cos^2\beta \, + $$
$$ + \,{\tt Im}\lambda_6 \, \ff{v^2\cos^3\beta\sin\beta\sin\xi}
{\cos^2\beta\mbox{ctg}\beta\sec\theta+\cos\xi\sin(2\beta)+
\sec\theta\sin^2\beta\mbox{tg}\beta} + $$
$$ + {\tt Re}\lambda_7 \,
 \left(\ff{v^2\cos^4\beta(4\cos(\theta+2\xi)-
2\cos(2\theta)\sec\theta)\mbox{tg}\beta}
{4(\cos^2\beta\mbox{ctg}\beta\sec\theta+\cos\xi\sin(2\beta)+
\sec\theta\sin^2\beta\mbox{tg}\beta)} + \right.$$
$$ \left. + \ff{v^2(2\sin^2(2\beta)\cos\xi+2\sec\theta\sin^4\beta-
\cos(2\theta+\xi)\sin^2(2\beta))\mbox{tg}\beta}
{4(\cos^2\beta\mbox{ctg}\beta\sec\theta+\cos\xi\sin(2\beta)+
\sec\theta\sin^2\beta\mbox{tg}\beta)}\right) + $$
$$ + {\tt Im}\lambda_7 \, \ff{v^2\sin(2\beta)(2\cos^2\beta\cos\xi
\sin(\theta+\xi)+\sin^2\beta(2\sin\xi+\sin(2\theta+\xi)))}
{2(\cos^2\beta\mbox{ctg}\beta\sec\theta+\cos\xi\sin(2\beta)+
\sec\theta\sin^2\beta\mbox{tg}\beta)} \, - $$
$$ - \, {\tt Im}\mu_{12}^2 \, \ff{\sin(2\beta)\sin\xi}
{\cos^2\beta\mbox{ctg}\beta\sec\theta+\cos\xi\sin(2\beta)+
\sec\theta\sin^2\beta\mbox{tg}\beta} + $$
$$ + m_A^2 \,
\ff{1}{\cos^2\beta\mbox{ctg}\beta\sec\theta+
\cos\xi\sin(2\beta)+\sec\theta\sin^2\beta\mbox{tg}\beta} \, . $$

If we set $\theta = 0$ and $\xi=0$, the formulas coincide with 
the special case of only the explicit $CP$ violation (\ref{diaglambda5}), (\ref{immu12}).
The substitution of the extremum conditions corresponding to
Tables 3 and 4 to (\ref{eq:comcos}) gives an identity independently
on the expression (\ref{eq:Remu12}) for $\mbox{\tt Re}\mu_{12}^2$.
The extremum is a minimum if the second derivative in $\theta$
is positively defined
\beq -
\sin\theta ( 2 {\tt Im}\mu_{12}^2 - v_1^2 {\tt Im}\lambda_6 -
v_2^2 {\tt Im}\lambda_7 )
  + 2v_1v_2 {\tt Im}\lambda_{\,5} \sin2\theta + \label{eq:extmin} \eeq
$$ + \cos\theta ( 2 {\tt Re}\mu_{12}^2 - v_1^2
{\tt Re}\lambda_6 - v_2^2 {\tt Re}\lambda_7 ) - 2v_1v_2
{\tt Re}\lambda_{\,5} \cos2\theta \, > \, 0 . $$
Numerical investigation shows that this condition is fullfilled
in a rather wide range of the MSSM parameter space. If for simplicity we 
set $\xi=$0 then the second
derivative is positively defined in any region of the parameter space,
so no restrictions on the phase of spontaneous $CP$ breaking appear in 
this special case from the minimization.

The diagonalization of the effective potential mass term
in the local minimum for the general case $\theta\not=0$
and $\xi\not=0$ is performed analogously to the procedure
described in section 3.1 using the following scheme:
(1) we define the four $\widetilde h$, $\widetilde H$, $\widetilde A$,
${\widetilde G}^0$ linear combinations of independent fields
$\eta_1$, $\eta_2$, $\chi_1$, $\chi_2$ that are contained in the 
two-doublet system (\ref{eq:Phi1}),~(\ref{eq:Phi2}), where
for the Goldstone field ${\widetilde G}^0$ we define zero row
of matrix elements and zero column of matrix elements in the symmetric mass
matrix 4$\times$4. In other words, the Goldstone mode is introduced
as the linear combination, orthogonal to the plane defined by
the "directions" in the complex scalar fields space, parallel to
the VEV's $v_1$ and $v_2\exp \, \{i(\xi+\zeta)\}$. Then the mass
matrix 4$\times$4 includes the symmetric 3$\times$3 block with zero
matrix elements in the power of the extremum conditions
from Tables 3 and 4; (2) we define an orthogonal transformation
for the 3$\times$3 submatrix fixing the mixing angle $\widetilde\alpha$
in the sector  $\widetilde h-\widetilde H$ to set to zero the
$\widetilde h \widetilde H$ nondiagonal term. In the framework of this
procedure for the case of nonzero phases $\xi \neq$0, $\theta \neq 0$ 
(when the fields are denoted by the symbol " $\; \widetilde { } \;$")
the limiting cases of zero phases $\xi=\theta=$0 (when the notation
for the fields does not contain the symbol " $\; \widetilde { } \;$")
and also the $CP$ conserving limit in the mass basis $h$, $H$, $A$,
are clearly seen.      
For the physical Higgs fields in the case {$\xi=0$, $\theta\not=0$}
we finally obtain the representation
\begin{eqnarray} 
\widetilde h &=&
-\eta_1\sin\widetilde\alpha +
(\chi_2\sin\theta+\eta_2\cos\theta)\cos\widetilde\alpha , \\ \nonumber
\widetilde H &=& \eta_1\cos\widetilde\alpha +
(\chi_2\sin\theta+\eta_2\cos\theta)\sin\widetilde\alpha , \\ \nonumber
\widetilde A &=& -\chi_1\sin\beta +
(\chi_2\cos\theta-\eta_2\sin\theta)\cos\beta , \\ \nonumber
{\widetilde G}^0 &=& \chi_1\cos\beta +
(\chi_2\cos\theta-\chi_2\sin\theta)\sin\beta . 
\end{eqnarray}
We checked explicitly, using the symbolic calculation packages, that 
direct substitution of these
fields to the potential (\ref{eq:genU}) gives the symmetric
4$\times$4 squared mass matrix with zero row and column,
corresponding to the Goldstone mode. The non-diagonal matrix elements
of the 3$\times$3 block, corresponding to the nondiagonal terms
$\widetilde h\widetilde A$ и $\widetilde H\widetilde A$ in the
local minimum, can be written in the form 
\begin{equation}
{\widetilde c}_1 =  - \frac{v^2}{2}(
\cos(\widetilde\alpha+\beta) \cos(2\theta){\tt Im} \,
{\lambda_{\,5}} - 2 \sin\widetilde\alpha \cos\beta \cos\theta {\tt
Im} \, {\lambda_6} + 2 \cos\widetilde\alpha \sin\beta \cos\theta
{\tt Im} \, {\lambda_7} - \eeq
$$ - \cos(\widetilde\alpha+\beta) \sin(2\theta) {\tt Re} \,
{\lambda_{\,5}} -  2 \sin\widetilde\alpha \cos\beta \sin\theta
{\tt Re} \, {\lambda_6} + 2 \cos\widetilde\alpha \sin\beta
\sin\theta{\tt Re} \, {\lambda_7}) , $$ \beq {\widetilde c}_2 =  -
\frac{v^2}{2}( \sin(\widetilde\alpha+\beta) \cos(2\theta){\tt Im}
\, {\lambda_{\,5}} - 2 \cos\widetilde\alpha \cos\beta \cos\theta
{\tt Im} \, {\lambda_6} + 2 \sin\widetilde\alpha \sin\beta
\cos\theta {\tt Im} \, {\lambda_7} + \eeq $$+
\cos(\widetilde\alpha+\beta) \sin(2\theta) {\tt Re} \,
{\lambda_{\,5}} -  2 \cos\widetilde\alpha \cos\beta \sin\theta{\tt
Re} \, {\lambda_6} + 2 \sin\widetilde\alpha \sin\beta
\sin\theta{\tt Re} \, {\lambda_7}) . $$
In the case $\theta=0$ they coincide with (\ref{c1c2}).

The same scheme is suitable for 
the case $\xi \neq$0, $\theta \neq$0 
when the relative phase $\xi$ between the $SU(2)$ doublets
appears in the mass eigenstates, which are obtained by the
replacement $\theta$ $\rightarrow$ $\theta-\xi$: 
\begin{eqnarray}
\widetilde h &=& -\eta_1\sin\widetilde\alpha +
(\chi_2\sin(\theta-\xi)+\eta_2\cos(\theta-\xi))\cos\widetilde\alpha
, \\  \nonumber
\widetilde H &=& \eta_1\cos\widetilde\alpha +
(\chi_2\sin(\theta-\xi)+\eta_2\cos(\theta-\xi))\sin\widetilde\alpha
, \\ \nonumber
\widetilde A &=& -\chi_1\sin\beta +
(\chi_2\cos(\theta-\xi)-\eta_2\sin(\theta-\xi))\cos\beta , \\ \nonumber
{\widetilde G}^0 &=& \chi_1\cos\beta +
(\chi_2\cos(\theta-\xi)-\chi_2\sin(\theta-\xi))\sin\beta . 
\end{eqnarray}

\section{Summary}

The potential of a two-Higgs-doublet model
in the general case is not $CP$ invariant and the parameters
$\mu^2_{12}$ and $\lambda_{5,6,7}$ of the two-doublet MSSM
Higgs sector should be taken complex. The choice of purely
real parameters implicitly assumes that the fine-tuning conditions
(\ref{conditions}) are additionally imposed without clear physical
motivation. In the MSSM the complex parameters naturally appear
if we allow the $CP$ invariance violating mixings in the squark-Higgs boson
sector of the MSSM, analogous to the CKM mixings for the three quark
generations in the charged current sector of the Standard Model.
If these mixings lead to a strong $CP$ parity violation\footnote{Recent
discussion of the weak $CP$ violation scenarios can be found in 
\cite{GinzV}.}
and the scalar sector of the MSSM is coupled strongly enough (i.e.
large imaginary parts of the parameters $\mu^2_{12}$ and $\lambda_{5,6,7}$
appear), the deviations of the observable effects in the scenario with $CP$ violation
from the phenomenology of the standard scenario can be substantial.
The deviations are particularly strong if the power terms 
$A_{t,b}/M_{SUSY}$, $\mu / M_{SUSY}$ are large and the charged Higgs boson
mass does not exceed 150-200 GeV, being rahter weakly dependent on the 
value of ${\tt tg}\beta$. 
Such models could lead in principle to a reconsideration of the 
experimental
priorities \cite{cmsnote} for the signals of Higgs bosons production
in the channels $\gamma \gamma$, $b \bar b$, $W^+ W^-$, $ZZ$, $ttH$, $bbH$ 
etc. at the LHC. The scenario with light Higgs boson $m_{h_1}\sim70-80$ 
GeV that could escape the detection at LEP2 \cite{LEP2}, the analysis
of $h_1$ signal at Tevatron and the high-luminosity linear colliders 
\cite{colliders} demonstrate that physical possibilities in the 
framework of $CP$ violating
scenarios could be considerably modified in comparison with the 
traditional $CP$ conserving limit.

The comparison of our results for the masses of scalars $m_{h_1}$,
$m_{h_2}$ and $m_{h_3}$ and their two-particle decay widths with
outputs of the CPsuperH \cite{CPsuperH} and the FeynHiggs 
\cite{feynh} packages demonstrates rather good qualitative agreement.
However, is some cases high
sensitivity of the observables to the magnitude of radiatively induced 
correction terms in the effective two-Higgs-doublet potential shows up,
so careful complementary analysis of the theoretical uncertainties is 
appropriate.

The relative phase of the $SU(2)$ scalar doublet $\zeta$
and the VEV phase $\xi$ (\ref{eq:vev}) could be constrained
on the basis of the conditions for the mass term diagonalization 
and the potential minimization (Section 3.3). In principle these 
conditions could lead
to some nontrivial relations between the $\zeta$, $\xi$ and the variables
of the MSSM parameter space. However, at the first sight it is questionable
to expect some direct relations of this type connecting the CKM phase
and the $\zeta$, $\xi$ phases of the THDM, which seem to describe the $CP$ violation of
different origin. Returning to the notations of the Introduction, we
can write the THDM type II Yukawa term as
\begin{eqnarray}
- \, L&=&\eta^u_{ij} \,{\bar{\psi}}^{i\prime}_L u^{j\prime}_R \Phi_1+
    \xi^d_{ij} \,{\bar{\psi}}^{i\prime}_L
d^{j\prime}_R\tilde{\Phi_2}+\mbox{h.c.},
\label{yukawaterms2}
\end{eqnarray}
where $\eta_{ij}^{u}$ и $\xi_{ij}^{d}$ "--- nondiagonal
complex $3\times 3$ matrices ($i,j=1,2,3$). As mentioned in the
Introduction, in order to define the quark fields mass eigenstates
the untary mixing matrix $V_{u^i,d^j}$ should be introduced in
the Lagrangian terms of the charged Higgs boson interaction with quarks
\beq
\ff{M_d\mbox{tg}\beta}{\sqrt{2}v}\,{\overline u^i_L}
V_{u^i,d^j}{d^j_R} H^{\,+} \, + \,
\ff{M_u}{\sqrt{2}v\mbox{tg}\beta}\,{\overline d^i_L}
V_{u^i,d^j}^\dagger {u^j_R} H^{\,-} \, . \eeq 
If we extract the universal phase factor from the mixing matrix
elements $V_{u^i,d^j} \rightarrow
e^{i \varphi} \left|V_{u^i,d^j}\right|$, $V_{u^i,d^j}^\dagger
\rightarrow e^{-i \varphi} \left|V_{u^i,d^j}\right|$, the Yukawa
interaction terms take the form
\beq
\ff{M_d\mbox{tg}\beta}{\sqrt{2}v}\,{\overline u^i_L}  e^{i
\varphi} \left| V_{u^i,d^j} \right|{d^j_R} H^{\,+} \, + \,
\ff{M_u}{\sqrt{2}v\mbox{tg}\beta}\,{\overline d^i_L} e^{-i
\varphi} \left| V_{u^i,d^j} \right| {u^j_R} H^{\,-}
,\label{eq:Yuk1} \eeq 
so we can identify the universal phase $\varphi$ as the relative
phase $\xi$ of the $SU(2)$ doublets. The structure of this sort, however,
does not look like the weak charged current sector mixing matrix,
where the universal complex factor is not suitable to
describe the effects of $CP$ violation in meson decays. 
\newpage
\vspace{3mm}
\begin{center}
{\large \bf Acknowledgments}
\end{center}
\vspace{3mm}
M.D. (MSU) is grateful to S.~Heinemeyer and J.S.~Lee for useful 
discussions. He also thanks very much
A.~Semenov for help with LanHEP calculations.
E. Akhmetzyanova thanks
the "Dynasty" foundation and ICPPM for partial financial support.
The work of M.Dolgopolov and M.Dubinin was partially supported by
RFBR grant 04-02-17448. The work of M.Dubinin was partially supported by
INTAS 03-51-4007, UR 02.03.028 and NS 1685.2003.2. 

\newpage

\section*{Appendix}

The decay width $h_i \rightarrow \gamma\gamma$ can be written as
\begin{eqnarray}
\Gamma(h_i\rightarrow
\gamma\gamma)=\frac{M_{h_i}^3\alpha^2}{256\pi^3\,v^2}
         \left[\,\left|S^\gamma_i(M_{h_i})\right|^2
              +\left|P^\gamma_i(M_{h_i})\right|^2\right]\,,
\end{eqnarray}
where the scalar and the pseudoscalar factors are given by 
\cite{CPsuperH,GGN}
\begin{eqnarray}
S^\gamma_i(M_{h_i})&=&2\sum_{f=b,t,\tilde{\chi}^\pm_1,\tilde{\chi}^\pm_2}
N_C\, Q_f^2\,
g^{S}_{h_i\bar{f}f}\,\frac{v}{m_f} F_{sf}(\tau_{if}) \nonumber \\
&& -
\sum_{\tilde{f}_j=\tilde{t}_1,\tilde{t}_2,\tilde{b}_1,\tilde{b}_2,
           \tilde{\tau}_1,\tilde{\tau}_2}
N_C\, Q_f^2g_{h_i\tilde{f}^*_j\tilde{f}_j}
\frac{v^2}{2m_{\tilde{f}_j}^2} F_0(\tau_{i\tilde{f}_j})
\nonumber \\
&&- g_{_{h_iVV}}F_1(\tau_{iW})- g_{_{h_iH^+H^-}}\frac{v}{2
M_{H^\pm}^2} F_0(\tau_{iH^\pm})
\,, \nonumber \\
P^\gamma_i(M_{h_i})&=&2\sum_{f=b,t,\tilde{\chi}^\pm_1,\tilde{\chi}^\pm_2}
N_C\,Q_f^2\,g^{P}_{h_i\bar{f}f} \,\frac{v}{m_f} F_{pf}(\tau_{if})
 \,.
\end{eqnarray}
$\tau_{ix}=M_{h_i}^2/4m_x^2$, $N_C=3$ for squarks and
$N_C=1$ for stau and chargino, respectively. The vertex factors
$g_{h_i f \bar f}$ can be easily extracted from Table 6, where
we list also the triple vertices with $h_i$ and gauge bosons. 
The threshold corrections induced by the exchanges of gluinos
and charginos \cite{LEP2,thresh} are not included in the following 
calculation.

The factors $F_{sf}$, $F_{pf}$, $F_0$ и $F_1$ \cite{factors} are expressed 
by means of the 
dimensionless function $f(\tau)$
\begin{eqnarray}
F_{sf}(\tau)&=&\tau^{-1}\,[1+(1-\tau^{-1}) f(\tau)]\,,~~
F_{pf}(\tau)=\tau^{-1}\,f(\tau)\,,\\
F_0(\tau)&=&\tau^{-1}\,[-1+\tau^{-1}f(\tau)]\,, \hspace{1.2 cm}
F_1(\tau)=2+3\tau^{-1}+3\tau^{-1} (2-\tau^{-1} )f(\tau)
\,,\nonumber \label{formfactor}
\end{eqnarray}
with an integral repesentation
\begin{eqnarray}
f(\tau)=-\frac{1}{2}\int_0^1\frac{{\rm d}y}{y}\ln\left[1-4\tau
y(1-y)\right]
       =\left\{\begin{array}{cl}
           {\rm arcsin}^2(\sqrt{\tau}) \,:   & \qquad \tau\leq 1\,, \\
   -\frac{1}{4}\left[\ln \left(\frac{\sqrt{\tau}+\sqrt{\tau-1}}{
                                     \sqrt{\tau}-\sqrt{\tau-1}}\right)
                    -i\pi\right]^2\,: & \qquad \tau\geq 1\,.
\end{array}\right.
\end{eqnarray}
QCD corrections in the large mass limit can be found in \cite{SDGZ}
\begin{equation}
J^\gamma_q=1-\frac{\alpha_s(M_{h_i}^2)}{\pi}\,, \hspace{1 cm}
J^\gamma_{\tilde{q}}=1+\frac{8\alpha_s(M_{h_i}^2)}{3\pi}\,.
\end{equation}
Chargino contributions depend on the couplings
\begin{eqnarray}
g_{h_1\tilde{\chi}^+_1\tilde{\chi}^-_1}^{S}&=&V_{11}U_{12}\,GS_1+V_{12}U_{11}\,GS_2\,,
\nonumber \\
g_{h_1\tilde{\chi}^+_1\tilde{\chi}^-_1}^{P} &=&
V_{11}U_{12}\,GP_1+V_{12}U_{11}\,GP_2\,,
\end{eqnarray}
\begin{eqnarray}
g_{h_1\tilde{\chi}^+_2\tilde{\chi}^-_2}^{S}&=&
V_{21}U_{22}\,GS_1+V_{22}U_{21}\,GS_2\,,
\nonumber \\
g_{h_1\tilde{\chi}^+_2\tilde{\chi}^-_2}^{P}
&=&V_{21}U_{22}\,GP_1+V_{22}U_{21}\,GP_2\,,
\end{eqnarray}
for $h_1$ we have $GS_1=-\sin\alpha\,a_{11} + \cos\alpha\,a_{21}$,
$GS_2=\cos\alpha\,a_{11}+ \sin\alpha\,a_{21}$, $GP_1=\sin\beta\, a_{31}$,
$GP_2=\cos\beta\,a_{31}$, and the matrix elements $U_{ij}$
\begin{eqnarray}
  & & U_{12} = U_{21} = \frac{1}{\sqrt{2}}\,
    \sqrt{1 + \frac{M^2_2 - \mu^2 - 2\,m_W^2\cos 2\beta}{W}} \\[2mm]
  & & U_{22} = -U_{11} = \frac{\varepsilon _{B}}{\sqrt{2}}\,
    \sqrt{1 - \frac{M^2_2 - \mu^2 - 2\,m_W^2\cos 2\beta}{W}} \\[2mm]
  & & V_{21} = -V_{12} = \frac{\varepsilon _{A}}{\sqrt{2}}\,
    \sqrt{1 + \frac{M^2_2 - \mu^2 + 2\,m_W^2\cos 2\beta}{W}} \\[2mm]
  & & V_{22} = V_{11} = \frac{4}{\sqrt{2}}\,
    \sqrt{1 - \frac{M^2_2 - \mu^2 + 2\,m_W^2\cos 2\beta}{W}}
\end{eqnarray}
where
\begin{equation}
  W = \sqrt{(M^2_2+\mu^2+2\,m_W^2)^2 - 4\,(M_2\!\cdot\!\mu -m_W^2\sin
  2\beta)^2},
\end{equation}
\begin{equation}
  \varepsilon_{A} = {\rm sign}(M_2 \sin\beta + \mu\,\cos\beta),
  \hspace{6mm}
  \varepsilon_{B} = {\rm sign}(M_2 \cos\beta + \mu\,\sin\beta).
\end{equation}
Chargino masses are given by 
$$
  m_{\tilde{\chi}_1^+}^2 = \frac{1}{2}|\sqrt{(M_2^2-\mu^2)^2+2m_W^2(1+\sin 2\beta)}
$$
\begin{equation}
  -\sqrt{(M_2^2+\mu^2)^2+2m_W^2(1-\sin 2\beta)}|,
\end{equation}
$$
  m_{\tilde{\chi}_2^+}^2 = \frac{1}{2}(\sqrt{(M_2^2-\mu^2)^2+2m_W^2(1+\sin
  2\beta)}
$$
\begin{equation}
 +\sqrt{(M_2^2+\mu^2)^2+2m_W^2(1-\sin 2\beta)}).
\end{equation}
Sfermion contributions depend on the couplings
$$ g_{h_1\tilde{f}^*_j\tilde{f}_j}
=\frac{1}{v}\left(\Gamma^{\alpha\tilde{f}^*\tilde{f}}\right)_{\beta\gamma}
a_{\alpha 1}U^{\tilde{f}*}_{\beta j} U^{\tilde{f}}_{\gamma j}\,,
$$
с $\alpha=(1,2,a)$, $\beta,\gamma = L, R$,
$i=(h_1,h_2,h_3)=(1,2,3)$ и $j,k=1,2$,
\begin{eqnarray}
U^{\tilde{f}}= \left( \begin{array}{cc}
\cos\theta_{\tilde{f}} & -\sin\theta_{\tilde{f}}\, {\rm e}^{-i\phi_{\tilde{f}}}\\
\sin\theta_{\tilde{f}}\, {\rm e}^{+i\phi_{\tilde{f}}}&
\cos\theta_{\tilde{f}}
       \end{array}
\right)\,,
\end{eqnarray}
$$
\Gamma^{1\tilde{f}^*\tilde{f}}=-\Gamma^{{\phi_1}\tilde{f}^*\tilde{f}}
\sin\alpha+\Gamma^{{\phi_2}\tilde{f}^*\tilde{f}} \cos\alpha,
$$
$$
\Gamma^{2\tilde{f}^*\tilde{f}}=\Gamma^{{\phi_1}\tilde{f}^*\tilde{f}}
\cos\alpha+\Gamma^{{\phi_2}\tilde{f}^*\tilde{f}} \sin\alpha,
$$
where
\begin{eqnarray}
\Gamma^{a\tilde{b}^*\tilde{b}} &=& \frac{1}{\sqrt{2}}\left(
\begin{array}{cc}
0 & i\,h_b^*(s_\beta A_b^*+c_\beta \mu) \\
-i\,h_b(s_\beta A_b+c_\beta \mu^*) & 0
\end{array} \right)\,,
\nonumber \\
\Gamma^{\phi_1\tilde{b}^*\tilde{b}} &=& \left(
\begin{array}{cc}
-|h_b|^2vc_\beta+ \frac{1}{4}\left(g_2^2+\frac{1}{3}g^{
2}_1\right)vc_\beta&
-\frac{1}{\sqrt{2}}h_b^*A_b^* \\
-\frac{1}{\sqrt{2}}h_bA_b & -|h_b|^2vc_\beta+
\frac{1}{6}g^{2}_1 vc_\beta
\end{array} \right)\,,
\nonumber \\
\Gamma^{\phi_2\tilde{b}^*\tilde{b}} &=& \left(
\begin{array}{cc}
- \frac{1}{4}\left(g^2_2+\frac{1}{3}g^{2}_1\right)vs_\beta&
\frac{1}{\sqrt{2}}h_b^*\mu \\
\frac{1}{\sqrt{2}}h_b\mu^* & -\frac{1}{6}g^{2}_1 vs_\beta
\end{array} \right)\,,
\nonumber \\
%
%
\Gamma^{a\tilde{t}^*\tilde{t}} &=& \frac{1}{\sqrt{2}}\left(
\begin{array}{cc}
0 & i\,h_t^*(c_\beta A_t^*+s_\beta \mu) \\
-i\,h_t(c_\beta A_t+s_\beta \mu^*) & 0
\end{array} \right)\,,
\nonumber \\
\Gamma^{\phi_1\tilde{t}^*\tilde{t}} &=& \left(
\begin{array}{cc}
- \frac{1}{4}\left(g^2_2-\frac{1}{3}g^{2}_1\right)vc_\beta&
\frac{1}{\sqrt{2}}h_t^*\mu \\
\frac{1}{\sqrt{2}}h_t\mu^* & -\frac{1}{3}g^{2}_1 vc_\beta
\end{array} \right)\,,
\nonumber \\
\Gamma^{\phi_2\tilde{t}^*\tilde{t}} &=& \left(
\begin{array}{cc}
-|h_t|^2vs_\beta+ \frac{1}{4}\left(g^2_2-\frac{1}{3}g^{
2}_1\right)vs_\beta&
-\frac{1}{\sqrt{2}}h_t^*A_t^* \\
-\frac{1}{\sqrt{2}}h_tA_t & -|h_t|^2vs_\beta+
\frac{1}{3}g^{2}_1 vs_\beta
\end{array} \right)\,,
\nonumber \\
%
%
\Gamma^{a\tilde{\tau}^*\tilde{\tau}} &=& \frac{1}{\sqrt{2}}\left(
\begin{array}{cc}
0 & i\,h_\tau^*(s_\beta A_\tau^*+c_\beta \mu) \\
-i\,h_\tau(s_\beta A_\tau+c_\beta \mu^*) & 0
\end{array} \right)\,,
\nonumber \\
\Gamma^{\phi_1\tilde{\tau}^*\tilde{\tau}} &=& \left(
\begin{array}{cc}
-|h_\tau|^2vc_\beta+ \frac{1}{4}\left(g^2_2-g^{2}_1\right)
vc_\beta&
-\frac{1}{\sqrt{2}}h_\tau^*A_\tau^* \\
-\frac{1}{\sqrt{2}}h_\tau A_\tau & -|h_\tau|^2vc_\beta+
\frac{1}{2}g^{2}_1 vc_\beta
\end{array} \right)\,,
\nonumber \\
\Gamma^{\phi_2\tilde{\tau}^*\tilde{\tau}} &=& \left(
\begin{array}{cc}
  -\frac{1}{4}\left(g^2_2- g^{2}_1\right) vs_\beta&
\frac{1}{\sqrt{2}}h_\tau^*\mu \\
\frac{1}{\sqrt{2}}h_\tau\mu^* & -\frac{1}{2}g^{2}_1 vs_\beta
\end{array} \right)\,.
\end{eqnarray}
In these formulas $h_{t,b,\tau}$ are real variables.\\
Sfermion masses are given by
\begin{equation}
  m_{\tilde{q}(\,\tilde{l}\,)_{1,2}}^2 = 
       \frac{1}{2} \left( m_{\tilde{q}(\,\tilde{l}\,)L}^2 +
      m_{\tilde{q}(\,\tilde{l}\,)R}^2
  \mp \sqrt{(m_{\tilde{q}(\,\tilde{l}\,)L}^2 
    - m_{\tilde{q}(\,\tilde{l}\,)R}^2)^2 + 4\, |a_{q(l)}|^2 m_{q(l)}^2 }
  \,\right),
\label{eq:sqmasseigenvalues}
\end{equation}
where
$$
m_{\tilde{q}L}^2 = M^2_{\tilde{Q}_3}\, +\, m^2_q\, +\, c_{2\beta}
m^2_Z\, ( T^q_z\, -\, Q_q s_W^2 ),
$$
$$
m_{\tilde{q}R}^2 = M^2_{\tilde{R}_3}\, +\, m^2_q\, +\, c_{2\beta}
m^2_Z\, Q_q s^2_W,
$$
$$
a_q m_q=h_q v_q (A_q - \mu^* R_q)/\sqrt{2},
$$
$$
m_{\tilde{l}L}^2 = M^2_{\tilde{L}_3}\, +\, m^2_\tau\, +\,
c_{2\beta} m^2_Z\, (s_W^2-1/2 ),
$$
$$
m_{\tilde{l}R}^2 = M^2_{\tilde{E}_3}\, +\, m^2_\tau\, -\,
c_{2\beta} m^2_Z\, s^2_W,
$$
$$
a_l m_l=h_\tau v_1 (A_\tau - \mu^* \tan\beta )/\sqrt{2}.
$$
Here the Yukawa couplings of quarks $h_q$, $q=t,b$, $R=U,D$, $T^t_z = - 
T^b_z = 1/2$, $Q_t = 2/3$, $Q_b
= -1/3$, $R_b = {\tt tg}\beta = v_2/v_1$, $R_t = {\tt ctg}\beta$, 
the mixing angles are
$$
  \cos\theta_{\tilde{q}(\,\tilde{l}\,)} =
  \frac{- |a_{q(l)}| 
m_{q(l)}}{\sqrt{(m_{\tilde{q}(\,\tilde{l}\,)L}^2-m_{\tilde{q}(\,\tilde{l}\,)_1}^2)^2 +
  |a_{q(l)}|^2 m_{q(l)}^2}},
$$
\begin{equation}
  \sin\theta_{\tilde{q}(\,\tilde{l}\,)}=
  \frac{m_{\tilde{q}(\,\tilde{l}\,)L}^2-m_{\tilde{q}(\,\tilde{l}\,)_1}^2}
       {\sqrt{(m_{\tilde{q}(\,\tilde{l}\,)L}^2
-m_{\tilde{q}(\,\tilde{l}\,)_1}^2)^2 + |a_{q(l)}|^2 m_{q(l)}^2}}.
\end{equation}
%

Charged Higgs boson contribution depends on
the effective triple self-couplings $g_{_{ H^+H^- h_i}}$ which can be 
written as 
\begin{eqnarray*} 
g_{H^+{}_{} \phantom{-} H^-{}_{} \phantom{-} h_1{}_{} \phantom{-}}  &=&
	-\frac{1}{2}\frac{1}{ s_{2\beta}{}^2  \cdot v}\big(4 ( 
s_{\alpha}  
    \cdot  c_{\beta} {}^3+ c_{\alpha}  \cdot  s_{\beta} {}^3) s_{2\beta} 
m_H{}^2  a_{21}
   -8 c_{\beta}{}^2  s_{\beta+\alpha} a_{21} {\tt Re} \mu^2_{12} \\[2mm]
 && -8 s_{\beta+\alpha} s_{\beta}{}^4  a_{21} {\tt Re} \mu^2_{12}-8 c_{\beta+\alpha} 
     c_{\beta}{}^2  a_{11} {\tt Re} \mu^2_{12}-8 c_{\beta+\alpha} s_{\beta}{}^4  a_{11} 
               {\tt Re} \mu^2_{12} \\[2mm]
  &&  - c_{\beta}{}^2  s_{2\beta}{}^2  s_{\beta+\alpha} a_{21} {\tt Re} \lambda_6 v{}^2 +4 
        c_{\beta}{}^2  s_{\alpha-\beta} a_{21} {\tt Re}\lambda_6 v{}^2  \\[2mm]
  && +4 c_{\beta}{}^4  s_{\alpha} s_{\beta}{}^3  a_{11} {\tt Re}\lambda_6 v{}^2 +4 c_{\alpha} 
       c_{\beta}{}^3  s_{\beta}{}^4  a_{11} {\tt Re}\lambda_6 v{}^2 \\
 & & +4 c_{\beta}{}^2  s_{\alpha} s_{\beta} a_{11} {\tt Re}\lambda_6 v{}^2 
+4 c_{\alpha} c_{\beta}{}^5  a_{11} {\tt Re}\lambda_6 v{}^2  \\[2mm]
 & & +4 ( c_{\alpha}  \cdot  c_{\beta} {}^3- s_{\alpha}  \cdot  s_{\beta} {}^3) s_{2\beta} 
   m_h{}^2  a_{11}- s_{2\beta}{}^2  s_{\beta+\alpha} s_{\beta}{}^2  a_{21} 
             {\tt Re}\lambda_7 v{}^2  \\[2mm]
 & & -4 s_{\alpha-\beta} s_{\beta}{}^2  a_{21} {\tt Re}\lambda_7 v{}^2 
              +4 c_{\beta}{}^2  s_{\alpha} 
               s_{\beta}{}^5  a_{11} {\tt Re}\lambda_7 v{}^2  \\[2mm]
 & & -4 c_{\alpha} c_{\beta} s_{\beta}{}^2  a_{11} {\tt Re}\lambda_7 v{}^2 
              -4 s_{\alpha} s_{\beta}{}^3  a_{11} {\tt Re}\lambda_7 v{}^2  \\[2mm]
 & & -4 c_{\alpha} c_{\beta}{}^3  s_{\beta}{}^4  a_{11} {\tt Re}\lambda_7 v{}^2 
          +4 c_{\beta-\alpha} s_{2\beta}{}^2  m_{H^\pm}{}^2  a_{21}
                 -4 s_{2\beta}{}^2  s_{\alpha-\beta} m_{H^\pm}{}^2  a_{11} 
\\[2mm]
 & & - s_{2\beta}{}^3  s_{\beta+\alpha} m_A{}^2  a_{21}
               - c_{\beta+\alpha} s_{2\beta}{}^3  m_A{}^2  a_{11}- 
s_{2\beta}{}^3  
              s_{\beta+\alpha} a_{21} {\tt Re}\lambda_5 v{}^2  \\[2mm]
 & & - c_{\beta+\alpha} s_{2\beta}{}^3  a_{11} {\tt Re}\lambda_5 v{}^2 
                   +8 c_{\beta}{}^3  s_{\beta}{}^3  a_{31} {\tt Im} \lambda_5 
   v{}^2 -8 c_{\beta}{}^2  s_{\beta}{}^4  a_{31} {\tt Im}\lambda_6 v{}^2  \\[2mm]
 & & -8 c_{\beta}{}^4  s_{\beta}{}^2  a_{31} {\tt Im}\lambda_7 v{}^2 \big)
\end{eqnarray*}
\begin{eqnarray*}
g_{H^+{}_{} \phantom{-} H^-{}_{} \phantom{-} h_2{}_{} \phantom{-}}  &=&
	-\frac{1}{2}\frac{1}{ s_{2\beta}{}^2  \cdot v}\big(4 ( 
s_{\alpha}  \cdot  c_{\beta} {}^3
+ c_{\alpha}  \cdot  s_{\beta} {}^3) s_{2\beta} m_H{}^2  a_{22}
-8 c_{\beta}{}^2  s_{\beta+\alpha} a_{22} {\tt Re} \mu^2_{12} \\[2mm]
  && -8 s_{\beta+\alpha} s_{\beta}{}^4  a_{22} {\tt Re} \mu^2_{12}
-8 c_{\beta+\alpha} c_{\beta}{}^2  a_{12} {\tt Re} \mu^2_{12}
-8 c_{\beta+\alpha} s_{\beta}{}^4  a_{12} {\tt Re} \mu^2_{12} \\[2mm]
  && - c_{\beta}{}^2  s_{2\beta}{}^2  s_{\beta+\alpha} a_{22} 
{\tt Re} \lambda_6 v{}^2 
+4 c_{\beta}{}^2  s_{\alpha-\beta} a_{22} {\tt Re}\lambda_6 v{}^2  \\[2mm]
  && +4 c_{\beta}{}^4  s_{\alpha} s_{\beta}{}^3  a_{12} {\tt Re}\lambda_6 v{}^2 
+4 c_{\alpha} c_{\beta}{}^3  s_{\beta}{}^4  a_{12} {\tt Re}\lambda_6 v{}^2 \\ 
  && +4 c_{\beta}{}^2  s_{\alpha} s_{\beta} a_{12} {\tt Re}\lambda_6 v{}^2 
+4 c_{\alpha} c_{\beta}{}^5  a_{12} {\tt Re}\lambda_6 v{}^2  \\[2mm]
  && +4 ( c_{\alpha}  \cdot  c_{\beta} {}^3
- s_{\alpha}  \cdot  s_{\beta} {}^3) s_{2\beta} m_h{}^2  a_{12}
- s_{2\beta}{}^2  s_{\beta+\alpha} s_{\beta}{}^2  a_{22} {\tt Re}\lambda_7 v{}^2  \\[2mm]
  && -4 s_{\alpha-\beta} s_{\beta}{}^2  a_{22} {\tt Re}\lambda_7 v{}^2 
+4 c_{\beta}{}^2  s_{\alpha} s_{\beta}{}^5  a_{12} {\tt Re}\lambda_7 v{}^2  \\[2mm]
  && -4 c_{\alpha} c_{\beta} s_{\beta}{}^2  a_{12} {\tt Re}\lambda_7 v{}^2 
-4 s_{\alpha} s_{\beta}{}^3  a_{12} {\tt Re}\lambda_7 v{}^2  \\[2mm]
  && -4 c_{\alpha} c_{\beta}{}^3  s_{\beta}{}^4  a_{12} {\tt Re}\lambda_7 v{}^2 
+4 c_{\beta-\alpha} s_{2\beta}{}^2  m_{H^\pm}{}^2  a_{22}
-4 s_{2\beta}{}^2  s_{\alpha-\beta} m_{H^\pm}{}^2  a_{12} \\[2mm]
  && - s_{2\beta}{}^3  s_{\beta+\alpha} m_A{}^2  a_{22}
- c_{\beta+\alpha} s_{2\beta}{}^3  m_A{}^2  a_{12}
- s_{2\beta}{}^3  s_{\beta+\alpha} a_{22} {\tt Re}\lambda_5 v{}^2  \\[2mm]
  && - c_{\beta+\alpha} s_{2\beta}{}^3  a_{12} {\tt Re}\lambda_5 v{}^2 
+8 c_{\beta}{}^3  s_{\beta}{}^3  a_{32} {\tt Im}\lambda_5 v{}^2 
-8 c_{\beta}{}^2  s_{\beta}{}^4  a_{32} {\tt Im}\lambda_6 v{}^2  \\[2mm]
  && -8 c_{\beta}{}^4  s_{\beta}{}^2  a_{32} {\tt Im}\lambda_7 v{}^2 \big)
\end{eqnarray*}
\begin{eqnarray*}
g_{H^+{}_{} \phantom{-}  H^-{}_{}  \phantom{-}  h_3{}_{}  \phantom{-}}  &=&
	 -\frac{1}{2}\frac{1}{ s_{2\beta}{}^2  \cdot v}\big(4 ( 
s_{\alpha}  
\cdot  c_{\beta} {}^3+ c_{\alpha}  \cdot  s_{\beta} {}^3) s_{2\beta} 
m_H{}^2  a_{23}
-8 c_{\beta}{}^2  s_{\beta+\alpha} a_{23} {\tt Re} \mu^2_{12}  \\[2mm]
 & &  -8 s_{\beta+\alpha} s_{\beta}{}^4  a_{23} {\tt Re} \mu^2_{12}
-8 c_{\beta+\alpha} c_{\beta}{}^2  a_{13} {\tt Re} 
\mu^2_{12}-8 c_{\beta+\alpha} s_{\beta}{}^4  a_{13} {\tt Re} \mu^2_{12}  \\[2mm]
 & &  - c_{\beta}{}^2  s_{2\beta}{}^2  s_{\beta+\alpha} a_{23} {\tt Re}\lambda_6 v{}^2 
+4 c_{\beta}{}^2  s_{\alpha-\beta} a_{23} {\tt Re}\lambda_6 v{}^2   \\[2mm]
 & &  +4 c_{\beta}{}^4  s_{\alpha} s_{\beta}{}^3  a_{13} {\tt Re}\lambda_6 v{}^2 
+4 c_{\alpha} c_{\beta}{}^3  s_{\beta}{}^4  a_{13} {\tt Re}\lambda_6 v{}^2  \\ 
 & &  +4 c_{\beta}{}^2  s_{\alpha} s_{\beta} a_{13} {\tt Re}\lambda_6 v{}^2 
+4 c_{\alpha} c_{\beta}{}^5  a_{13} {\tt Re}\lambda_6 v{}^2   \\[2mm]
 & &  +4 ( c_{\alpha}  \cdot  c_{\beta} {}^3
      - s_{\alpha}  \cdot  s_{\beta} {}^3) s_{2\beta} m_h{}^2  a_{13}-  
      s_{2\beta}{}^2  s_{\beta+\alpha} s_{\beta}{}^2  a_{23} {\tt Re}\lambda_7 v{}^2   \\[2mm]
 & &  -4 s_{\alpha-\beta} s_{\beta}{}^2  a_{23} {\tt Re}\lambda_7 v{}^2 
+4 c_{\beta}{}^2  s_{\alpha} s_{\beta}{}^5  a_{13} {\tt Re}\lambda_7 v{}^2   \\[2mm]
 & &  -4 c_{\alpha} c_{\beta} s_{\beta}{}^2  a_{13} {\tt Re}\lambda_7 v{}^2 
            -4 s_{\alpha} s_{\beta}{}^3  a_{13} {\tt Re}\lambda_7 v{}^2   \\[2mm]
 & &  -4 c_{\alpha} c_{\beta}{}^3  s_{\beta}{}^4  a_{13} {\tt Re}\lambda_7 v{}^2 
+4 c_{\beta-\alpha} s_{2\beta}{}^2  m_{H^\pm}{}^2  a_{23}-4 s_{2\beta}{}^2  
s_{\alpha-\beta} m_{H^\pm}{}^2  a_{13}  \\[2mm]
 & &  - s_{2\beta}{}^3  s_{\beta+\alpha} m_A{}^2  a_{23}
- c_{\beta+\alpha} s_{2\beta}{}^3  m_A{}^2  a_{13}
- s_{2\beta}{}^3  s_{\beta+\alpha} a_{23} {\tt Re}\lambda_5 v{}^2   \\[2mm]
 & &  - c_{\beta+\alpha} s_{2\beta}{}^3  a_{13} {\tt Re}\lambda_5 v{}^2 
   +8 c_{\beta}{}^3  s_{\beta}{}^3  a_{33} {\tt Im}\lambda_5 
    v{}^2 -8 c_{\beta}{}^2  s_{\beta}{}^4  a_{33} {\tt Im}\lambda_6 v{}^2   \\[2mm]
 & &  -8 c_{\beta}{}^4  s_{\beta}{}^2  a_{33} {\tt Im}\lambda_7 v{}^2 \big) 
\end{eqnarray*}
This representation uses the mass basis for $CP$ even/odd Higgs fields 
$(h,H,A)$ then rotated by matrix $a_{ij}$ in the three-dimensional 
$(h,H,A)$ isospace, and 
for this reason includes $m_h$, $m_H$, $m_A$ and $m_{H^\pm}$ of the
$CP$ conserving limit, calculated with one-loop MSSM corrections from the 
squark sector.
In this sense the vertices above are MSSM effective one-loop Higgs 
self-interaction vertices. If the imaginary parts in these vertices are
set to zero they are reduced to the self-interaction vertices of the
$CP$ conserving limit, when $m_h$, $m_H$, $m_A$ and $m_{H\pm}$ are the
masses of physical states. Various extremal cases (decoupling limits)
are clearly seen.
Equivalent representation of the triple couplings can be written in the 
$\lambda_i$ basis (see details on the representations in mass and 
$\lambda_i$ basis in \cite{Dubinin02}). For example
\begin{eqnarray*}
g_{_{h_1H^+H^-}} & =& -\, v \, \sum_{\alpha =1}^3 a_{\alpha 1}\,
g_{_{\alpha H^+H^-}}\,,
\end{eqnarray*}
\vspace{-4mm}
where
\begin{eqnarray*}
g_{_{1 H^+H^-}}&=& 
   \, {\tt Re}\Delta \lambda_5\, s_\beta c_\beta  \,
                    c_{\alpha+\beta}  -
  \, {\tt Re} \Delta \lambda_6\, c_\alpha 
    \, s^2_\beta \, c_\beta                                             \\
&& +  \, {\tt Re} \Delta \lambda_6 \,
   s_\alpha \, s^3_\beta  +
   \, {\tt Re}\Delta \lambda_7\, c_\beta \,
   \left(  
        s_\alpha \, s_\beta \, c_\beta \, \right.              
  -      \left. c_\alpha \,\left(  c^2_\beta  -
        2\,  s^2_\beta  \right)  \right)                    \\ 
&& -  \, {\tt Re}\Delta \lambda_6 \,
   s_\alpha \, s_{2\,\beta} \, c_\beta
-   2\, s_\alpha \, s^2_\beta \, c_\beta \,
    {{\lambda }_1} + 2\, c_\alpha \, s_\beta \,
     {c^2_\beta }\, {{\lambda }_2} \\
&& -     \,{c^3_\beta }\, s_\alpha \,
     {{\lambda }_3}                               
+   \, c_\alpha \,{s^3_\beta }\,
   {{\lambda }_3} - \, c_\alpha \, c^2_\beta \,
   s_\beta \,{{\lambda }_4} +
  \, c_\beta \, s_\alpha \, s^2_\beta \,
   {{\lambda }_4}, \\
g_{_{2 H^+H^-}}&=&
   \, {\tt Re}\Delta \lambda_5\, s_\beta c_\beta  \,
                    s_{\alpha+\beta}
+  2 \, {\tt Re} \Delta \lambda_6\, c_\alpha
    \, s_\beta \, c^2_\beta                                    
 -  \, {\tt Re} \Delta \lambda_6 \,
   c_\alpha \, s^3_\beta                                  \\
&& -  \, {\tt Re} \Delta \lambda_6 \,
   s_\alpha \, s^2_\beta c_\beta                             
 -\,   \, {\tt Re}\Delta \lambda_7\, c_\beta \,
   \left(
        c_\alpha \, s_\beta \, c_\beta \, \right.
  +     \left. s_\alpha \,\left(  c^2_\beta  -
        2\,  s^2_\beta  \right)  \right)                    \\
&& +2 \,  \, c_\alpha \, s^2_\beta \, c_\beta \lambda_1 \,
   +2 \,  \, s_\alpha \, s_\beta \, c^2_\beta \lambda_2 \,
   +      \, c_\alpha \, c^3_\beta \lambda_3 \,             \\
&&   +      \, s_\alpha \, s^3_\beta \lambda_3 \,
   -      \, c_\alpha s^2_\beta c_\beta \lambda_4 \,
   -      \, s_\alpha s_\beta c^2_\beta \lambda_4,       \\
g_{_{3 H^+H^-}} & = & 
  \,  c^2_\beta \, {\tt Im} \Delta \lambda_7 \,
         - \, s_\beta \, c_\beta \, {\tt Im} \Delta \lambda_5 \,       
         + \, s^2_\beta \, {\tt Im} \Delta \lambda_6 \, 
\end{eqnarray*}
In this representation the scalar masses of the $CP$ conserving limit do 
not explicitly participate. The magnitude of the coupling $g_{H^+ \, H^- 
\, h_1}$  
is shown in Fig.\ref{fg:ggfi}. 
\begin{figure}[ht]
\begin{center}
\hspace{-0.8cm} \epsfxsize=0.9\textwidth
\centerline{\psfig{figure=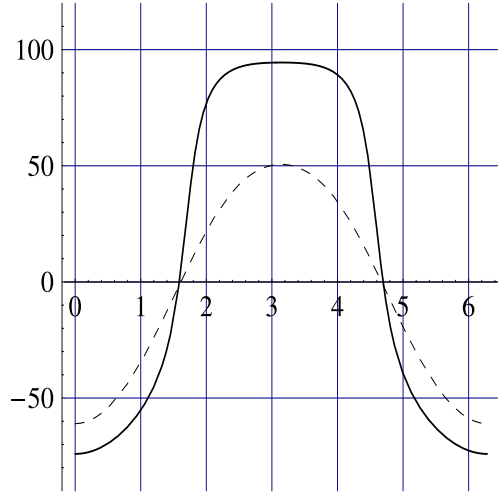,height=6.5cm,width=6.5cm}}
\end{center}
\vspace{4mm} 
\caption{ \label{fg:ggfi}
Triple Higgs boson interaction vertex $g_{H^+ \, H^- \, h_1}$ (GeV) {\it 
vs} the phase 
${\tt Arg}(\mu A)$ at the parameter values $M_{SUSY}=$500 GeV, ${\tt 
tg}\beta=$5, $A_{t,b}=$1000 GeV, $\mu=$2000 GeV. Dashed line 
$m_{H^\pm}=$300 GeV,
solid line $m_{H^\pm}=$200 GeV. }
\end{figure}

The decay width $h_i\rightarrow gg$ has the form
\begin{eqnarray}
\Gamma(h_i\rightarrow gg)\ =\
\frac{M_{h_i}^3\alpha^2_S}{32\pi^3\,v^2}
         \left[\,K^g_{H}\, \left|S^g_i(M_{h_i})\right|^2\:
              +\: K^g_{A}\, \left|P^g_i(M_{h_i})\right|^2\right]\,,
\end{eqnarray}
where
\begin{eqnarray}
S^g_i(M_{h_i})&=&\sum_{f=b,t}
g^{S}_{h_if\bar{f}}\,\frac{v}{m_f} F_{sf}(\tau_{if})
-\sum_{\tilde{f}_j=\tilde{t}_1,\tilde{t}_2,\tilde{b}_1,\tilde{b}_2}
g_{h_i\tilde{f}^*_j\tilde{f}_j}
\frac{v^2}{4m_{\tilde{f}_j}^2} F_0(\tau_{i\tilde{f}_j}) \,, \nonumber \\
P^g_i(M_{h_i})&=&\sum_{f=b,t}
g^{P}_{h_if\bar{f}}\,\frac{v}{m_f} F_{pf}(\tau_{if}) \, 
\end{eqnarray}
and QCD $K$-factors are
\begin{eqnarray}
  \label{KgHA}
K^g_H &=& 1\ +\ \frac{\alpha_S (M^2_{h_i})}{\pi}\,
\bigg(\,\frac{95}{4} \: -\: \frac{7}{6}\,N_F\,\bigg)\nonumber\\
K^g_A &=& 1\ +\ \frac{\alpha_S (M^2_{h_i})}{\pi}\,
\bigg(\,\frac{97}{4} \: -\: \frac{7}{6}\,N_F\,\bigg)\,,
\end{eqnarray}
$N_F=5$ is the number of quark flavors with masses less than $m_{h_1}$.

The decay width of Higgs boson to the two fermions
$h_1\rightarrow f\bar{f}$ can be written as
\beq \Gamma_{h_1 \rightarrow f \bar{f}} = \ff{N_C g_f^2
m_{h_1} \beta_k^{\ff{3}{2}}}{8 \pi} \, \left\{\begin{array}{l}
(s_\alpha a_{21}- c_\alpha a_{11})^2 \ff{1}{s_\beta^2}
+{\tt ctg}^2\beta\, a_{31}^2\, ,
\quad f\equiv u, c, t,\\
(c_\alpha a_{21}- s_\alpha a_{11})^2 \ff{1}{c_\beta^2}+
{\tt tg}^2\beta\, a_{31}^2\, , \quad f\equiv b, d, s, e, \mu, \tau ,
\end{array}
\right. \label{eq:decferm} 
\eeq 
where $\beta_k=1-4 k$,
$k=\frac{m_f^2}{m_{h_1}^2}$, $g_f=\ff{g m_f}{2 m_W}$ and
$N_C=$3 (1) for quarks (leptons).

In the following Table 5 we list the Higgs boson masses
$m_{h_1}$, $m_{h_2}$, $m_{h_3}$ which are calculated 
using the  
effective $\lambda_i$ parameters (\ref{eq:lambda1})-(\ref{eq:lambda7}), 
Section 2,
and the mass term diagonalization method described in Section 3.1. 
The 
decay widths $\Gamma_{h_1\to
gg}$, $\Gamma_{h_1\to\gamma\gamma}$ (unprimed) include only the leading 
one-loop
contributions of $t$, $b$ quarks and $W^\pm$ bosons. For an illustration 
of 
the sensitivity
of $m_{h_1}$, $m_{h_2}$, $m_{h_3}$ and their decay widths to the
values of $\lambda_i$ we computed 
Higgs boson masses $m'_{h_1,\,h_2,\,h_3}$ and the leading one-loop
decay widths $\Gamma'_{h_1\to gg}$,
$\Gamma'_{h_1\to\gamma\gamma}$ (include $t$, $b$ and $W$ contributions 
only) 
using
the effectife potential parametrization with both the one-loop and 
two-loop contributions to $\lambda_i$ from the paper 
\cite{PilaftsisWagner}.
Finally, the decay widths
$\Gamma''_{h_1\to gg}$, $\Gamma''_{h_1\to\gamma\gamma}$ are found 
using the effective parameters (\ref{eq:lambda1})-(\ref{eq:lambda7}) and 
taking into accout all possible
one-loop fermion ($t$, $b$), gauge boson $W^\pm$, 
sfermion ($\tilde t$,
$\tilde b$), chargino and charged Higgs boson contributions,
with $K$-factors introduced in the expressions for decay widths.

Table 5 contains also the output of the CPsuperH \cite{CPsuperH} package
and the FeynHiggs \cite{feynh} package with the input parameter 
values taken the same as used in our 
parameter set. The two-loop evaluation in the CPsuperH
and the one-loop evaluation in the FeynHiggs 2.1beta
has been performed.
Note that physical Higgs bosons $H_1$, $H_2$, $H_3$ of the
CPsuperH and FeynHiggs are evaluated in the way that is technically
different from the construction of our 
mixed states $h_1$, $h_2$, $h_3$,
however a difference of numbers (which is
from several percent to 40\% in the majority of cases) is
caused mainly by theoretical uncertainties of the effective two-doublet 
potential representation, 
not by different definitions of the Higgs boson eigenstates in the 
generic basis of scalar doublets, as demonstrated explicitly in section 
3.1.  

In Fig.5 and Figs.6-9 we show the variation of the light Higgs boson
mass and the variations of $\Gamma({h_1\to gg})$, $\Gamma({h_1\to 
\gamma \gamma})$ decay widths in different regions of the parameter
space ($\varphi$, $m_{H^\pm}$, $A_{t,b}$, $\mu$, ${\tt tg}\beta$). 
At the parameter set ( 0, 300 GeV, 1000 GeV, 2000 GeV, 5 )
the decay widths of $h_1$ to $\gamma \gamma$ and $gg$ are not far from 
the decay widths of the SM Higgs boson with $m_H=$120 GeV. Largest 
sensitivity
of the widths to the charged Higgs mass is observed. At $m_{H^\pm}$ 
around 200 GeV (Fig.6a, Fig.8a) we observe the suppression of the 
branchings of $h_1$
to $gg$ and $\gamma \gamma$ of more than 10 times at $\varphi\sim \pi$,
which takes place in CPsuperH and FeynHiggs at higher 
masses of $m_{H^\pm}$ around 300 GeV.   

Our approach is algorithmized in the form of the model in CompHEP 41.10
format \cite{comphep}, where the symbolic expressions for vertices are a
starting level for calculation of the complete tree-level sets of
diagrams with the following cross section/decay width calculations
and the generation of unweighted events.


\newpage
\begin{table}[h!]
\begin{center}
\begin{tabular}{|c||c|c|c|c|c|c|c|}
\hline & $\varphi=0$ & $\pi/6$ & $\pi/3$ & $\pi/2$ &
$2\pi/3$ & $5\pi/6$ &
$\pi$ \\
\hline\hline
$m_{h_1}$,~ГэВ & 115.4 & 118.7 & 125.9 & 131.4 & 130.7 & 125.2 & 122.0 \\
$m'_{h_1}$ & 112.1 & 114.4 & 119.7 & 124.2 & 125.0 & 123.0 & 121.6 \\
$m_{H_1}$ \cite{feynh} & 115.8 & 118.8 & 125.5 & 130.2 & 123.2
& 98.2 & 78.0 \\
$m_{H_1}$ \cite{CPsuperH} & 106.8 & 109.0 & 113.9 & 117.4 & 114.9
& 105.7 & 99.4
\\
\hline
$m_{h_2}$ & 295.5 & 289.6 & 279.7 & 269.3 & 262.2 & 259.8 & 259.6 \\
$m'_{h_2}$ & 294.4 & 291.0 & 283.9 & 276.2 & 270.6 & 268.1 & 267.6 \\
$m_{H_2}$ \cite{feynh} & 295.6 & 290.0 & 279.1 & 264.3 & 249.2
& 239.7 & 236.9\\
$m_{H_2}$ \cite{CPsuperH} & 302.2 & 297.8 & 290.9 & 282.2 & 273.9
& 268.3 & 264.4 \\ \hline
$m_{h_3}$ & 297.1 & 299.5 & 300.4 & 299.9 & 298.8 & 297.6 & 297.1 \\
$m'_{h_3}$ & 298.2 & 299.1 & 299.2 & 298.2 & 296.7 & 295.1 & 294.4 \\
$m_{H_3}$ \cite{feynh} & 297.6 & 300.0 & 301.1 & 301.3 & 300.9
& 300.4 & 300.2\\
$m_{H_3}$ \cite{CPsuperH} & 302.3 & 304.4 & 305.0 & 304.5 & 303.5
& 302.4 &
302.0 \\
\hline \hline $\Gamma_{h_1\to gg}\times 10^4$ & 1.378 & 1.529 &
1.907 & 2.220 & 2.101 & 1.707
& 1.516 \\
$\Gamma'_{h_1\to gg}\times 10^4$ & 1.283 & 1.381 & 1.624 & 1.841 &
1.846 & 1.687
& 1.597 \\
$\Gamma''_{h_1\to gg}\times 10^4$ & 2.103 & 2.355 & 3.024 & 3.643
& 3.397 & 2.412
& 1.889 \\
$\Gamma_{H_1\to gg}\times 10^4$ \cite{feynh} & 2.040 & 2.187 &
2.462 & 2.225
& 0.863 & 0.037 & 0.110 \\
$\Gamma_{H_1\to gg}\times 10^4$ \cite{CPsuperH} & 1.878 & 1.964 &
2.107 & 1.961
& 1.262 & 0.503 & 0.263 \\
\hline $\Gamma_{h_1\to\gamma\gamma}\times 10^6$ & 7.703 & 8.593 &
10.981 & 13.313 &
12.953 & 10.645 & 9.508 \\
$\Gamma'_{h_1\to\gamma\gamma}\times 10^6$ & 6.887 & 7.447 & 8.896
& 10.369 &
10.683 & 9.935 & 9.460 \\
$\Gamma''_{h_1\to\gamma\gamma}\times 10^6$ & 7.470 & 8.371 &
10.832 & 13.321 &
12.945 & 10.274 & 8.887 \\
$\Gamma_{H_1\to\gamma\gamma}\times 10^6$ \cite{feynh} & 6.373 &
7.058 & 9.038 & 11.217 &
9.983 & 5.336 & 3.021 \\ 
$\Gamma_{H_1\to\gamma\gamma}\times 10^6$ \cite{CPsuperH} & 5.796 &
6.287 & 7.605 & 8.996 &
8.969 & 7.223 & 6.101 \\ \hline
\hline &&&&&&
\\[-3mm]
$\Gamma_{h_1\to \mu \bar{\mu}}\times 10^{-5}$  & 0.212 &
0.204 & 0.179 & 0.166
& 0.218 & 0.304 & 0.341  \\
$\Gamma_{H_1\to \mu \bar{\mu}}\times 10^{-5}$ \cite{CPsuperH} &
0.157 &
0.152 & 0.141 & 0.137 & 0.175 & 0.240 & 0.269 \\
\hline $\Gamma_{h_1\to \tau \bar{\tau}}\times 10^{-3}$  & 0.591 &
0.567 & 0.498 & 0.461
& 0.607 & 0.848 & 0.950  \\
$\Gamma_{H_1\to \tau \bar{\tau}}\times 10^{-3}$ \cite{CPsuperH}&
0.435 & 0.423 & 0.391 & 0.382
& 0.485 & 0.668 & 0.746 \\
 \hline \hline & & & & & &
&\\[-3mm]
$\Gamma_{h_1\to d \bar{d}}\times 
10^{-7}$ & 0.202 & 0.194 &
0.170 & 0.158
& 0.208 & 0.290 & 0.325 \\
$\Gamma_{H_1\to d \bar{d}}\times 10^{-7}$ \cite{CPsuperH}& 0.193 &
0.187 & 0.171 & 0.167
& 0.212 & 0.297 & 0.335 \\
\hline $\Gamma_{h_1\to s \bar{s}}\times 10^{-5}$ & 0.744 & 0.713 &
0.626 & 0.580
& 0.764 & 1.066 & 1.195 \\
$\Gamma_{H_1\to s \bar{s}}\times 10^{-5}$ \cite{CPsuperH} & 0.709
& 0.687 & 0.629 & 0.612
& 0.780 & 1.089 & 1.230 \\
\hline $\Gamma_{h_1\to c \bar{c}}\times 10^{-3}$ & 0.083 & 0.086 &
0.093 & 0.097
& 0.095 & 0.088 & 0.083 \\
$\Gamma_{H_1\to c \bar{c}}\times 10^{-3}$ \cite{CPsuperH} & 0.101
& 0.103 & 0.108 & 0.111
& 0.107 & 0.096 & 0.089 \\
\hline
 $\Gamma_{h_1\to b \bar{b}}\times 10^{-2}$ &
0.504 & 0.483 & 0.424 & 0.393 & 0.518 & 0.724 &
0.810 \\
$\Gamma_{H_1\to b \bar{b}}\times 10^{-2}$ \cite{CPsuperH}  & 0.481
& 0.469 & 0.426 & 0.414
& 0.528 & 0.737 & 0.832 \\
\hline 
\end{tabular}
\label{tab:MandD1}
\end{center}
\vspace{-3mm}
\caption{{\scriptsize Higgs boson mases and their two-particle decay 
widths. The 
parameter set $\alpha_{EM}(m_Z)=0.7812\cdot$10$^{-2}$,
$\alpha_S(m_Z)=$0.1172, $G_F=1.174\cdot10^{-5}$\,GeV$^{-2}$, $m_b=$3 GeV,
${\tt tg} \beta=$5, $M_{SUSY}=500$\,GeV,
$|A_{\,t}|=|A_{\,b}|=A=1000$\,GeV, $|\mu|=2000$\,GeV, $m_{H^{\pm}}
= 300$\,GeV. Our results together with CPsuperH [17] and FeynHiggs [18]
with options 2003011100 (the one-loop regime). $m_{h_i}$, $\Gamma$ denote 
our results with the $\lambda_i$ at one-loop, $m'_{h_i}$, $\Gamma'$ 
our results with the two-loop terms \cite{PilaftsisWagner} introduced to 
$\lambda_i$,
$\Gamma$/$\Gamma''$ are the decay widths in our case when sparticles 
are not involved/included.}
}
\end{table}

\newpage

\begin{table}[h!]
\begin{center}
\begin{tabular}{|l|l|} \hline
Fields in the vertex & Vertex factor \\ \hline
$\bar{b}{}_{a p }$ \phantom{-} $b{}_{b q }$ \phantom{-} $h_1{}_{}$ \phantom{-}  &
	$-\frac{ M_b}{ c_{\beta} \cdot v}\delta_{p q} \big( c_{\alpha}\cdot  a_{21}\cdot \delta_{a b} - s_{\alpha}\cdot  a_{11}\cdot \delta_{a b} - s_{\beta}\cdot  i\cdot  a_{31}\cdot \gamma_{a b}^5 \big)$\\[2mm]
$\bar{b}{}_{a p }$ \phantom{-} $b{}_{b q }$ \phantom{-} $h_2{}_{}$ \phantom{-}  &
	$-\frac{ M_b}{ c_{\beta} \cdot v}\delta_{p q} \big( c_{\alpha}\cdot  a_{22}\cdot \delta_{a b} - s_{\alpha}\cdot  a_{12}\cdot \delta_{a b} - s_{\beta}\cdot  i\cdot  a_{32}\cdot \gamma_{a b}^5 \big)$\\[2mm]
$\bar{b}{}_{a p }$ \phantom{-} $b{}_{b q }$ \phantom{-} $h_3{}_{}$ \phantom{-}  &
	$-\frac{ M_b}{ c_{\beta} \cdot v}\delta_{p q} \big( c_{\alpha}\cdot  a_{23}\cdot \delta_{a b} - s_{\alpha}\cdot  a_{13}\cdot \delta_{a b} - s_{\beta}\cdot  i\cdot  a_{33}\cdot \gamma_{a b}^5 \big)$\\[2mm]
$\bar{t}{}_{a p }$ \phantom{-} $b{}_{b q }$ \phantom{-} $H^+{}_{}$ \phantom{-}  &
	$-\frac{ i \cdot\sqrt{2} \cdot Vtb}{ s_{2\beta} \cdot v}\delta_{p q} \big( s_{\beta}{}^2 \cdot  M_b\cdot (1+\gamma^5)_{a b} + c_{\beta}{}^2 \cdot  M_t\cdot (1-\gamma^5)_{a b} \big)$\\[2mm]
$\bar{t}{}_{a p }$ \phantom{-} $t{}_{b q }$ \phantom{-} $h_1{}_{}$ \phantom{-}  &
	$-\frac{ M_t}{ s_{\beta} \cdot v}\delta_{p q} \big( s_{\alpha}\cdot  a_{21}\cdot \delta_{a b} + c_{\alpha}\cdot  a_{11}\cdot \delta_{a b} - c_{\beta}\cdot  i\cdot  a_{31}\cdot \gamma_{a b}^5 \big)$\\[2mm]
$\bar{t}{}_{a p }$ \phantom{-} $t{}_{b q }$ \phantom{-} $h_2{}_{}$ \phantom{-}  &
	$-\frac{ M_t}{ s_{\beta} \cdot v}\delta_{p q} \big( s_{\alpha}\cdot  a_{22}\cdot \delta_{a b} + c_{\alpha}\cdot  a_{12}\cdot \delta_{a b} - c_{\beta}\cdot  i\cdot  a_{32}\cdot \gamma_{a b}^5 \big)$\\[2mm]
$\bar{t}{}_{a p }$ \phantom{-} $t{}_{b q }$ \phantom{-} $h_3{}_{}$ \phantom{-}  &
	$-\frac{ M_t}{ s_{\beta} \cdot v}\delta_{p q} \big( s_{\alpha}\cdot  a_{23}\cdot \delta_{a b} + c_{\alpha}\cdot  a_{13}\cdot \delta_{a b} - c_{\beta}\cdot  i\cdot  a_{33}\cdot \gamma_{a b}^5 \big)$\\[2mm]
$H^+{}_{}$ \phantom{-} $W^-{}_{\mu }$ \phantom{-} $h_1{}_{}$ \phantom{-}  &
	$-\frac{1}{2}\frac{ e}{ s_w}\big( s_{\alpha-\beta}\cdot  i\cdot  a_{21}\cdot p_3^\mu + c_{\beta-\alpha}\cdot  i\cdot  
a_{11}\cdot p_3^\mu - s_{\alpha-\beta}\cdot  i\cdot  a_{21}\cdot p_1^\mu $ \\[2mm]
  & $- c_{\beta-\alpha}\cdot  i\cdot  a_{11}\cdot p_1^\mu + a_{31}\cdot p_3^\mu - a_{31}\cdot p_1^\mu \big)$\\[2mm]
$H^+{}_{}$ \phantom{-} $W^-{}_{\mu }$ \phantom{-} $h_2{}_{}$ \phantom{-}  &
	$-\frac{1}{2}\frac{ e}{ s_w}\big( s_{\alpha-\beta}\cdot  i\cdot  a_{22}\cdot p_3^\mu + c_{\beta-\alpha}\cdot  i\cdot  
a_{12}\cdot p_3^\mu - s_{\alpha-\beta}\cdot  i\cdot  a_{22}\cdot p_1^\mu $\\[2mm]
&$- c_{\beta-\alpha}\cdot  i\cdot  a_{12}\cdot p_1^\mu 
  + a_{32}\cdot p_3^\mu - a_{32}\cdot p_1^\mu \big)$\\[2mm]
$H^+{}_{}$ \phantom{-} $W^-{}_{\mu }$ \phantom{-} $h_3{}_{}$ \phantom{-}  &
	$-\frac{1}{2}\frac{ e}{ s_w}\big( s_{\alpha-\beta}\cdot  i\cdot  a_{23}\cdot p_3^\mu + c_{\beta-\alpha}\cdot  i\cdot  
a_{13}\cdot p_3^\mu - s_{\alpha-\beta}\cdot  i\cdot  a_{23}\cdot p_1^\mu $\\[2mm]
&$- c_{\beta-\alpha}\cdot  i\cdot  a_{13}\cdot p_1^\mu + a_{33}\cdot p_3^\mu - a_{33}\cdot p_1^\mu \big)$\\[2mm]
$W^+{}_{\mu }$ \phantom{-} $W^-{}_{\nu }$ \phantom{-} $h_1{}_{}$ \phantom{-}  &
	$\frac{1}{2}\frac{ e{}^2  \cdot v}{ s_w{}^2 }g^{\mu \nu} \big( c_{\beta-\alpha} a_{21}- s_{\alpha-\beta} a_{11}\big)$\\[2mm]
$W^+{}_{\mu }$ \phantom{-} $W^-{}_{\nu }$ \phantom{-} $h_2{}_{}$ \phantom{-}  &
	$\frac{1}{2}\frac{ e{}^2  \cdot v}{ s_w{}^2 }g^{\mu \nu} \big( c_{\beta-\alpha} a_{22}- s_{\alpha-\beta} a_{12}\big)$\\[2mm]
$W^+{}_{\mu }$ \phantom{-} $W^-{}_{\nu }$ \phantom{-} $h_3{}_{}$ \phantom{-}  &
	$\frac{1}{2}\frac{ e{}^2  \cdot v}{ s_w{}^2 }g^{\mu \nu} \big( c_{\beta-\alpha} a_{23}- s_{\alpha-\beta} a_{13}\big)$\\[2mm]
${Z}_{\mu }$ \phantom{-} ${Z}_{\nu }$ \phantom{-} $h_1{}_{}$ \phantom{-}  &
	$2\frac{ e{}^2  \cdot v}{ s_{2w}{}^2 }g^{\mu \nu} \big( c_{\beta-\alpha} a_{21}- s_{\alpha-\beta} a_{11}\big)$\\[2mm]
${Z}_{\mu }$ \phantom{-} ${Z}_{\nu }$ \phantom{-} $h_2{}_{}$ \phantom{-}  &
	$2\frac{ e{}^2  \cdot v}{ s_{2w}{}^2 }g^{\mu \nu} \big( c_{\beta-\alpha} a_{22}- s_{\alpha-\beta} a_{12}\big)$\\[2mm]
${Z}_{\mu }$ \phantom{-} ${Z}_{\nu }$ \phantom{-} $h_3{}_{}$ \phantom{-}  &
	$2\frac{ e{}^2  \cdot v}{ s_{2w}{}^2 }g^{\mu \nu} \big( c_{\beta-\alpha} a_{23}- s_{\alpha-\beta} a_{13}\big)$\\ \hline
\end{tabular}
\end{center}
\caption{Vertex factors of $h_1$, $h_2$, $h_3$. This is a part
of the complete set of vertices generated by LanHEP package 
\cite{lanhep}.}
\end{table}



\newpage
\unitlength=1.0cm
\begin{figure}[t]
\begin{center}
\begin{picture}(10,10)
\put(9,-1.9){\mbox{$A_t=A_b$}}
\put(1,-1.9){\mbox{$A_t=A_b$}}
\put(9,5.6){\mbox{$m_{H^\pm}$}}
\put(1,5.6){\mbox{$m_{H^\pm}$}}
\put(-2.8,5.8){\epsfxsize=7.0cm
         \epsfysize=7.0cm \leavevmode \epsfbox{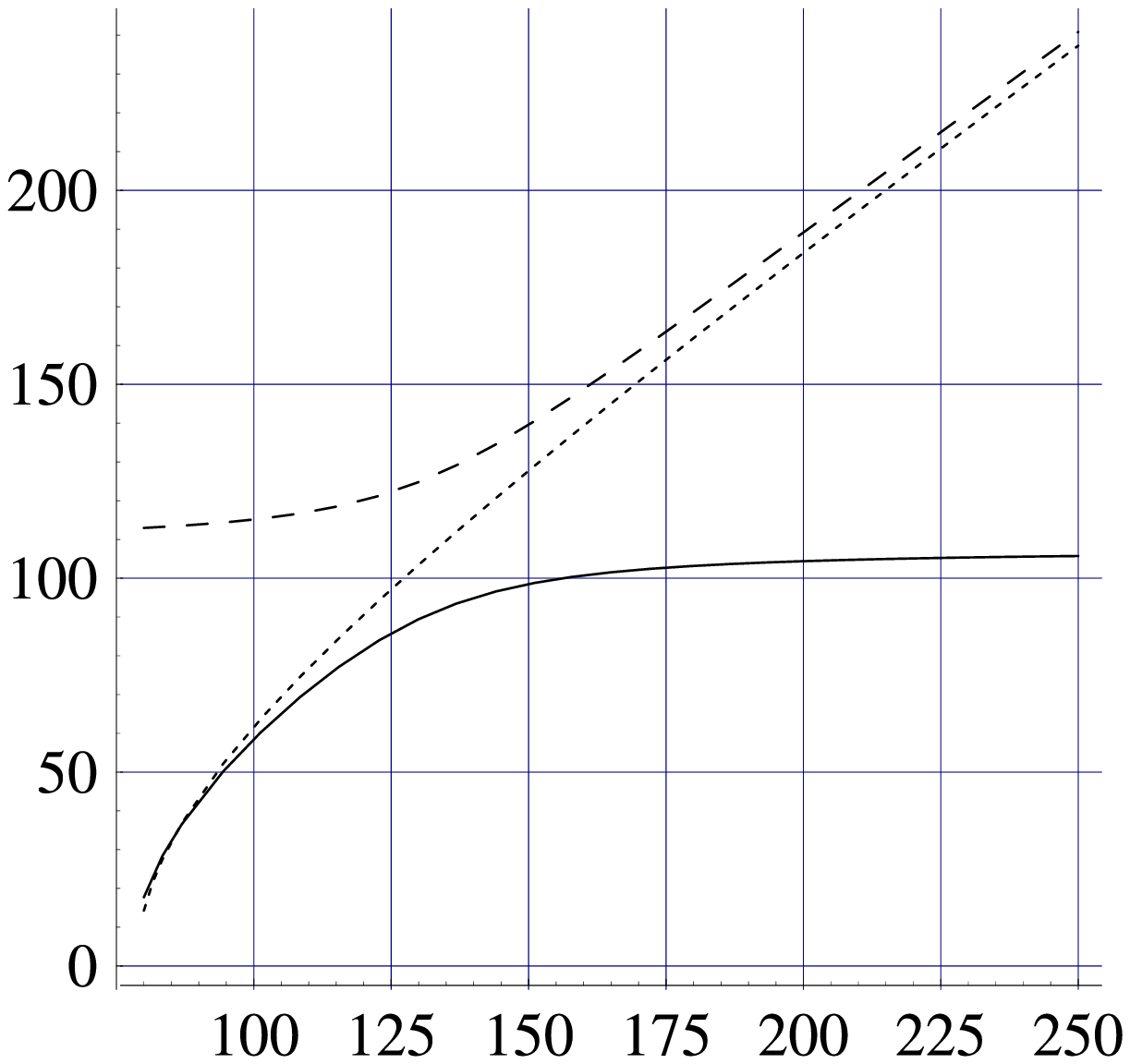}}
\put(-1.9,11.8){\mbox{(a)}}
\put(5.5,5.8){\epsfxsize=7.0cm
         \epsfysize=7.0cm \leavevmode \epsfbox{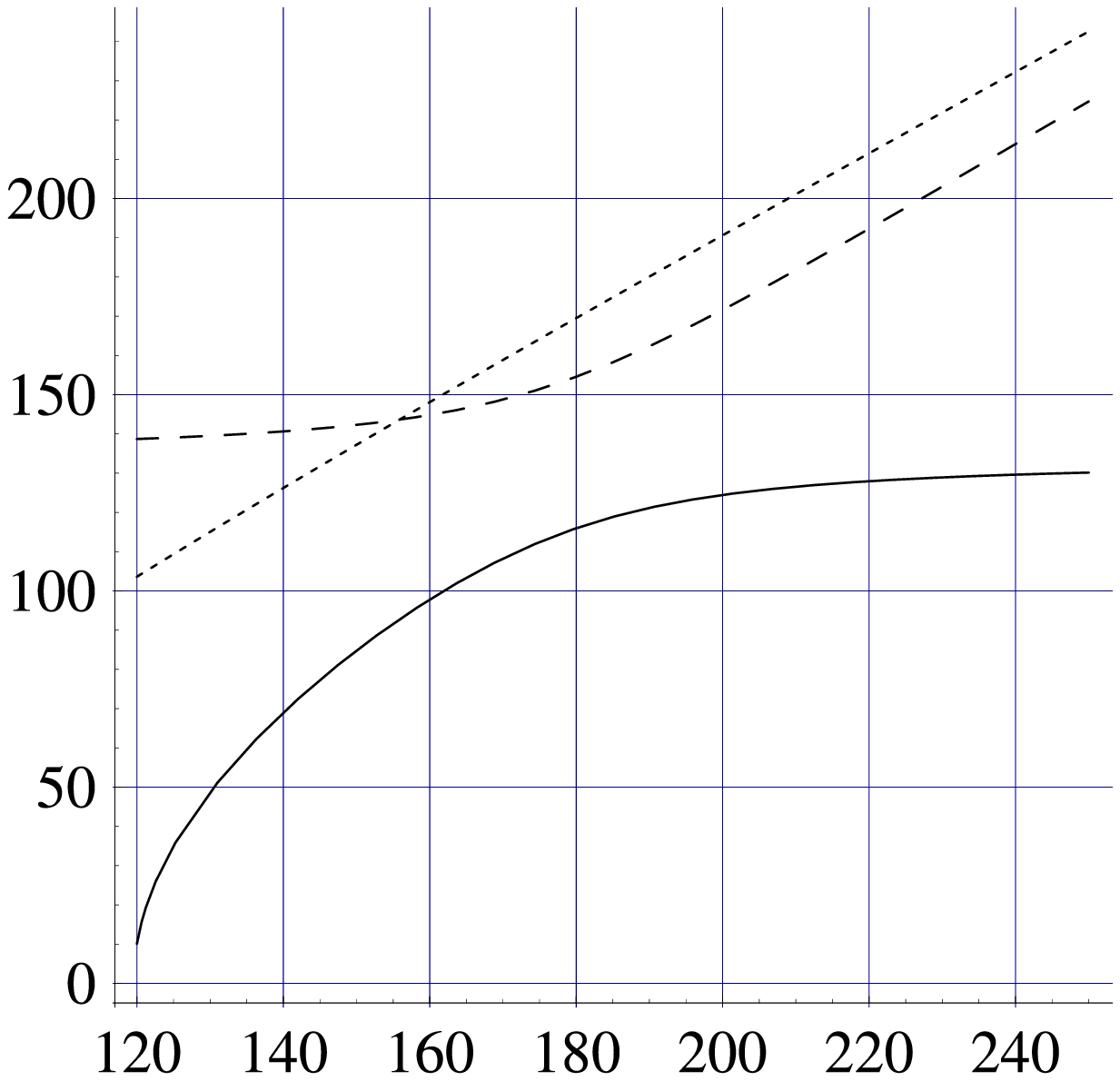}}
\put(6.6,11.9){\mbox{(b)}}
\put(-2.8,-1.5){\epsfxsize=7.0cm
         \epsfysize=7.0cm \leavevmode \epsfbox{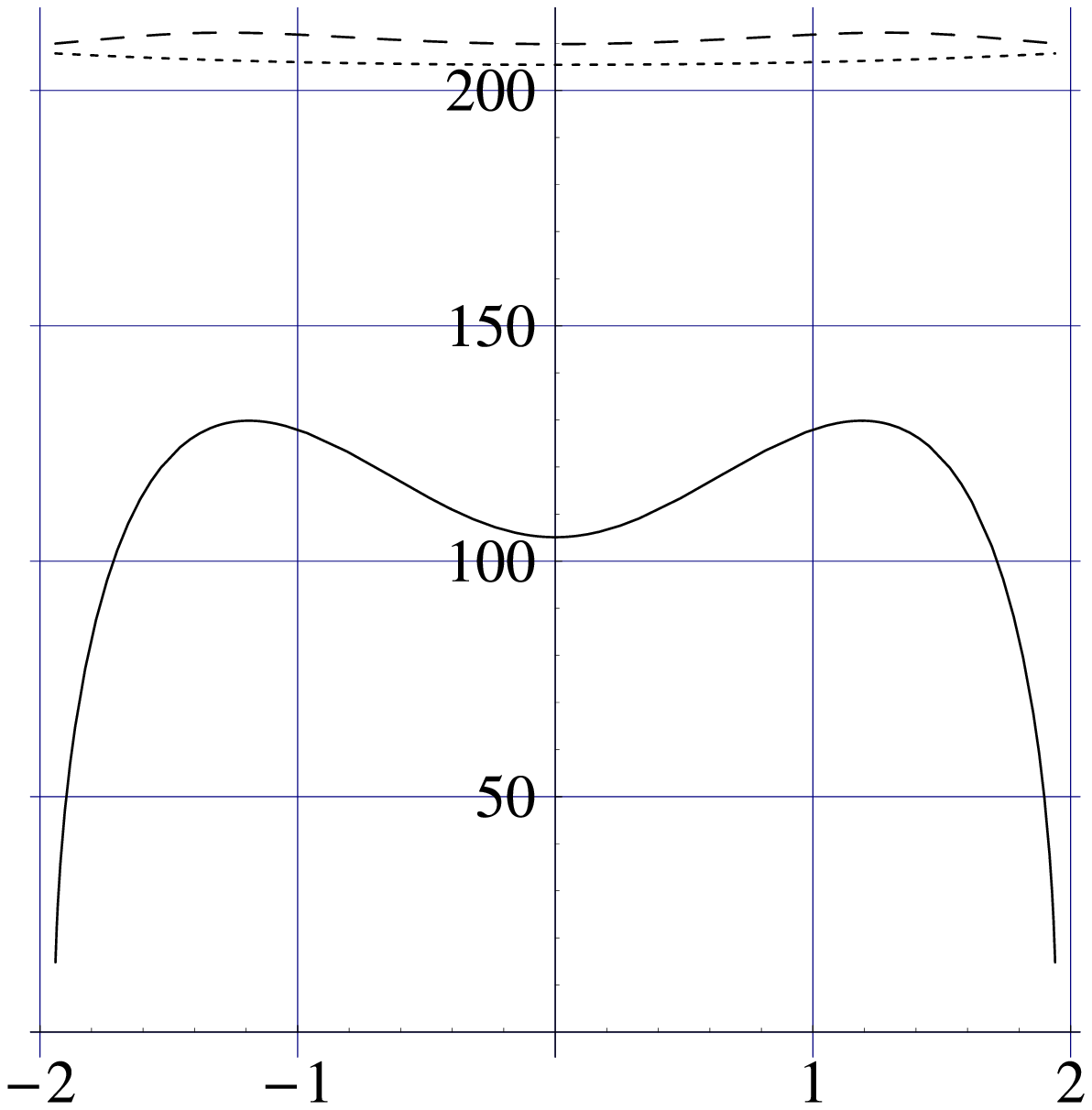}}
\put(-1.8,4.3){\mbox{(c)}}
\put( 5.5,-1.5){\epsfxsize=7.0cm
         \epsfysize=7.0cm \leavevmode \epsfbox{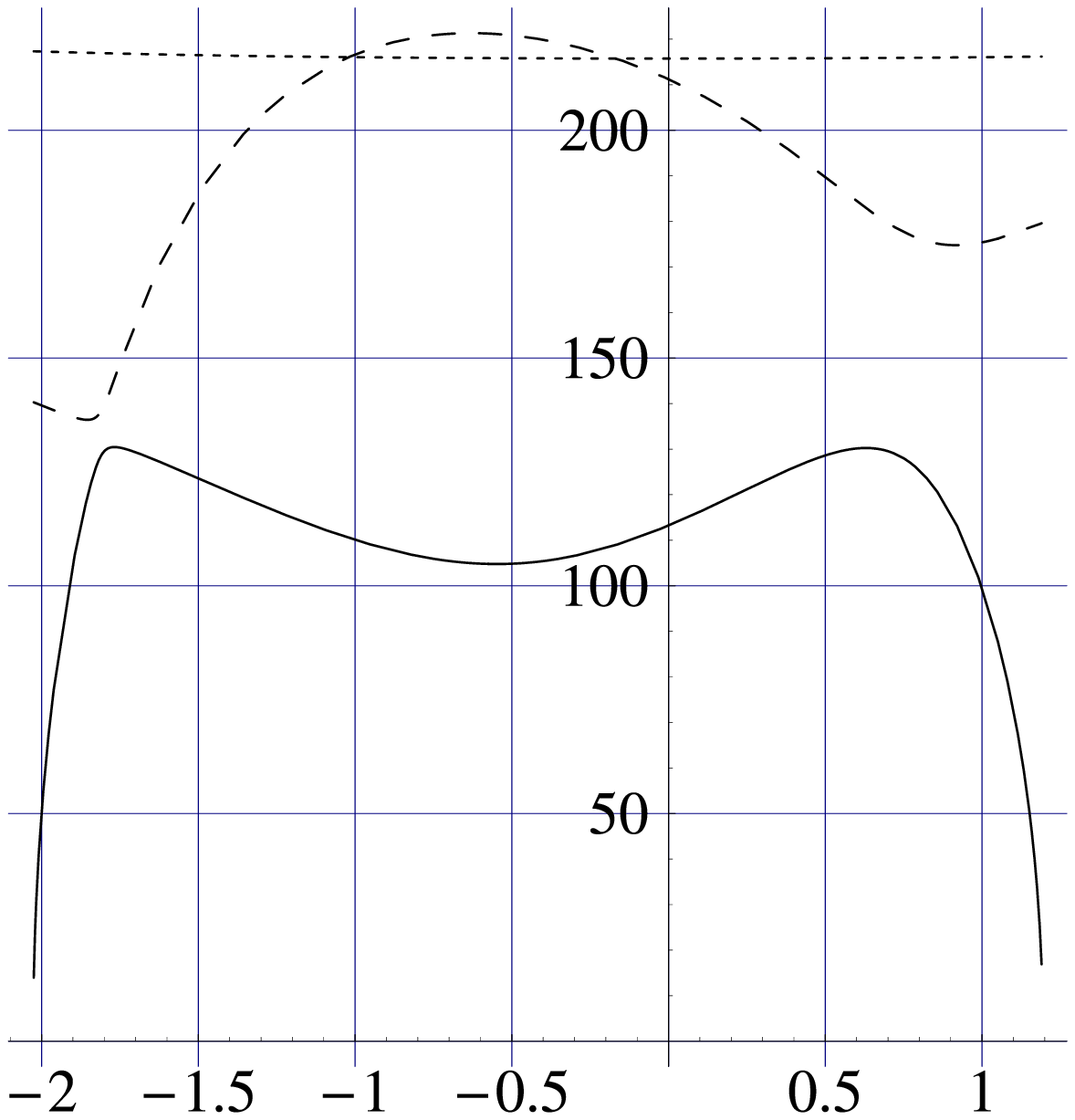}}
\put(6.2,4.3){\mbox{(d)}}
\end{picture}
\end{center}
\vspace{15mm}
\caption[]{\label{fg:charge1}
Neutral Higgs boson masses $h,H,A$ versus
$m_{H^{\pm}}$ and the trilinear parameters $A_t$, $A_b$ in the $CP$
conserving limit. Solid line denotes the
$m_h$ mass, short dashed -- $m_A$, long-dashed
 -- $m_H$. (a) ${\tt tg} \beta=$5, $M_{SUSY}=$0.5 TeV,
$A_t=A_b=\mu=$0; (b) ${\tt tg} \beta=$5, $M_{SUSY}=$0.5 TeV,
$A_t=A_b=$0.9 TeV, $\mu=-1.5$ TeV; (c) ${\tt tg} \beta=$5,
$M_{SUSY}=$0.5 TeV, $m_{H^{\pm}}=$220 GeV, $\mu=$0, $A_t=A_b$; (d)
${\tt tg} \beta=$5, $M_{SUSY}=$0.5 TeV, $m_{H^{\pm}}=$220 GeV,
$\mu=-$2 TeV, $A_t=A_b$.}
\end{figure}

\newpage

\newpage
\unitlength=1.0cm
\begin{figure}[t]
\begin{center}
\begin{picture}(10,10)
\put(9,-1.7){\mbox{${\tt arg}(\mu A)$}}
\put(-2.8,5.8){\epsfxsize=7.0cm
         \epsfysize=7.0cm \leavevmode \epsfbox{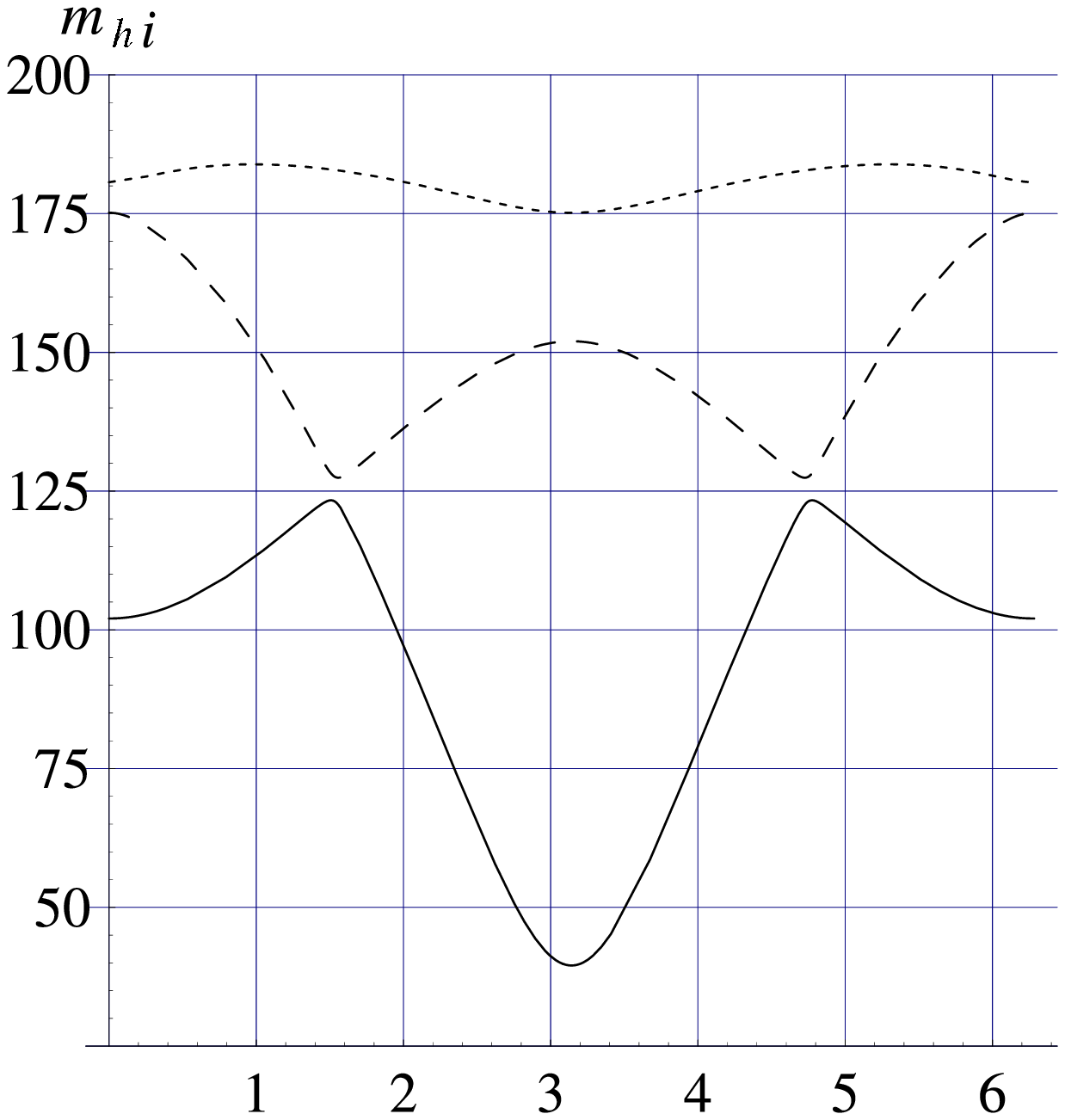}}
\put(-1.9,11.8){\mbox{(a)}}
\put(5.5,5.8){\epsfxsize=7.0cm
         \epsfysize=7.0cm \leavevmode \epsfbox{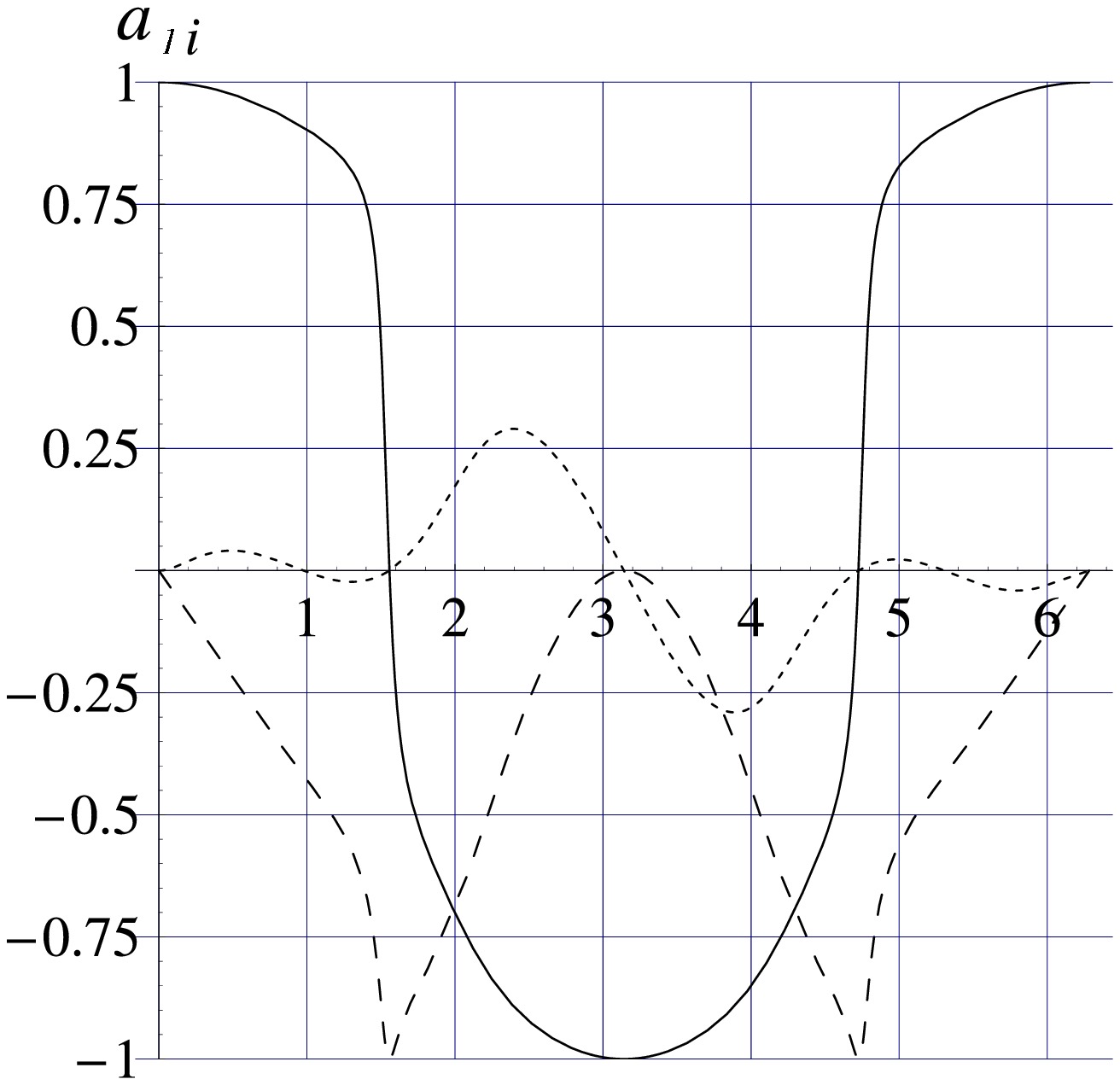}}
\put(6.6,11.6){\mbox{(b)}}
\put(-2.8,-1.5){\epsfxsize=7.0cm
         \epsfysize=7.0cm \leavevmode \epsfbox{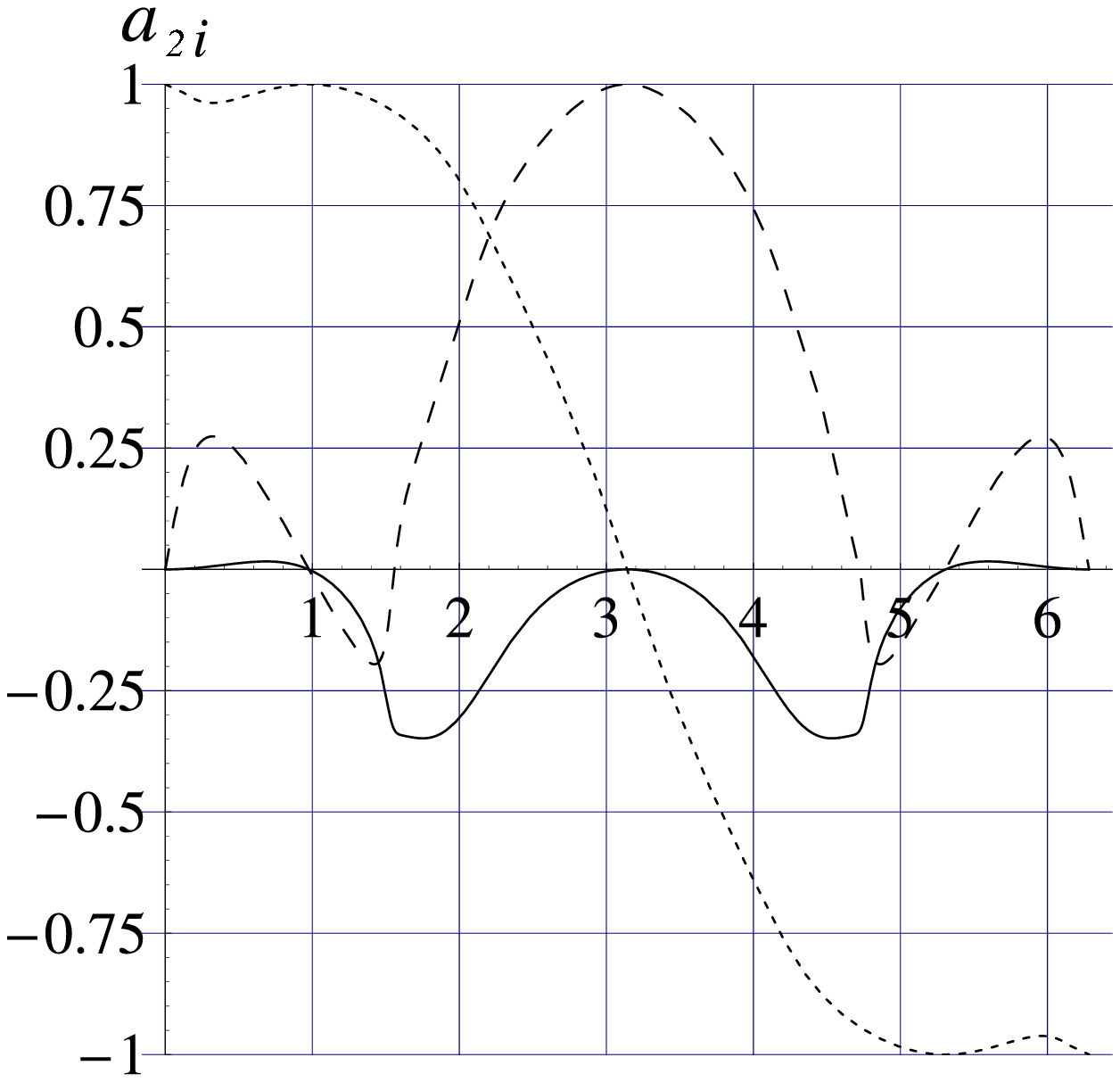}}
\put(-1.8,4.3){\mbox{(c)}}
\put( 5.5,-1.5){\epsfxsize=7.0cm
         \epsfysize=7.0cm \leavevmode \epsfbox{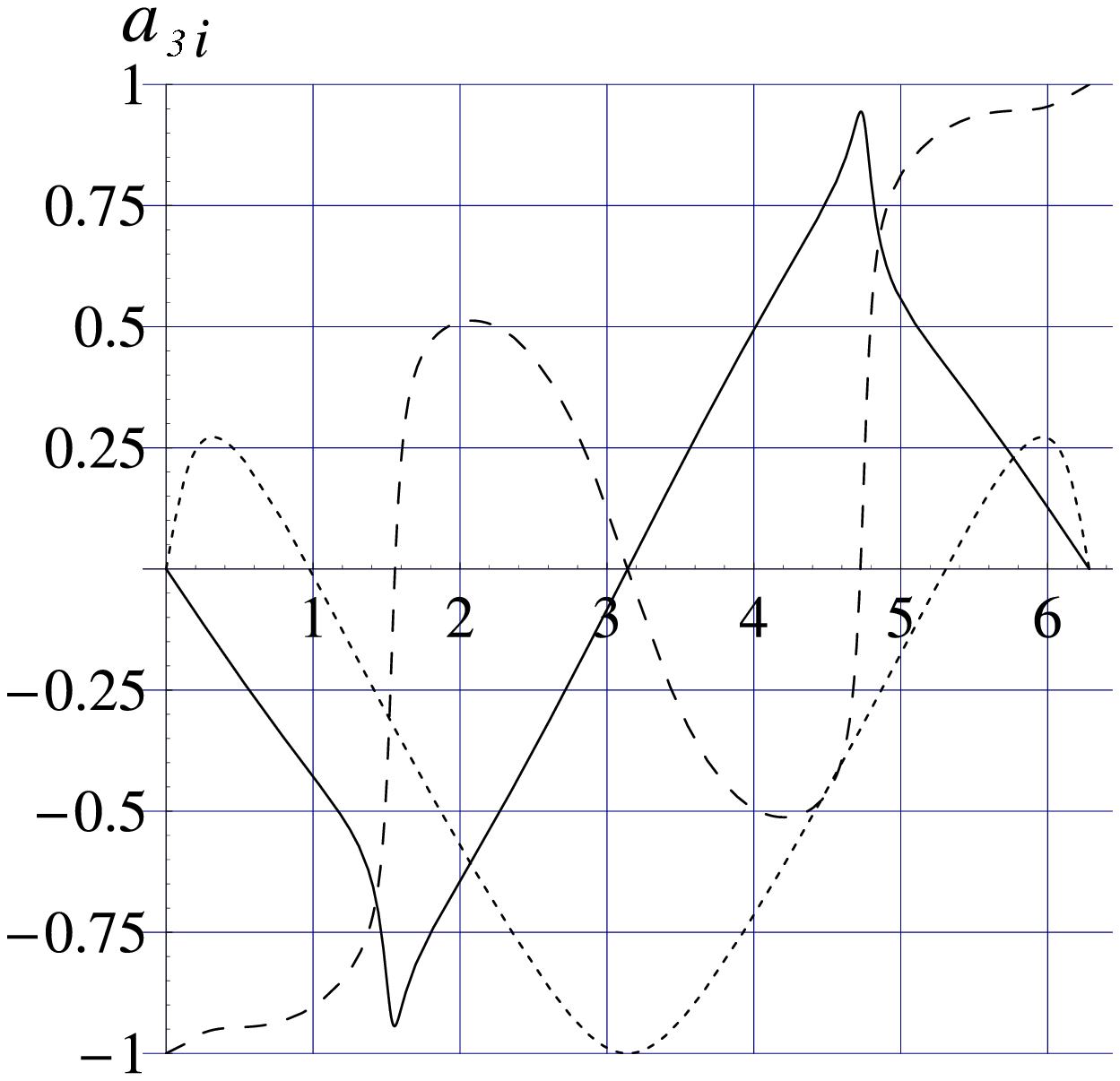}}
\put(6.6,4.3){\mbox{(d)}}
\end{picture}
\end{center}
\vspace{15mm}
\caption{ \label{fg:massi}
(a) Neutral Higgs boson masses, (b),(c),(d) the matrix
elements
$a_{ij}$ versus the phase $\varphi={\tt arg}(\mu A)$
at the parameter values ${\rm tg}\beta=$5, $m_{H^{\pm}}=$180 GeV,
$M_{SUSY}=$0.5 TeV, $A_t=A_b=$1 TeV, $\mu=$2 TeV. Solid line denotes
$i=1$, long dashed -- $i=2$ and short dashed
-- $i=3$. }
\end{figure}

\newpage
\unitlength=1.0cm
\begin{figure}[t]
\begin{center}
\begin{picture}(10,10)
\put(9,-1.7){\mbox{${\tt arg}(\mu A)$}}
\put(-2.8,5.8){\epsfxsize=7.0cm
         \epsfysize=7.0cm \leavevmode \epsfbox{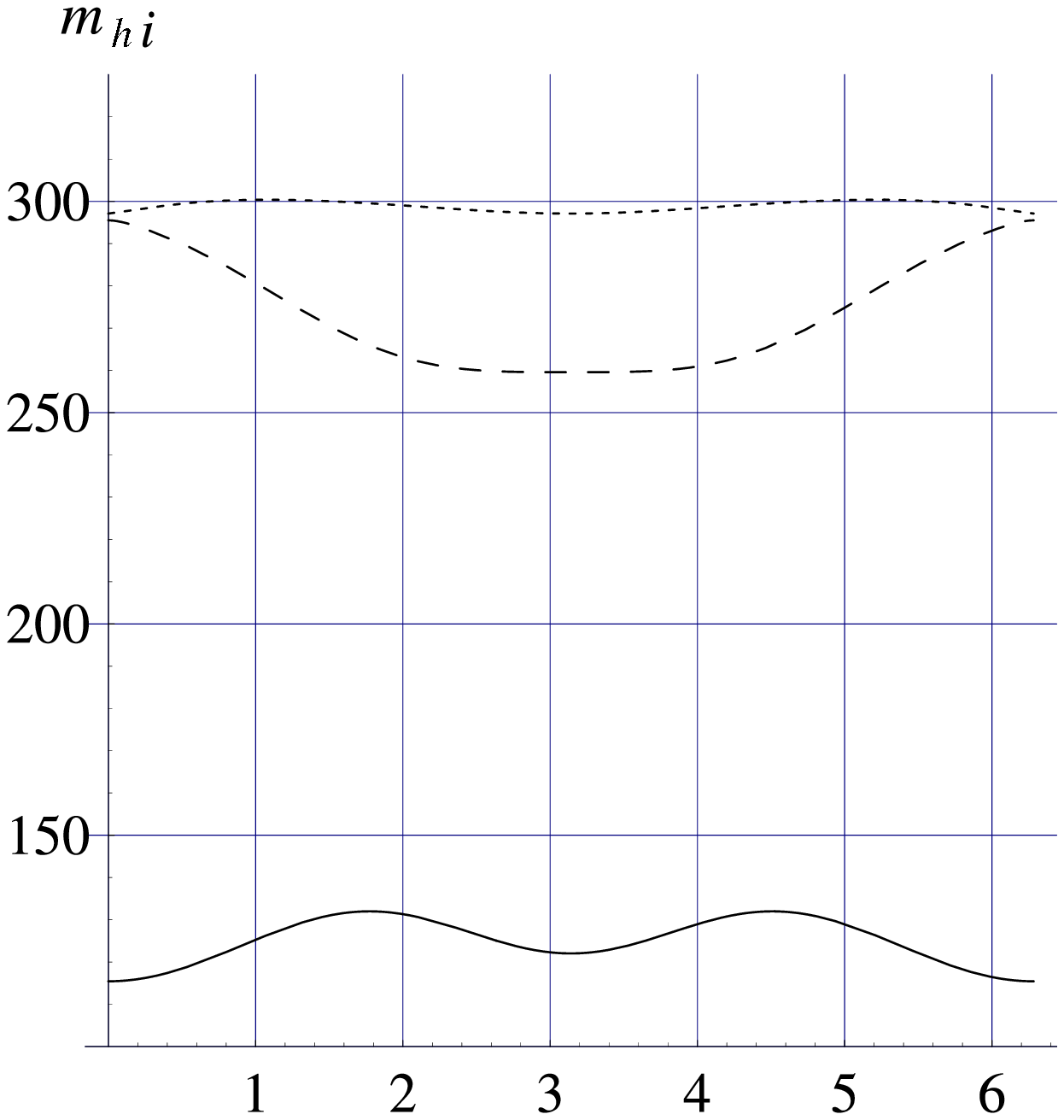}}
\put(-1.9,11.8){\mbox{(a)}}
\put(5.5,5.8){\epsfxsize=7.0cm
         \epsfysize=7.0cm \leavevmode \epsfbox{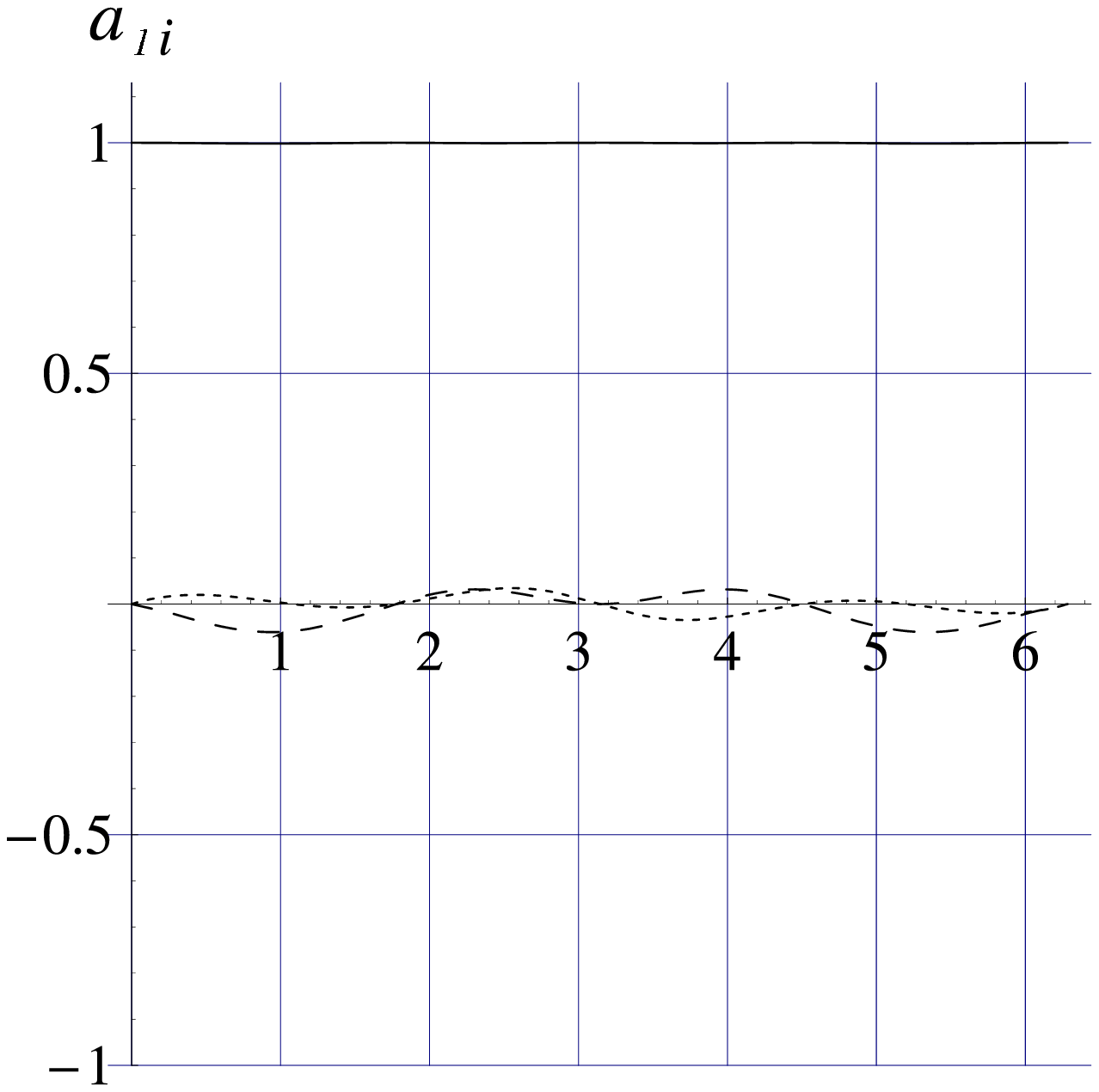}}
\put(6.6,11.6){\mbox{(b)}}
\put(-2.8,-1.5){\epsfxsize=7.0cm
         \epsfysize=7.0cm \leavevmode \epsfbox{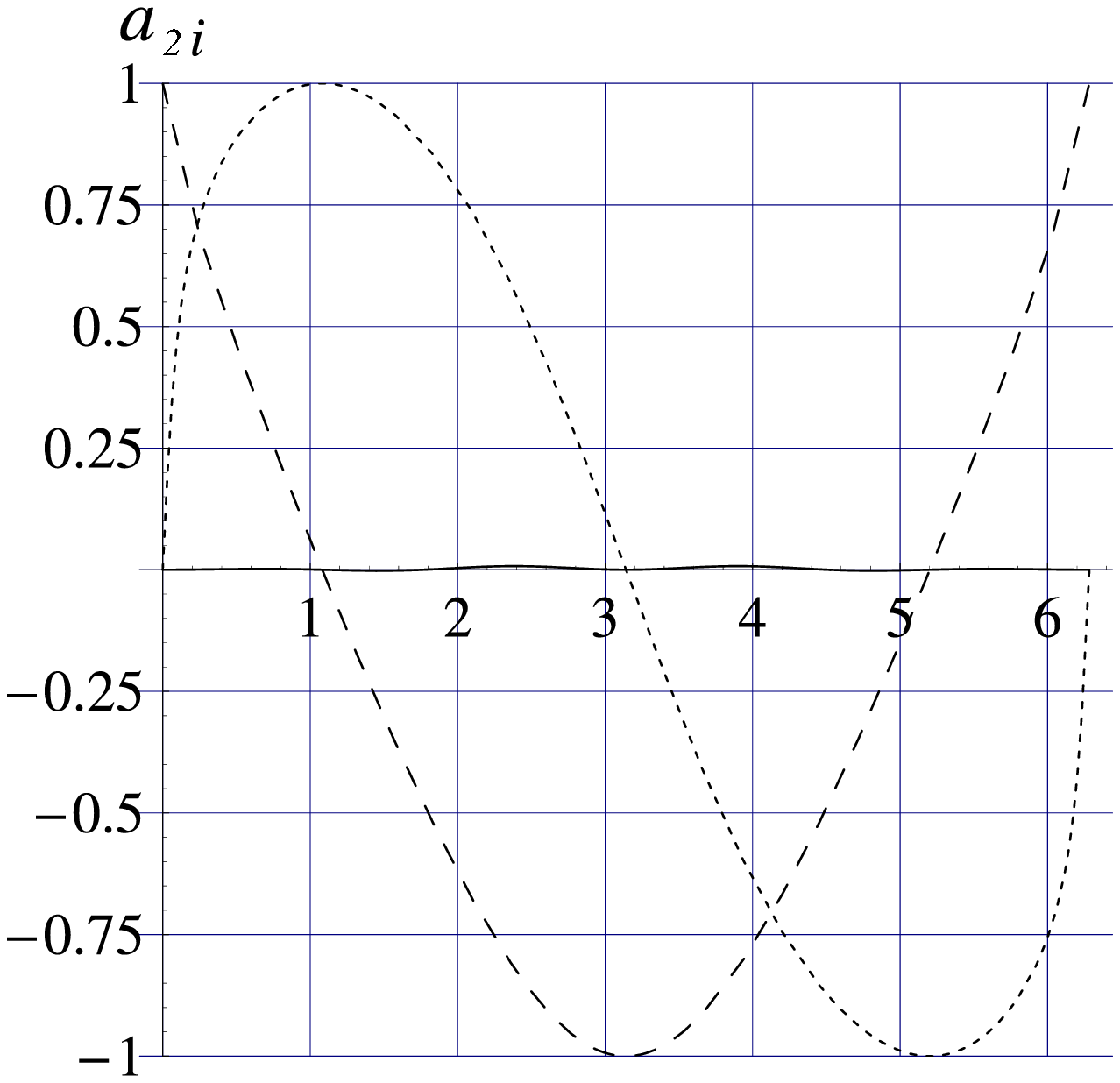}}
\put(-1.8,4.3){\mbox{(c)}}
\put( 5.5,-1.5){\epsfxsize=7.0cm
         \epsfysize=7.0cm \leavevmode \epsfbox{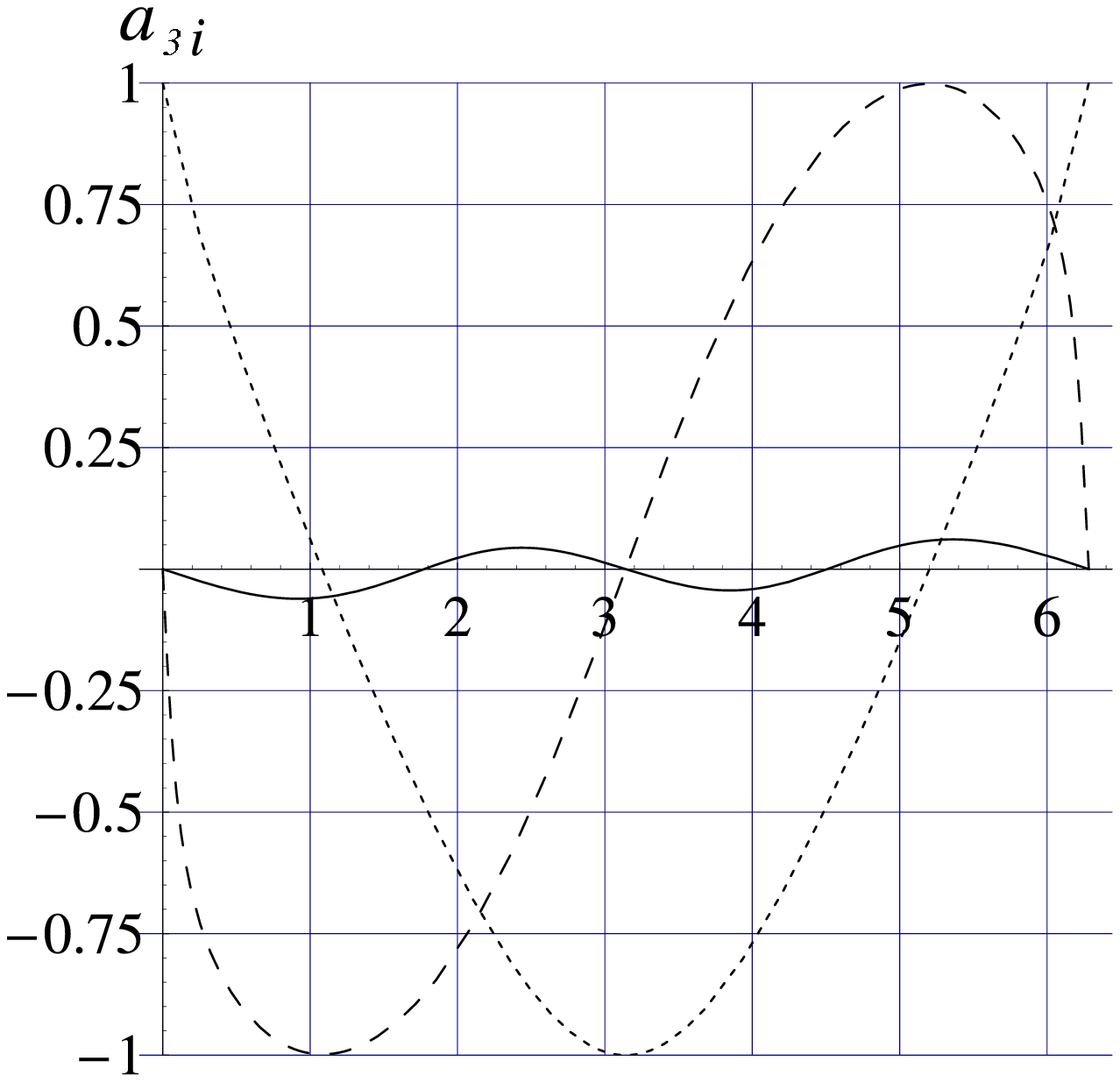}}
\put(6.6,4.3){\mbox{(d)}}
\end{picture}
\end{center}
\vspace{15mm}
\caption{
(a) Neutral Higgs boson masses, (b),(c),(d) the matrix
elements
$a_{ij}$ versus the phase $\varphi={\tt arg}(\mu A)$
at the parameter values ${\rm tg}\beta=$5, $m_{H^{\pm}}=$300 GeV,
$M_{SUSY}=$0.5 TeV, $A_t=A_b=$1 TeV, $\mu=$2 TeV. Solid line denotes
$i=1$, long dashed -- $i=2$ and short dashed
-- $i=3$. }
\end{figure}

\unitlength=1.0cm
\begin{figure}[t]
\begin{center}
\begin{picture}(10,10)
\put(-3.9,12.0){\mbox{\Large \bf $m_{h_1}$}}
\put(9,-1.7){\mbox{${\tt arg}(\mu A)$}}
\put(-2.8,5.8){\epsfxsize=7.0cm
         \epsfysize=7.0cm \leavevmode \epsfbox{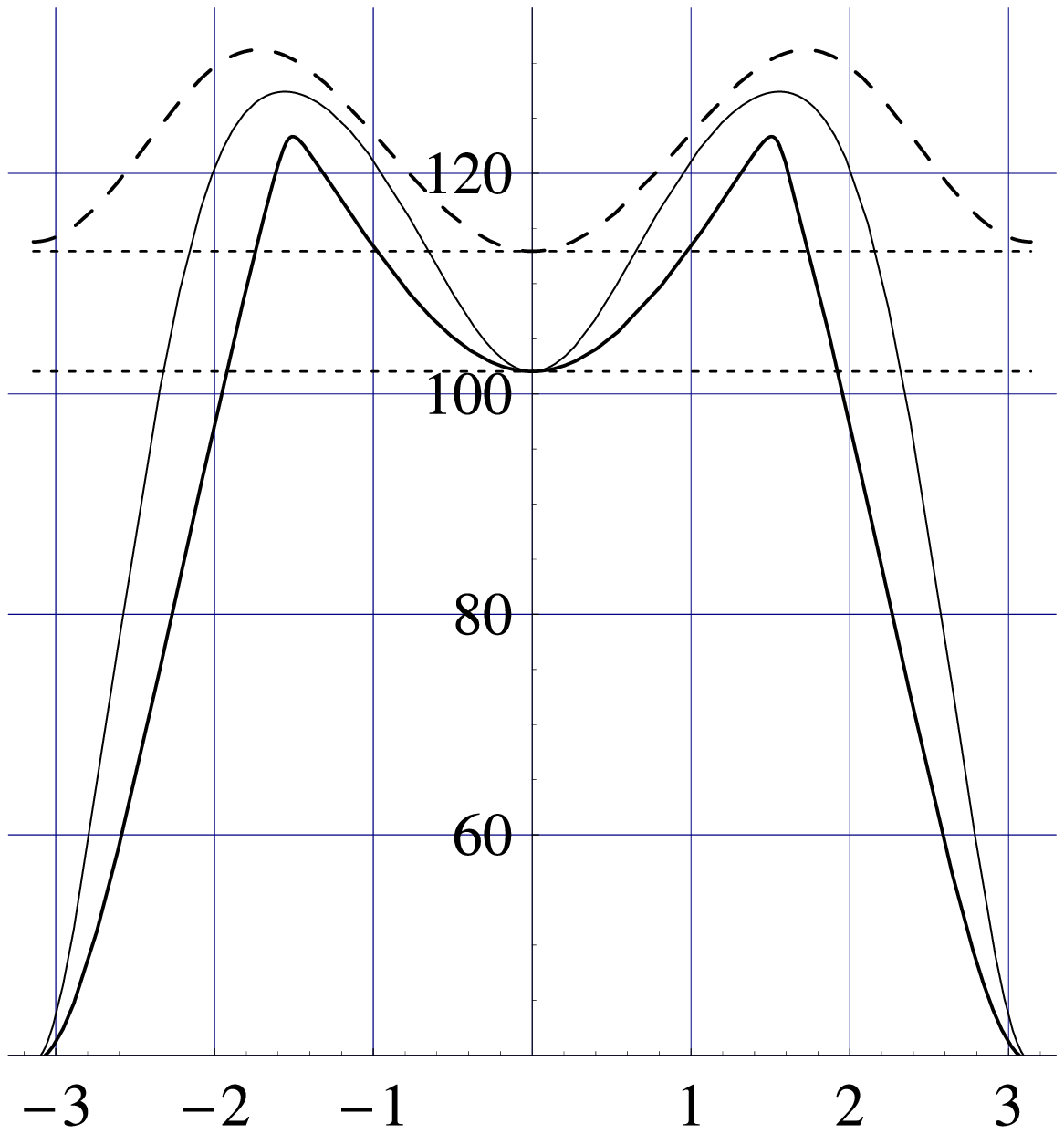}}
\put(-2.0,12.5){\mbox{(a)}}
\put(5.5,5.8){\epsfxsize=7.0cm
         \epsfysize=7.0cm \leavevmode \epsfbox{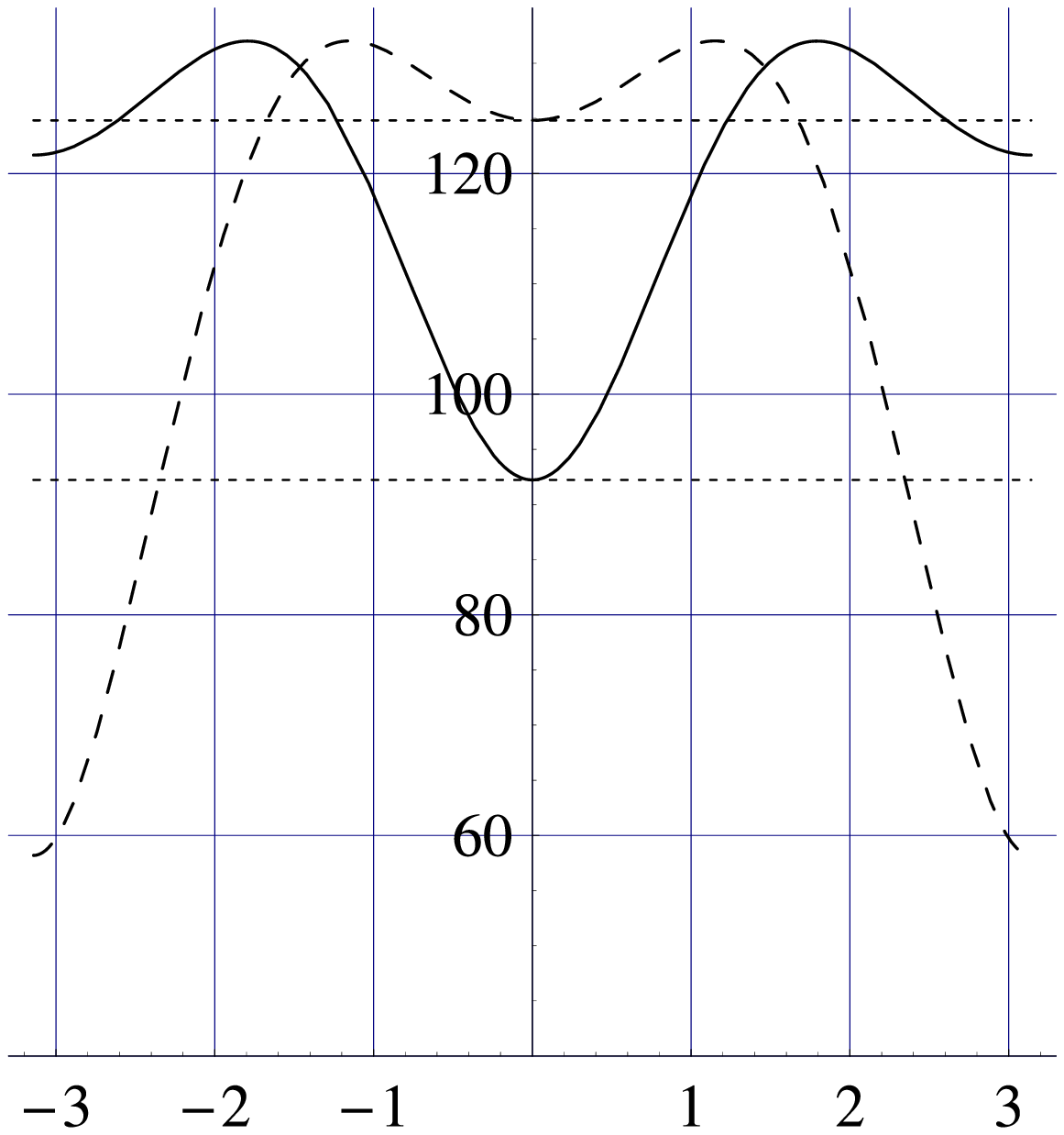}}
\put(6.4,12.5){\mbox{(b)}}
\put(-2.8,-1.5){\epsfxsize=7.0cm
         \epsfysize=7.0cm \leavevmode \epsfbox{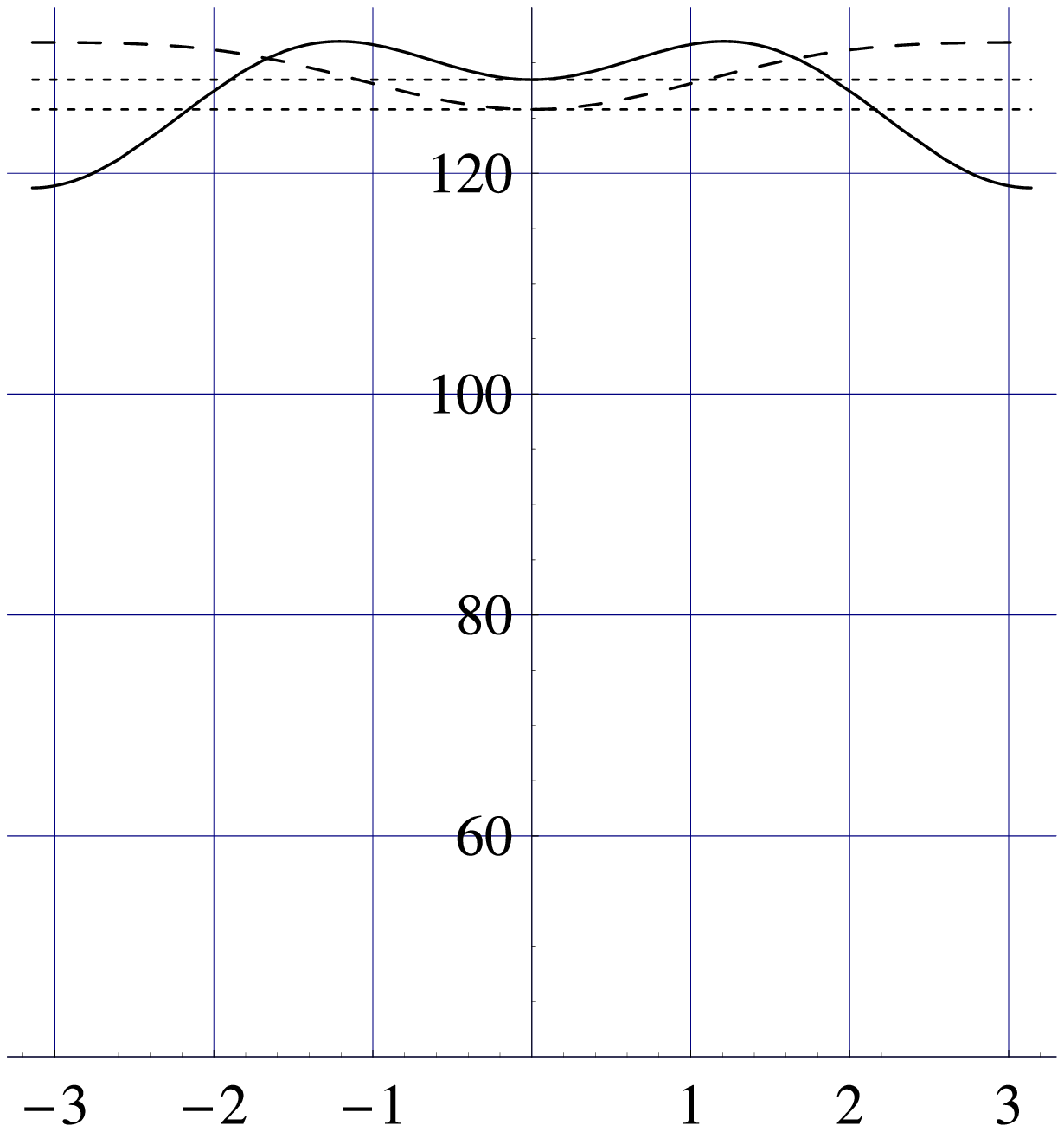}}
\put(-2.0,3.7){\mbox{(c)}}
\put( 5.5,-1.5){\epsfxsize=7.0cm
         \epsfysize=7.0cm \leavevmode \epsfbox{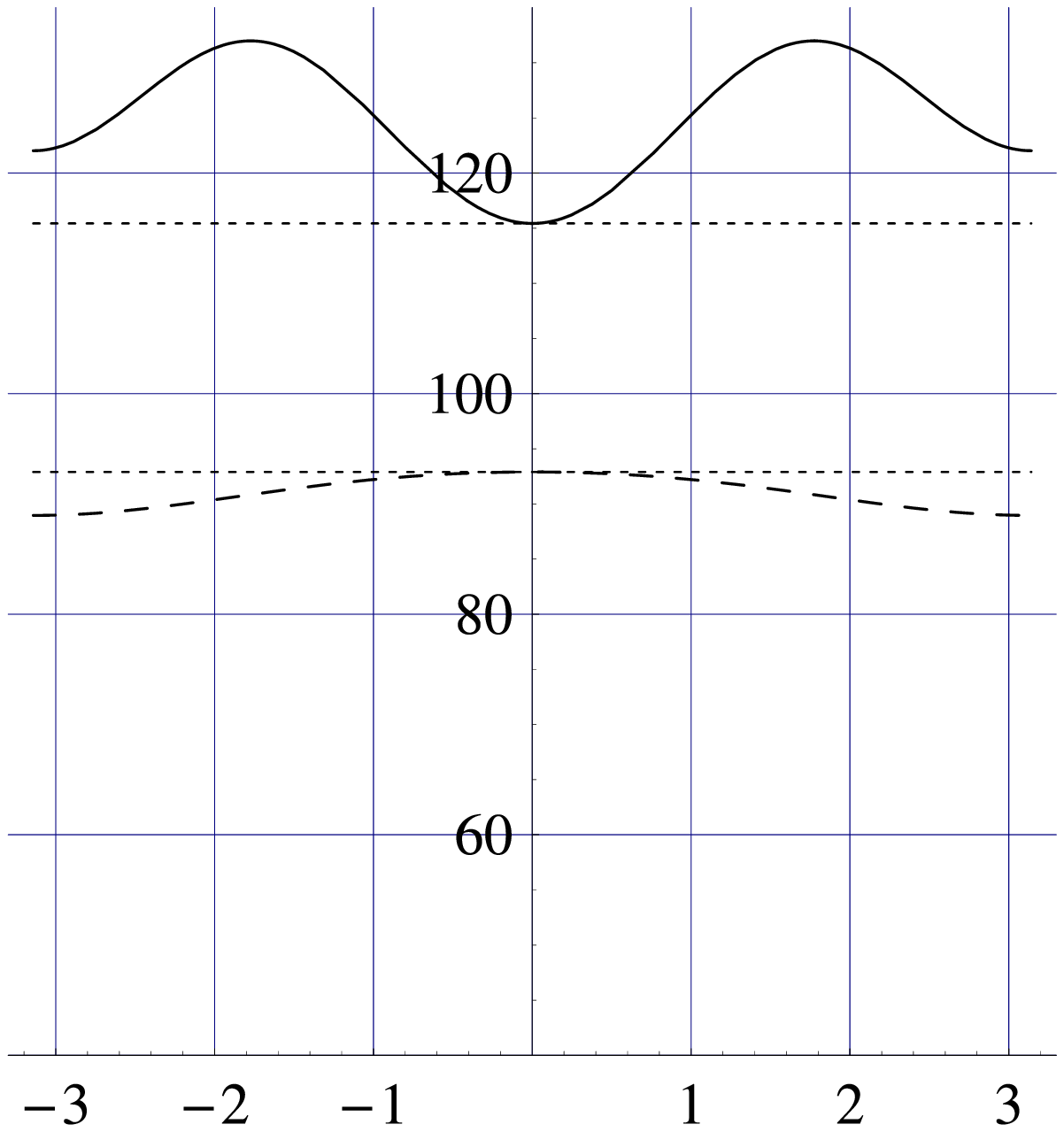}}
\put(6.4,3.7){\mbox{(d)}}
\end{picture}
\end{center}
\vspace{15mm}
\caption{ Light Higgs boson mass $m_{h_1}(\varphi)$ (GeV) vs ${\tt 
arg}({\mu
A})$ in various regions
of the MSSM parameter space. Horizontal dotted lines indicate the
$h_1$ mass in the $CP$-conserving limit ($m_{h_1}=m_h$).
(a) ${\rm tg}\beta=$5,
$M_{SUSY}=$0.5 TeV, $A_t=A_b=$1 TeV, $\mu=$2 TeV. Solid
line $m_{H^{\pm}}=$180 GeV, dashed $m_{H^{\pm}}=$250 GeV. Thin solid line
denotes $m_h(\varphi)$.
(b) ${\rm tg}\beta=$5, $m_{H^{\pm}}=$300 GeV, $\mu=$2 TeV,
solid line $A_t=A_b=$-1.2 TeV, dashed $A_t=A_b=$1.3 TeV.
(c) ${\rm tg}\beta=$5, $m_{H^{\pm}}=$300 GeV, $A_t=A_b=$1 TeV,
solid line $\mu=$-1.6 TeV, dashed $\mu=$0.7 TeV
(d) $\mu=$2 TeV, $m_{H^{\pm}}=$300 GeV, $A_t=A_b=$1 TeV,
solid line ${\rm tg}\beta=$5, dashed ${\rm tg}\beta=$40.
 }
\end{figure}

\unitlength=1.0cm
\begin{figure}[t]
\begin{center}
\begin{picture}(10,10)
\put(-4.9,12.0){\mbox{\bf $\Gamma(h_1\rightarrow gg$)}}
\put(9,-1.7){\mbox{${\tt arg}(\mu A)$}}
\put(-2.8,5.8){\epsfxsize=7.0cm
         \epsfysize=7.0cm \leavevmode \epsfbox{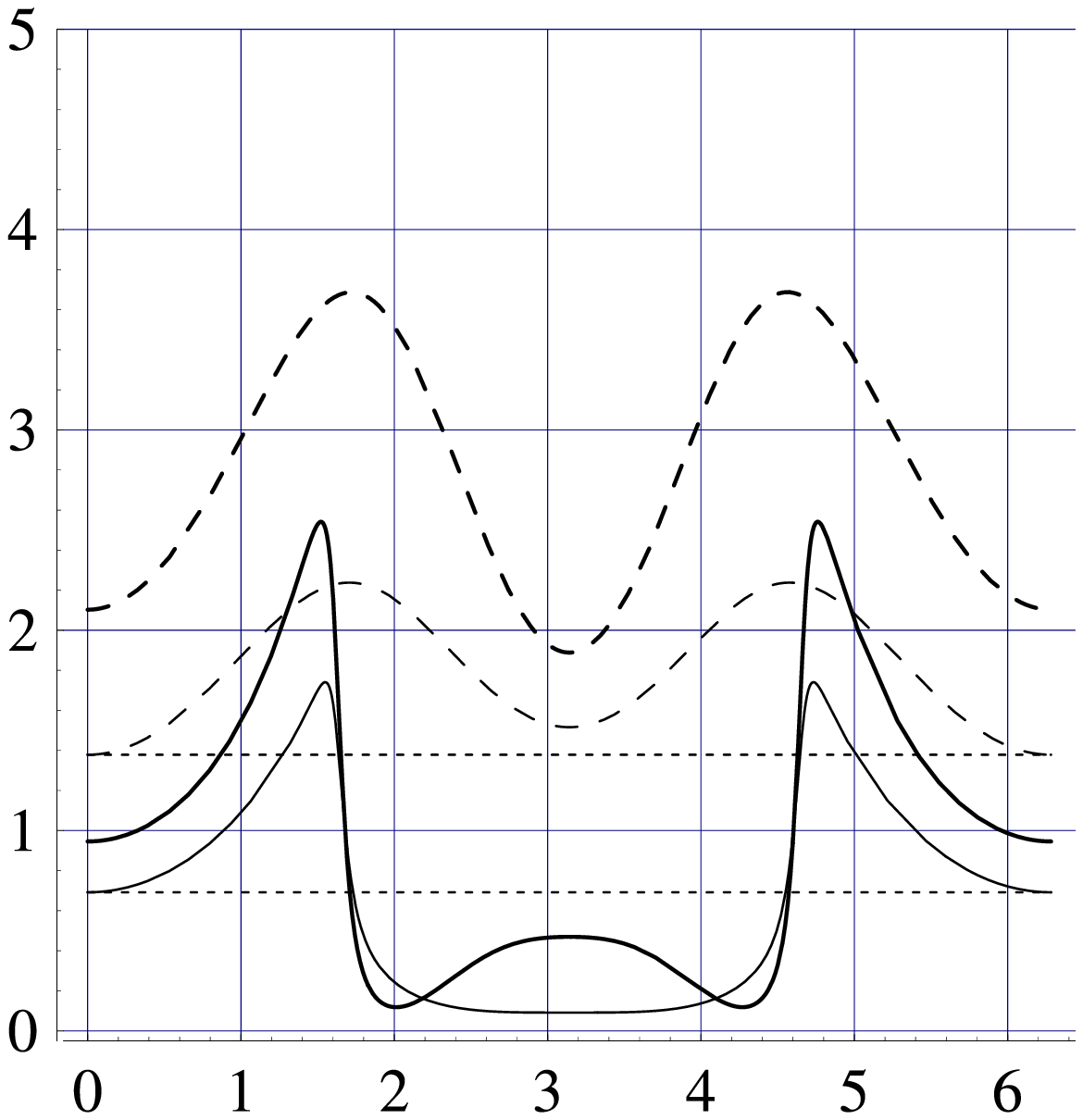}}
\put(-1.8,12.0){\mbox{(a)}}
\put(5.5,5.8){\epsfxsize=7.0cm
         \epsfysize=7.0cm \leavevmode \epsfbox{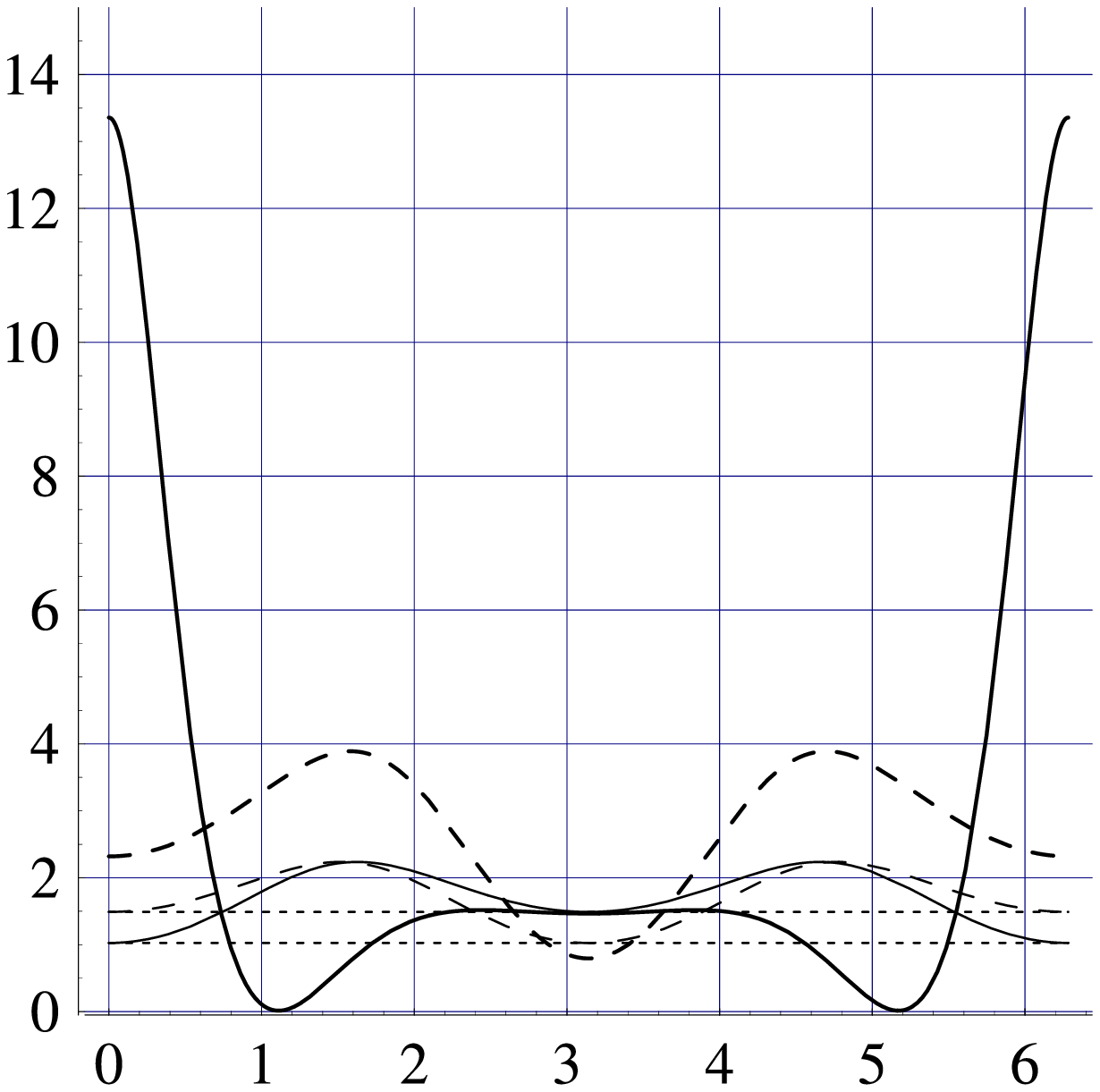}}
\put(6.6,12.0){\mbox{(b)}}
\put(-2.8,-1.5){\epsfxsize=7.0cm
         \epsfysize=7.0cm \leavevmode \epsfbox{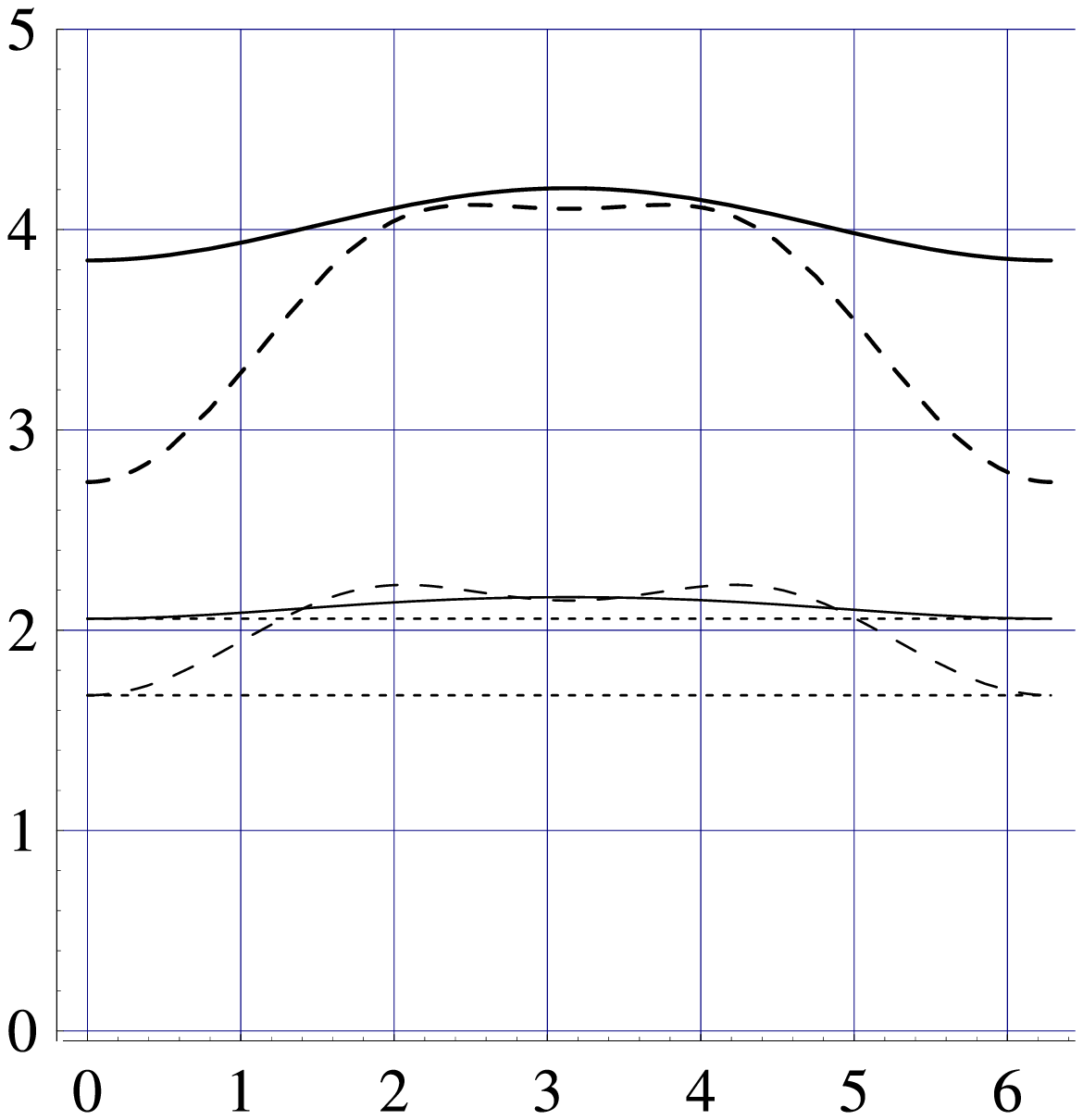}}
\put(-1.8,4.7){\mbox{(c)}}
\put( 5.5,-1.5){\epsfxsize=7.0cm
         \epsfysize=7.0cm \leavevmode \epsfbox{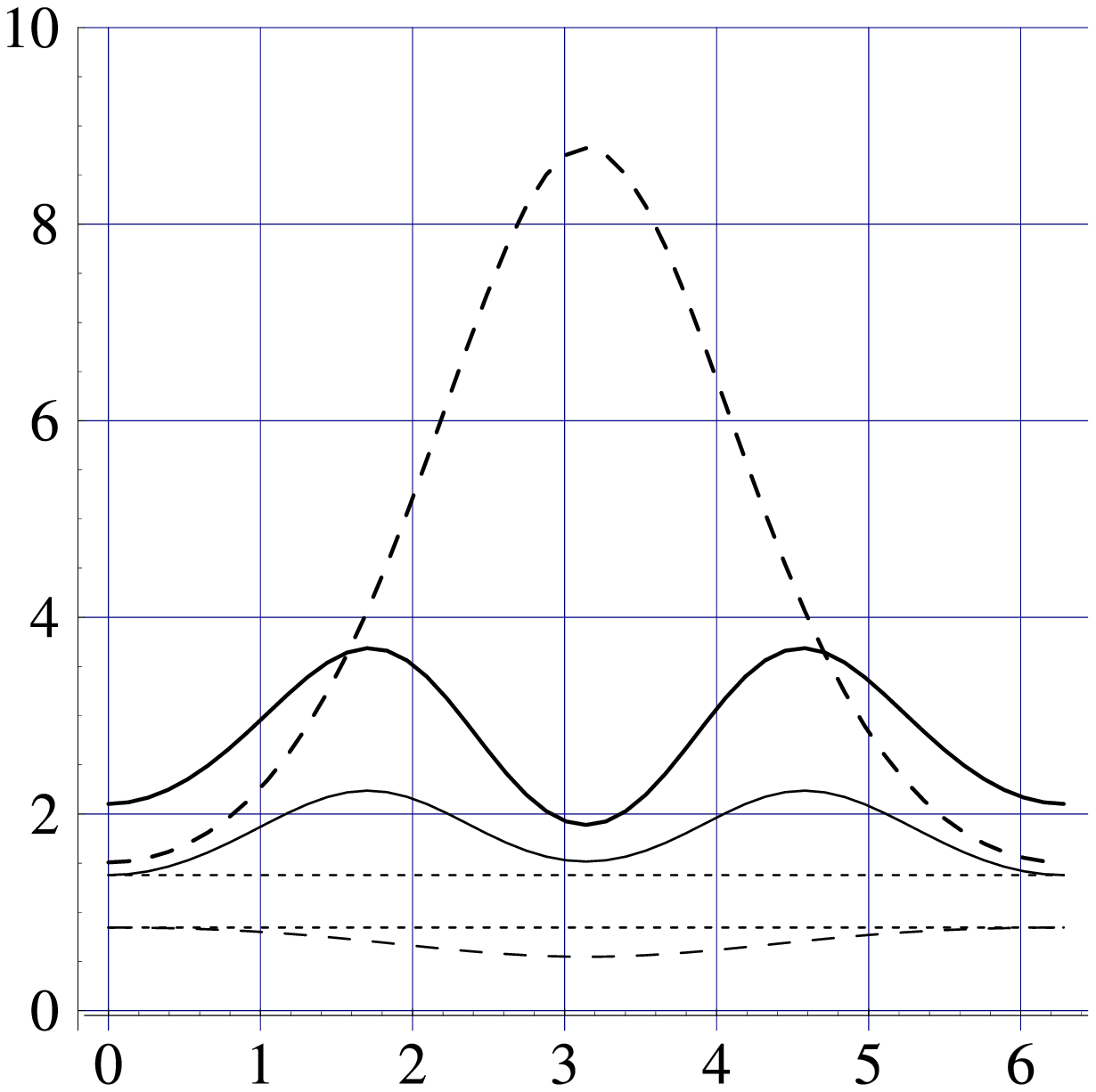}}
\put(6.6,4.7){\mbox{(d)}}
\end{picture}
\end{center}
\vspace{15mm}
\caption{ 
The decay width $\Gamma(h_1\rightarrow gg)\times 10^{4}$ (GeV)
at $M_{SUSY}=$500 GeV.
Dotted lines show $\Gamma_h$ in the
$CP$ conserving limit $\varphi=0$.
Thin solid or dashed lines are only for SM contributions, thick 
solid or dashed 
lines show SM and sparticle contrbutios with K-factor included.
(a) ${\rm tg}\beta=$5, $A_t=A_b=$1 TeV, $\mu=$2 TeV, solid line
$m_{H^\pm}=$190 GeV, dashed $m_{H^\pm}=$300 GeV, 
(b) ${\rm tg}\beta=$5, $m_{H^\pm}=$300 GeV, $\mu=$2 TeV, solid line
$A_t=A_b=$-1.1 TeV, dashed $A_t=A_b=$1.1 TeV,
(c) ${\rm tg}\beta=$5, $m_{H^\pm}=$300 GeV, $A_t=A_b=$1 TeV, solid line
$\mu=$0.2 TeV, dashed $\mu=$1.2 TeV,
(d) $\mu=$2 TeV, $m_{H^\pm}=$300 GeV, $A_t=A_b=$1 TeV, solid line
${\rm tg}\beta=$5, dashed ${\rm tg}\beta=$40   }

\end{figure}


\newpage

\unitlength=1.0cm
\begin{figure}[t]
\begin{center}
\begin{picture}(10,10)
\put(-4.9,12.0){\mbox{\bf $\Gamma(h_1\rightarrow gg$)}}
\put(2.8,5.6){\mbox{$m_{H^\pm}$}}
\put(12.2,5.7){\mbox{$A_t=A_b$}}
\put(2.8,-1.7){\mbox{$\mu$}}
\put(11,-1.7){\mbox{${\tt tg}(\beta)$}}
\put(-2.8,5.8){\epsfxsize=7.0cm
         \epsfysize=7.0cm \leavevmode \epsfbox{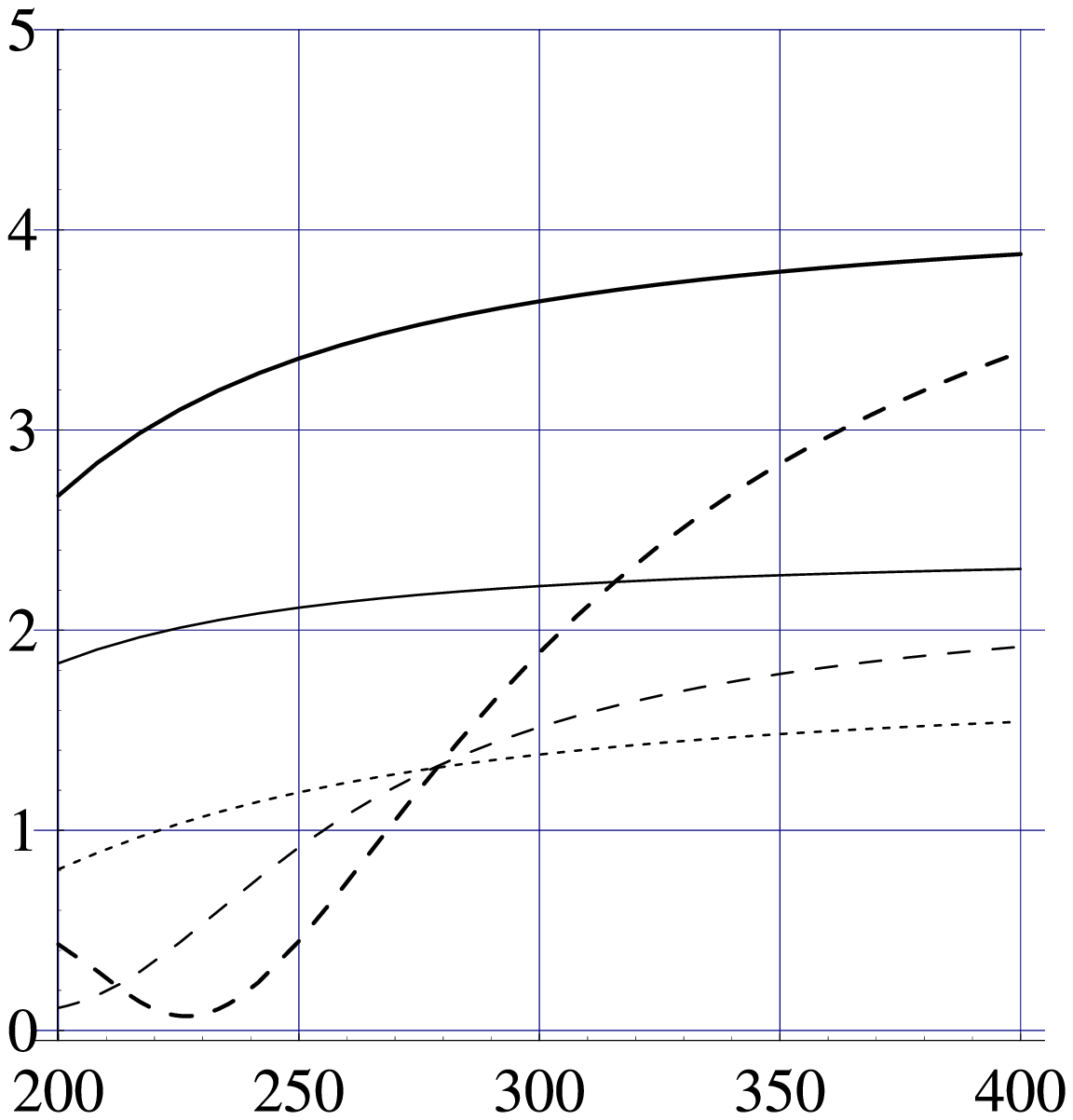}}
\put(-1.8,12.0){\mbox{(a)}}
\put(5.5,5.8){\epsfxsize=7.0cm
         \epsfysize=7.0cm \leavevmode \epsfbox{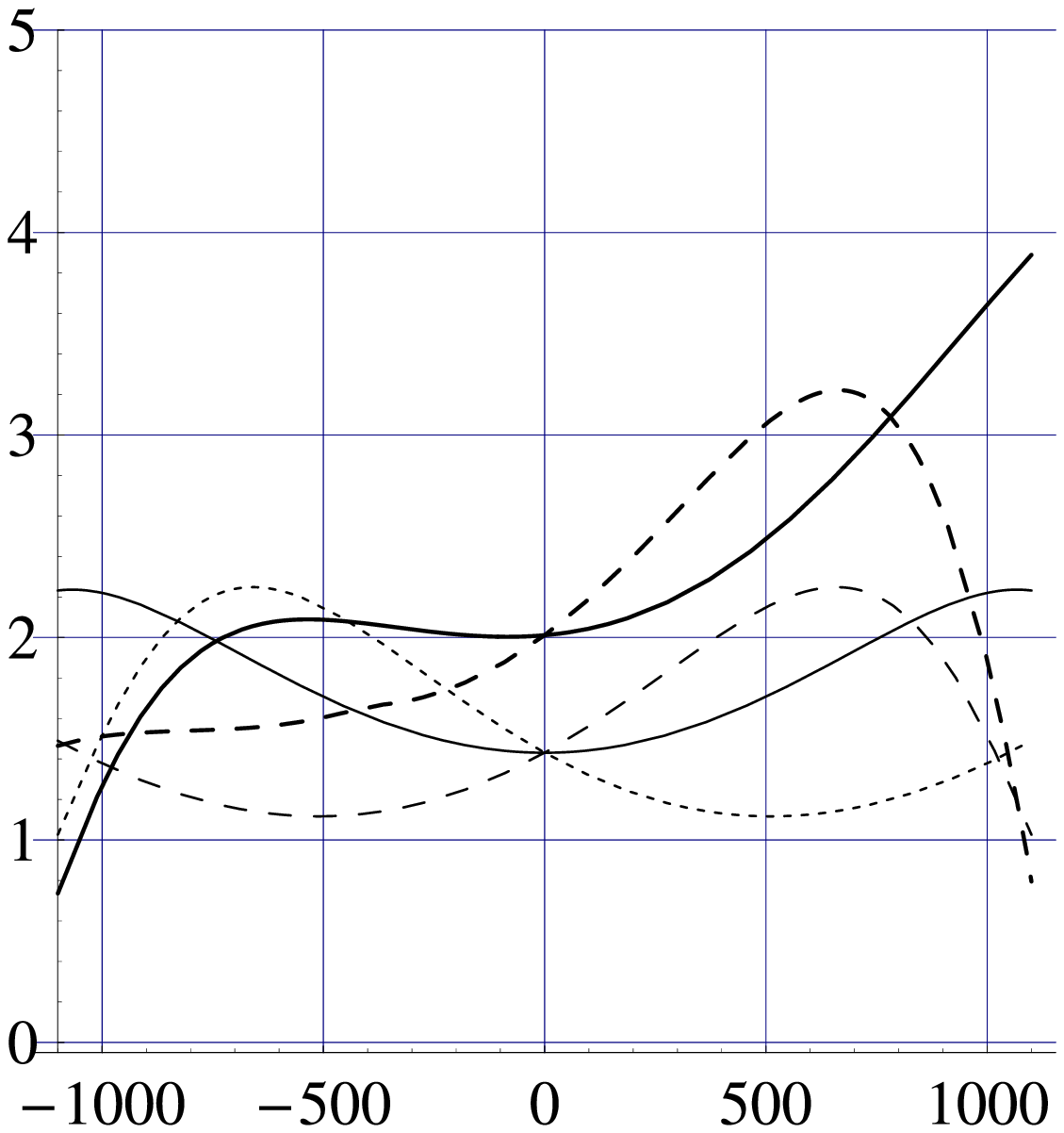}}
\put(6.2,12.0){\mbox{(b)}}
\put(-2.8,-1.5){\epsfxsize=7.0cm
         \epsfysize=7.0cm \leavevmode \epsfbox{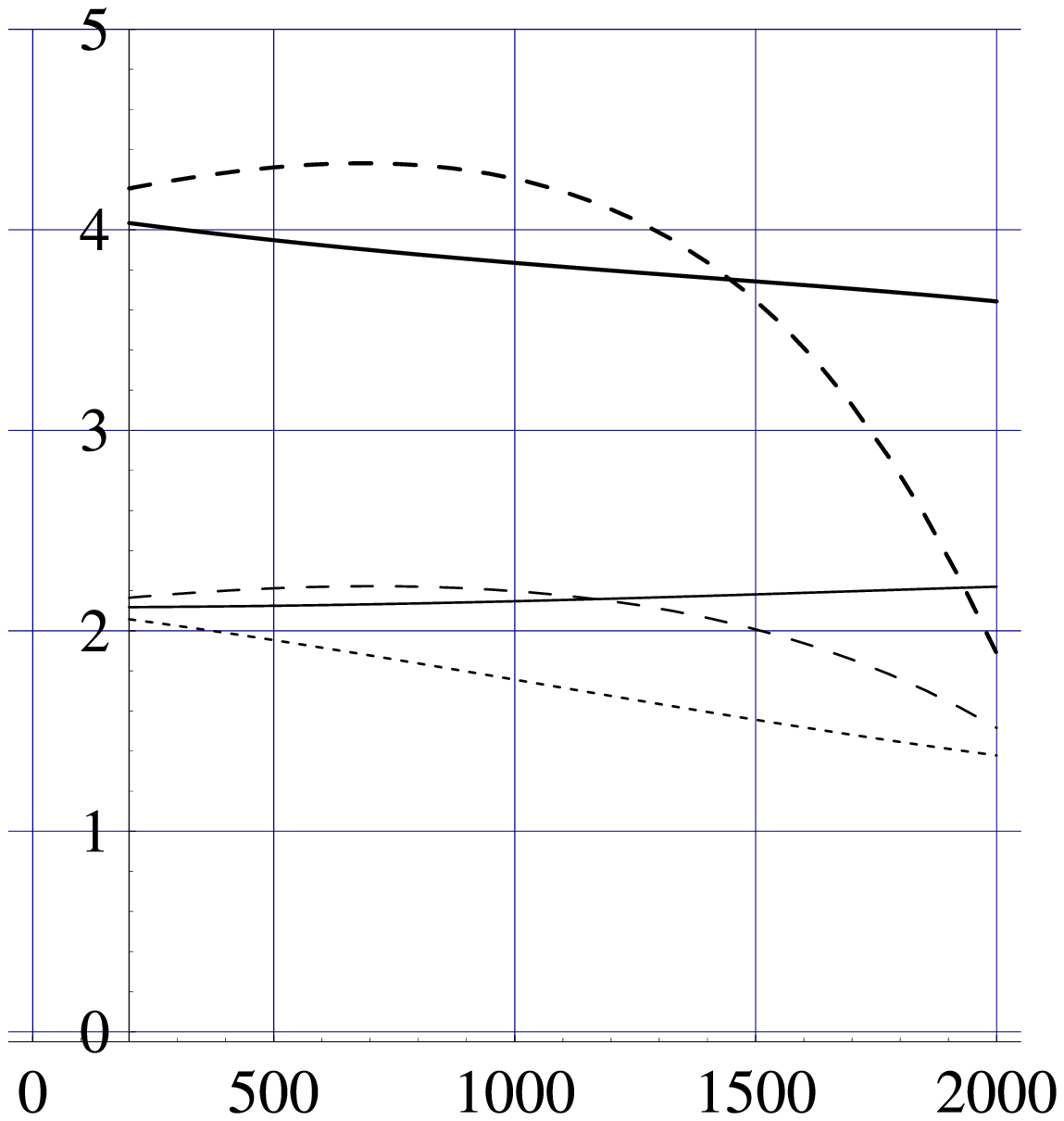}}
\put(-1.8,4.7){\mbox{(c)}}
\put( 5.5,-1.5){\epsfxsize=7.0cm
         \epsfysize=7.0cm \leavevmode \epsfbox{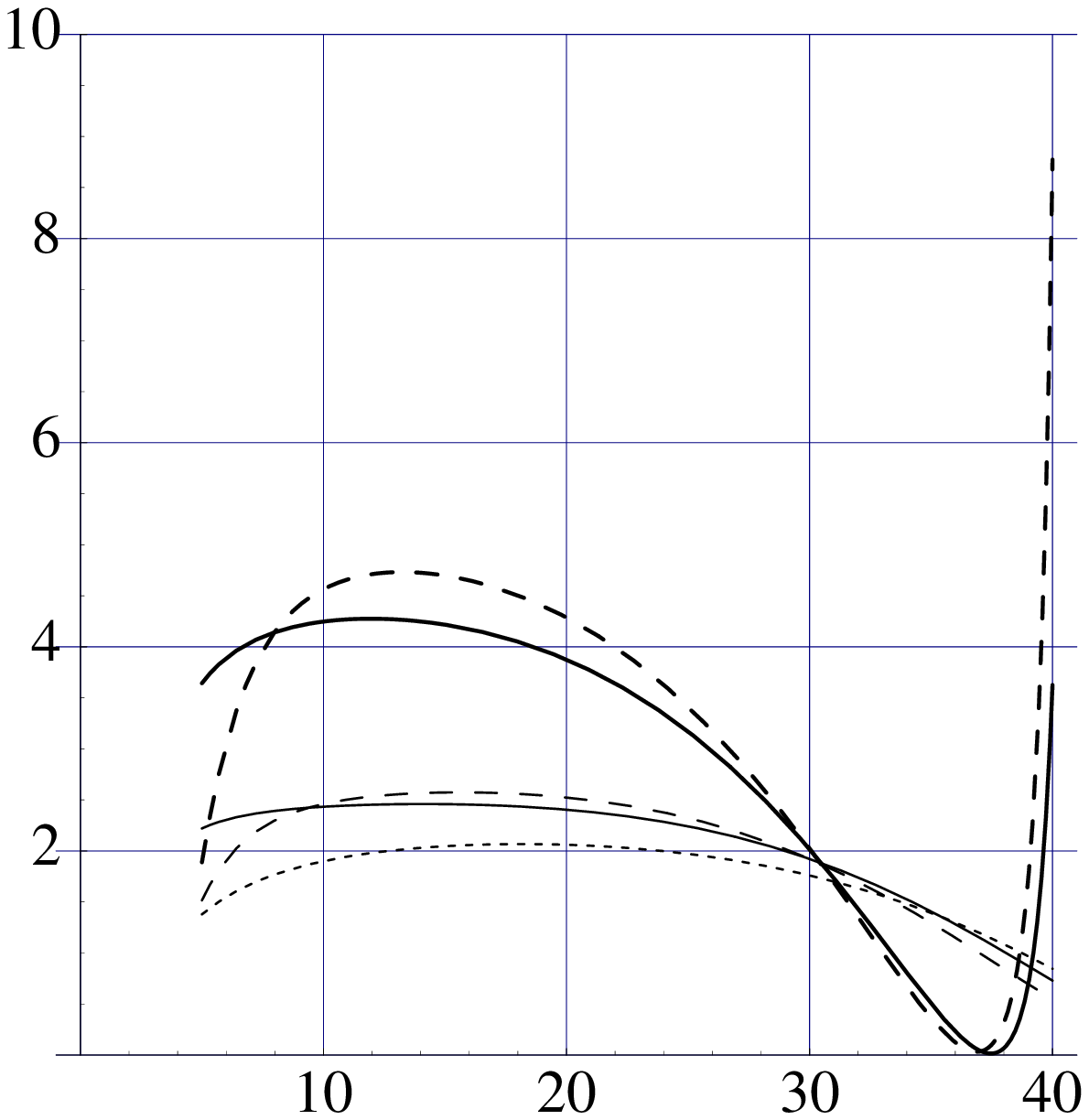}}
\put(6.6,4.7){\mbox{(d)}}
\end{picture}
\end{center}
\vspace{15mm}
\caption{ 
The decay width $\Gamma(h_1\rightarrow gg)\times 10^{4}$ (GeV)
at $M_{SUSY}=$500 GeV.
Dotted lines show $\Gamma_h$ in the
$CP$ conserving limit $\varphi=0$.
Thin solid or dashed lines are only for SM contributions, thick 
solid or dashed
lines show SM and sparticle contributions with K-factor included.
(a) ${\rm tg}\beta=$5, $A_t=A_b=$1 TeV, $\mu=$2 TeV, solid line
$\varphi=\pi/2$, dashed $\varphi=\pi$,
(b) ${\rm tg}\beta=$5, $m_{H^\pm}=$300 GeV, $\mu=$2 TeV, solid line
$\varphi=\pi/2$, dashed $\varphi=\pi$,
(c) ${\rm tg}\beta=$5, $m_{H^\pm}=$300 GeV, $A_t=A_b=$1 TeV, solid line
$\varphi=\pi/2$, dashed $\varphi=\pi$,
(d) $\mu=$2 TeV, $m_{H^\pm}=$300 GeV, $A_t=A_b=$1 TeV, solid line
$\varphi=\pi/2$, dashed $\varphi=\pi$.   }

\end{figure}


\newpage

\unitlength=1.0cm
\begin{figure}[t]
\begin{center}
\begin{picture}(10,10)
\put(-4.9,12.0){\mbox{\bf $\Gamma(h_1\rightarrow \gamma \gamma$)}}
\put(9,-1.7){\mbox{${\tt arg}(\mu A)$}}
\put(-2.8,5.8){\epsfxsize=7.0cm
         \epsfysize=7.0cm \leavevmode \epsfbox{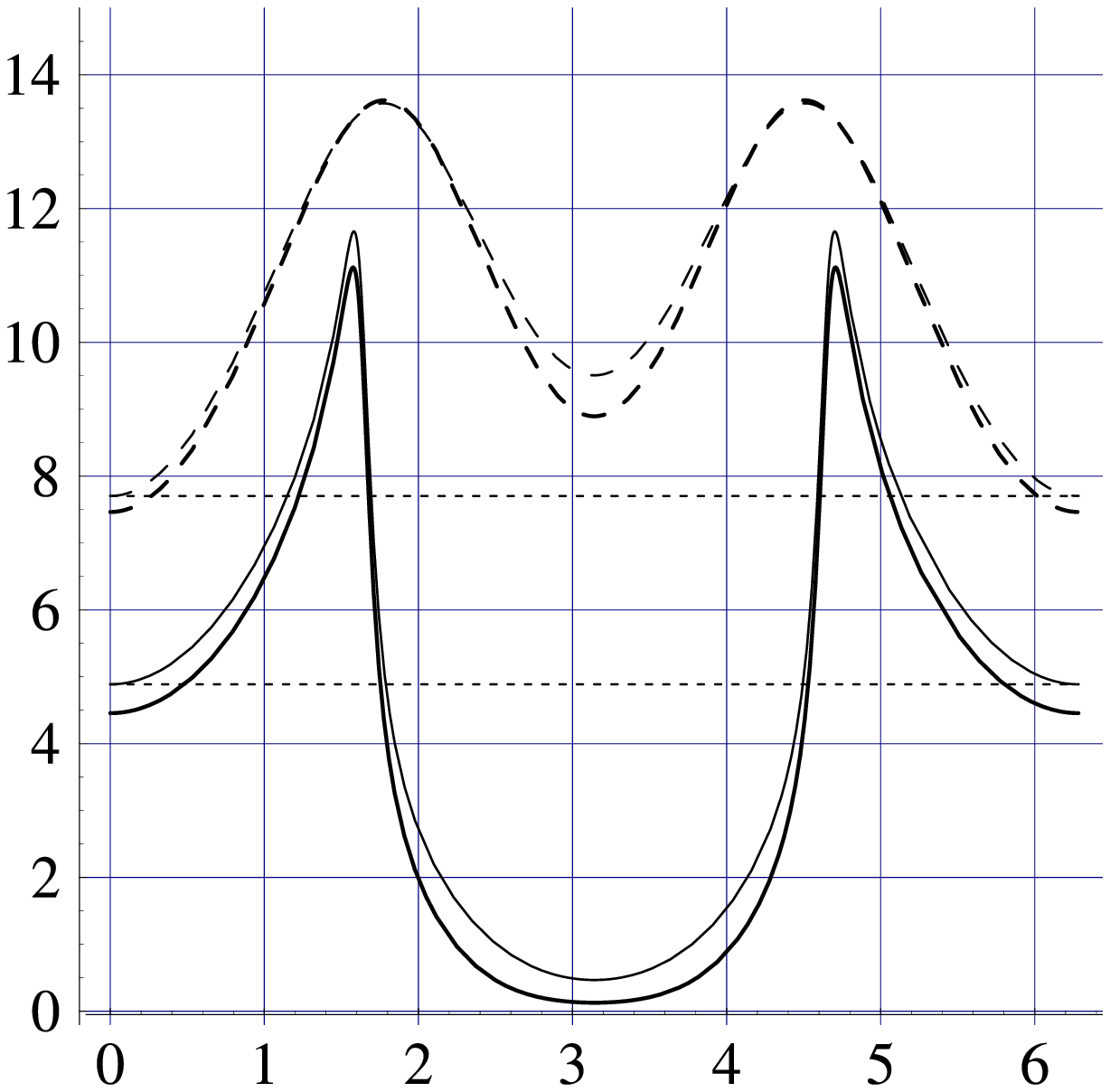}}
\put(-1.8,12.0){\mbox{(a)}}
\put(5.5,5.8){\epsfxsize=7.0cm
         \epsfysize=7.0cm \leavevmode \epsfbox{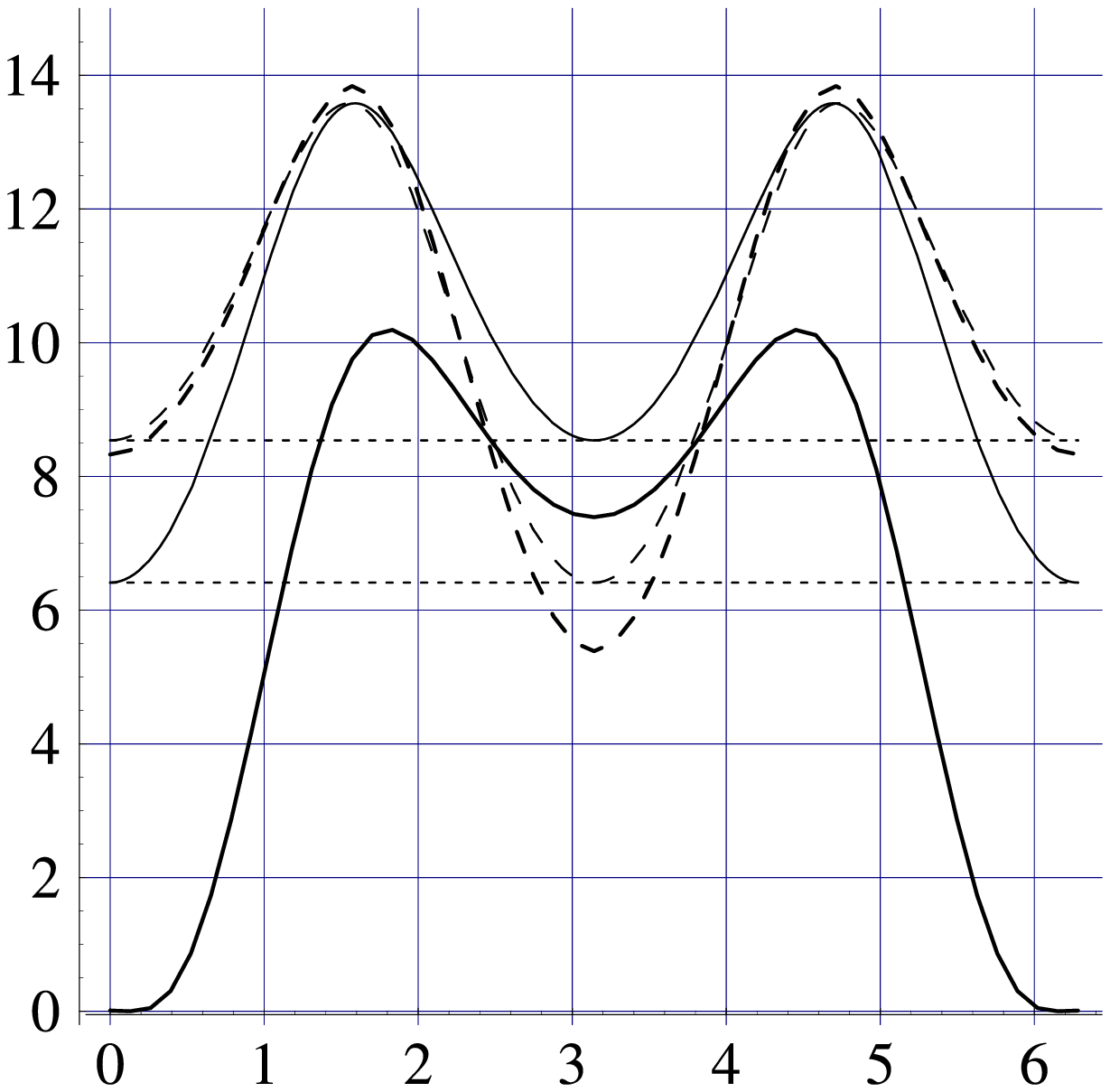}}
\put(6.6,12.0){\mbox{(b)}}
\put(-2.8,-1.5){\epsfxsize=7.0cm
         \epsfysize=7.0cm \leavevmode \epsfbox{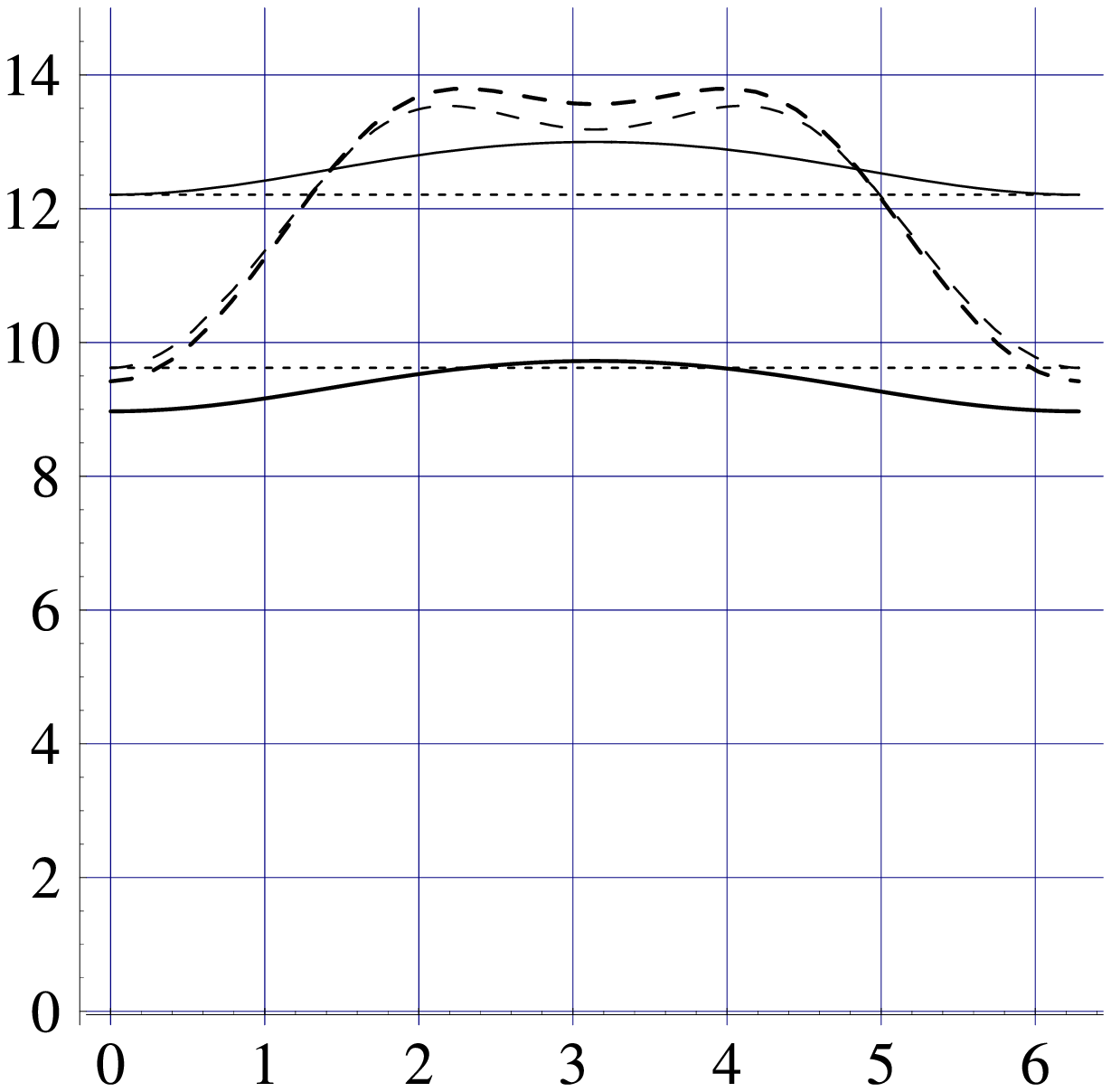}}
\put(-1.8,4.7){\mbox{(c)}}
\put( 5.5,-1.5){\epsfxsize=7.0cm
         \epsfysize=7.0cm \leavevmode \epsfbox{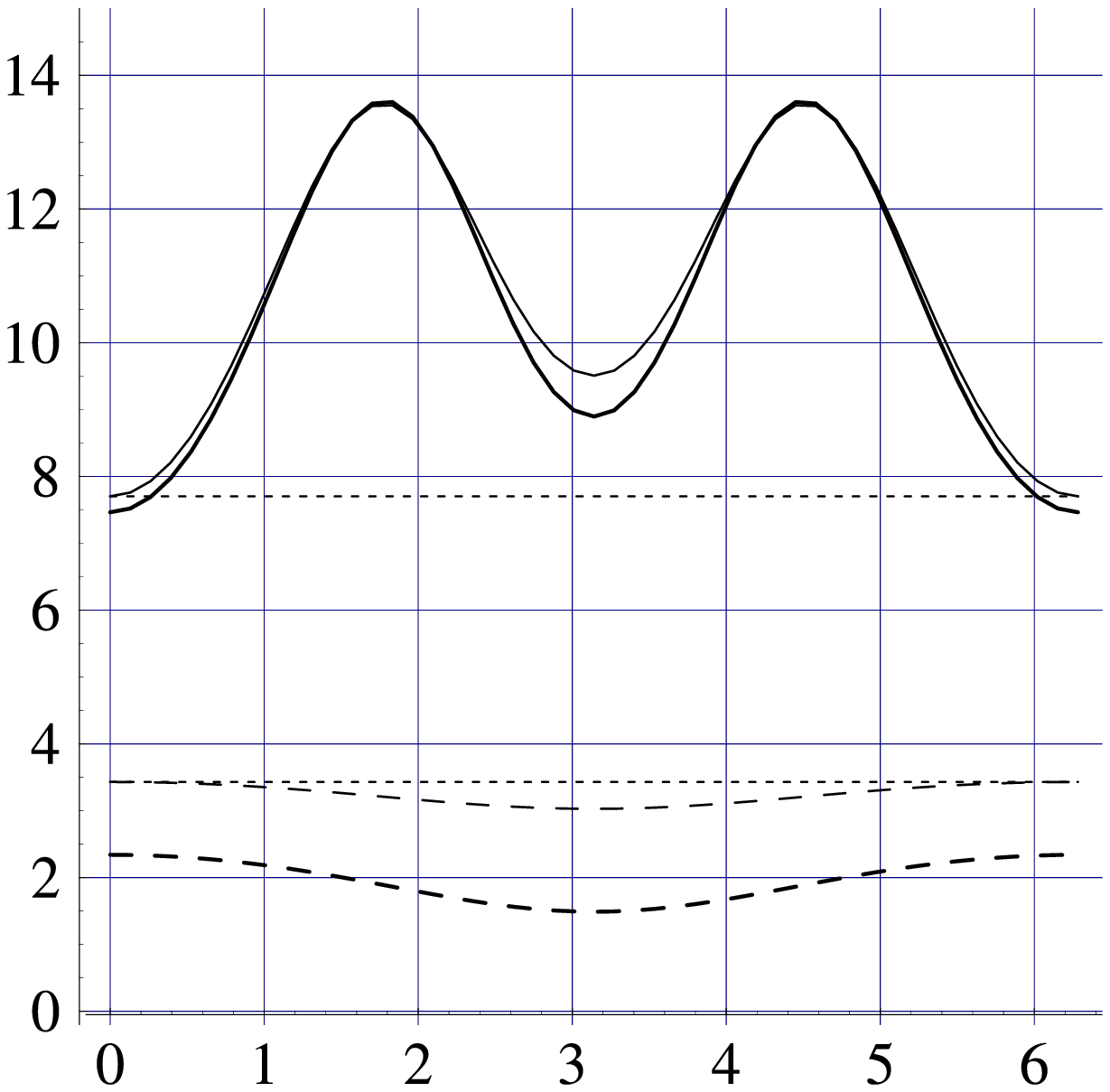}}
\put(6.6,4.7){\mbox{(d)}}
\end{picture}
\end{center}
\vspace{15mm}

\caption{ 
The decay width $\Gamma(h_1\rightarrow \gamma \gamma)\times 10^{6}$ (GeV)
at $M_{SUSY}=$500 GeV.
Dotted lines show $\Gamma_h$ in the
$CP$ conserving limit $\varphi=0$.
Thin solid or dashed lines are only for SM contributions, thick 
solid or dashed
lines show SM and sparticle contributions with J-factor included.
(a) ${\rm tg}\beta=$5, $A_t=A_b=$1 TeV, $\mu=$2 TeV, solid line
$m_{H^\pm}=$190 GeV, dashed $m_{H^\pm}=$300 GeV,
(b) ${\rm tg}\beta=$5, $m_{H^\pm}=$300 GeV, $\mu=$2 TeV, solid line
$A_t=A_b=$-1.1 TeV, dashed $A_t=A_b=$1.1 TeV,
(c) ${\rm tg}\beta=$5, $m_{H^\pm}=$300 GeV, $A_t=A_b=$1 TeV, solid line
$\mu=$0.2 TeV, dashed $\mu=$1.2 TeV,
(d) $\mu=$2 TeV, $m_{H^\pm}=$300 GeV, $A_t=A_b=$1 TeV, solid line
${\rm tg}\beta=$5, dashed ${\rm tg}\beta=$40   }

\end{figure}


\newpage

\unitlength=1.0cm
\begin{figure}[t]
\begin{center}
\begin{picture}(10,10)
\put(-4.9,12.0){\mbox{\bf $\Gamma(h_1\rightarrow \gamma \gamma$)}}
\put(2.8,5.6){\mbox{$m_{H^\pm}$}}
\put(12.2,5.7){\mbox{$A_t=A_b$}}
\put(2.8,-1.7){\mbox{$\mu$}}
\put(11,-1.7){\mbox{${\tt tg}(\beta)$}}
\put(-2.8,5.8){\epsfxsize=7.0cm
         \epsfysize=7.0cm \leavevmode \epsfbox{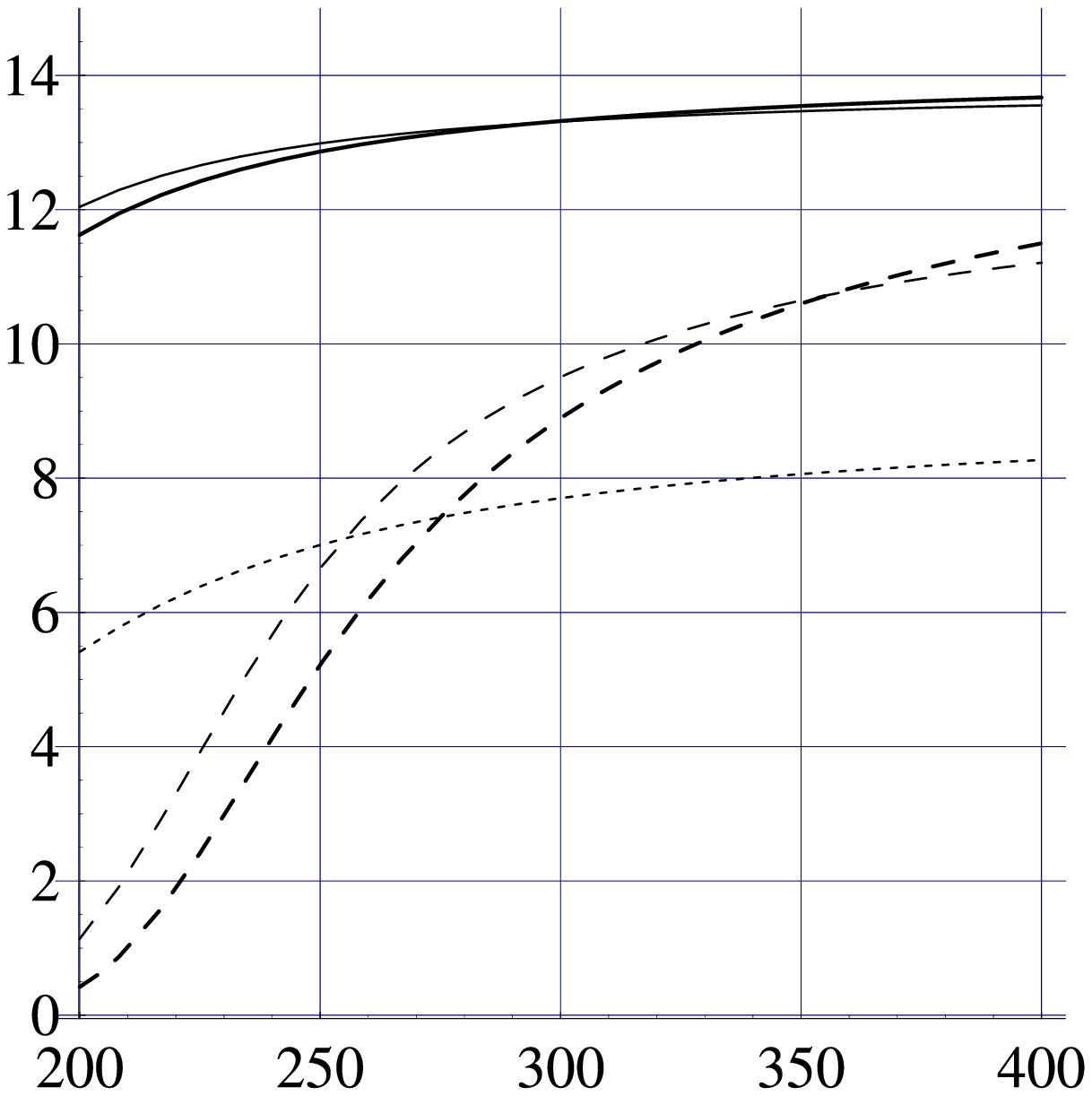}}
\put(-1.8,12.0){\mbox{(a)}}
\put(5.5,5.8){\epsfxsize=7.0cm
         \epsfysize=7.0cm \leavevmode \epsfbox{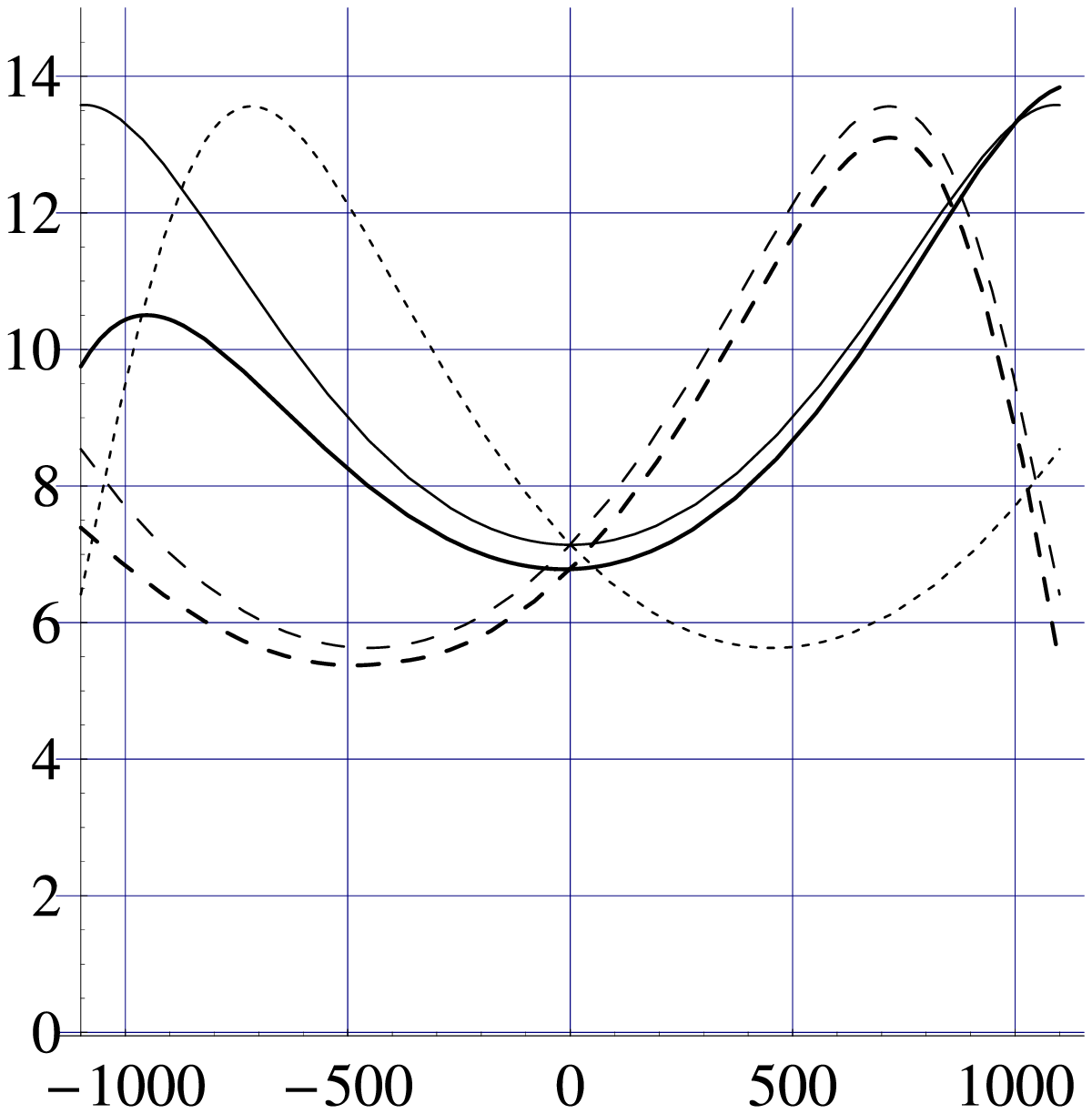}}
\put(6.4,12.3){\mbox{(b)}}
\put(-2.8,-1.5){\epsfxsize=7.0cm
         \epsfysize=7.0cm \leavevmode \epsfbox{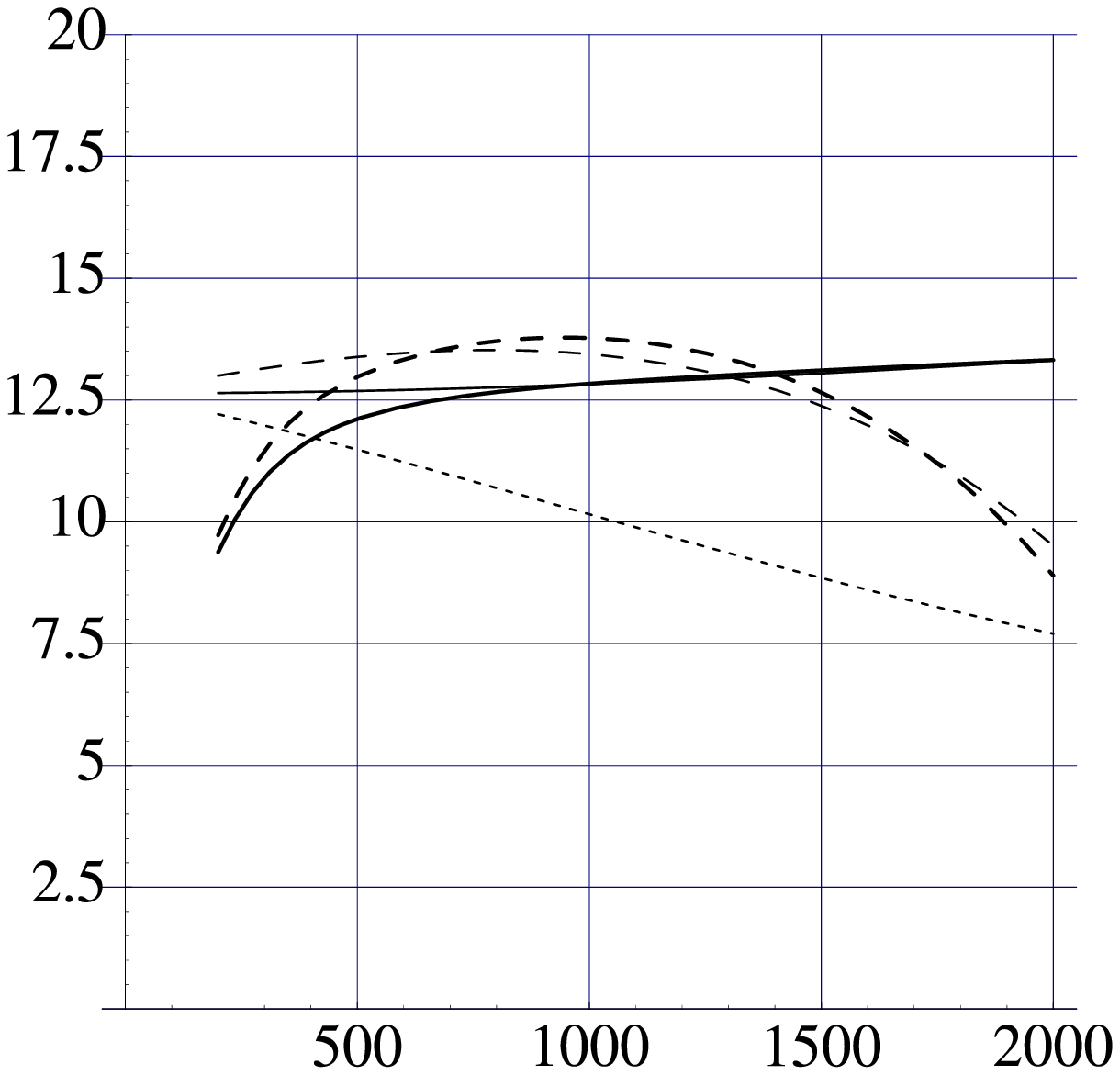}}
\put(-1.8,4.7){\mbox{(c)}}
\put( 5.5,-1.5){\epsfxsize=7.0cm
         \epsfysize=7.0cm \leavevmode \epsfbox{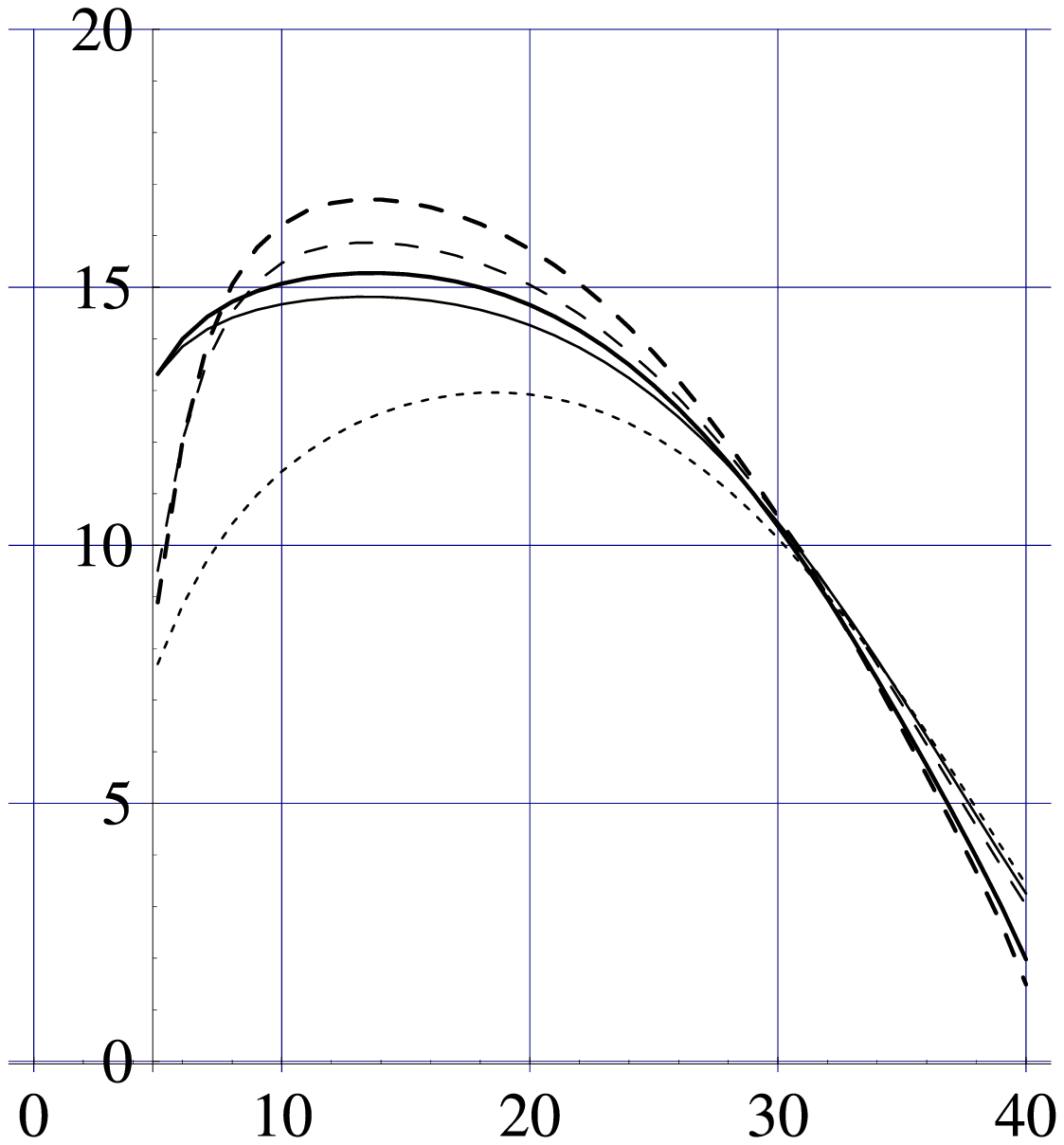}}
\put(6.6,4.7){\mbox{(d)}}
\end{picture}
\end{center}
\vspace{15mm}
\caption{ 
The decay width $\Gamma(h_1\rightarrow \gamma \gamma)\times 10^{6}$ (GeV)
at $M_{SUSY}=$500 GeV.
Dotted lines show $\Gamma_h$ in the
$CP$ conserving limit $\varphi=0$.
Thin solid or dashed lines denote the SM contributions, thick 
solid or dashed
lines show SM and sparticle contributions with J-factor included.
(a) ${\rm tg}\beta=$5, $A_t=A_b=$1 TeV, $\mu=$2 TeV, solid line
$\varphi=\pi/2$, dashed $\varphi=\pi$,
(b) ${\rm tg}\beta=$5, $m_{H^\pm}=$300 GeV, $\mu=$2 TeV, solid line
$\varphi=\pi/2$, dashed $\varphi=\pi$,
(c) ${\rm tg}\beta=$5, $m_{H^\pm}=$300 GeV, $A_t=A_b=$1 TeV, solid line
$\varphi=\pi/2$, dashed $\varphi=\pi$,
(d) $\mu=$2 TeV, $m_{H^\pm}=$300 GeV, $A_t=A_b=$1 TeV, solid line
$\varphi=\pi/2$, dashed $\varphi=\pi$.   }

\end{figure}


\newpage


\begin{thebibliography}{99}

\bibitem{[2]} H.~Georgi, Hadr. J. Phys. {\bf 1}, 155 (1978).

\bibitem{[3]} T.D.~Lee, Phys. Rev. {\bf D8}, 1226 (1973).

\bibitem{PilaftsisWagner}
A.~Pilaftsis, C.E.M.~Wagner, Nucl. Phys. {\bf B553}, 3 (1999)
(hep-ph/9902371)


\bibitem{overall}
M.~Carena, J.~Ellis, A.~Pilaftsis, C.E.M.~Wagner, Nucl. Phys. {\bf B625}, 
345 (2002) (hep-ph/0111245)\\
S.~Heinemeyer, Eur. Phys. J. {\bf C22}, 521 (2001) 
(hep-ph/0108059) \\
S.Y.~Choi, J.S.~Lee, Phys. Rev. {\bf D61}, 015003 (2000)
(hep-ph/9907496) \\
S.Y.~Choi, M.~Drees, J.S.~Lee, Phys. Lett. {\bf B481}, 57 (2000)
(hep-ph/0002287)
D.A.~Demir, Phys. Rev. {\bf D60}, 055006 (1999) (hep-ph/9901389) 

\bibitem{Dubinin02}
M.N.~Dubinin, A.V.~Semenov,
Eur.~J.~Phys. {\bf C28}, 223 (2003) (hep-ph/0206205)

\bibitem{W87} J.~Liu, L.~Wolfenstein,
 Nucl.~Phys. {\bf B289}, 1 (1987)\\
Y.L.~Wu, L.~Wolfenstein,
Phys.~Rev.~Lett. {\bf 73}, 1762 (1994);
Phys.~Rev.~Lett. {\bf 73}, 2809 (1994)

\bibitem{ginzkraw}
I.~Ginzburg, M.~Krawczyk, hep-ph/0408011

\bibitem{Inoue}
K.~Inoue, A.~Kakuto, H.~Komatsu, S.~Takeshita,
Progr.~Theor.~Phys.  {\bf 67}, 1889 (1982);
Progr.~Theor.~Phys. \textbf{68}, 927 (1982)\\
R.A.~Flores, M.~Sher, Ann.~Phys. (N.Y.) \textbf{148}, 95 (1983)

\bibitem{RGE91}
Y.~Okada, M.~Yamaguchi, T.~Yanagida, Phys. Lett. {\bf B262}, 54 (1991)\\
J.~Ellis, G.~Ridolfi, F.~Zwirner, Phys. Lett. {\bf B257}, 83 (1991)\\ 
H.E.~Haber, R.~Hempfling, Phys. Rev. Lett. {\bf 66}, 1815 (1991)\\
R.~Barbieri, M.~Frigeni, F.~Caravaglios, Phys. Lett. {\bf B258}, 167 
(1991)

\bibitem{HH1993}
H.E.~Haber, R.~Hempfling, Phys.~Rev. {\bf D48}, 4280 (1993)

\bibitem{CEPW0003180}
M.~Carena, J.R.~Ellis, A.~Pilaftsis, C.E.M.~Wagner, Nucl.~Phys. {\bf 
B586}, 92 (2000) (hep-ph/0003180)

\bibitem{Quiros97} M.~Quiros,
in: Perspectives on Higgs physics II, Ed. by G.L.~Kane (World
Scientific. 1998), p.148 (hep-ph/9703412)

\bibitem{qfthep1}
E.~Akhmetzyanova, M.~Dolgopolov, M.~Dubinin
in: Proceedings of the International Workshop "Supersymmetries and
Quantum Symmetries - SQS'03", Dubna, 2003; in:
Proceedings of XVII Workshop on High Energy Physics and Quantum
Field Theory (QFTHEP 2003), Samara-Saratov,
2003.

\bibitem{CEQW} M.~Carena, J.R.~Espinosa, M.~Quiros, C.E.M.~Wagner, 
Phys.Lett. {\bf B355}, 209 (1995)

\bibitem{Dubinin01}
M.~Dubinin, A.~Semenov, SNUTP report 98-140, hep-ph/9812246\\
F.~Boudjema, A.~Semenov, Phys. Rev. {\bf D66}, 095007 (2002)
(hep-ph/0201219)

\bibitem{GunionHaber} J.F.~Gunion, H.E.~Haber, Phys.~Rev. {\bf D67}, 
075019 (2003) (hep-ph/0207010)

\bibitem{CPsuperH} J.S.~Lee, A.~Pilaftsis, M.~Carena, S.Y.~Choi, M.~Drees,
J.R.~Ellis, C.E.M.~Wagner, Comput.Phys.Commun. {\bf 156}, 283 (2004)
(hep-ph/0307377) 

\bibitem{feynh}
M. Frank, S. Heinemeyer, W. Hollik, G. Weiglein, in: Hamburg 2002, 
Supersymmetry and unification of fundamental interactions (SUSY02), 
vol.2, p.637 (hep-ph/0212037)

\bibitem{conciliate}
M.~Carena, H.E.~Haber, S.~Heinemeyer, W.~Hollik, C.E.M.~Wagner, 
G.~Weiglein,
Nucl.Phys. {\bf B580}, 29 (2000) (hep-ph/0001002)\\ 
J.~R.~Espinosa, R.~J.~Zhang, JHEP {\bf 0003} 026 (2000)
(hep-ph/9912236)

\bibitem{cpx}
M.~Carena, J.~Ellis, A.~Pilaftsis, C.~Wagner, Phys.Lett. {\bf} B495 155 
(2000) (hep-ph/0009212)

\bibitem{Veltman}
M.J.G. Veltman, CERN report 97-05, 1997

\bibitem{GinzV} I.F.~Ginzburg, M.V.~Vychugin,
in: Proceedings of XVI Workshop on High Energy Physics and Quantum
Field Theory (QFTHEP 2001), Moscow, 2002, p.~64 (hep-ph/0201117)

\bibitem{cmsnote}
S.~Abdullin et.al., CMS Note 2003/033.

\bibitem{LEP2}
M.~Carena, J.~Ellis, S.~Mrenna, A.~Pilaftsis, C.E.M.~Wagner,
Nucl. Phys. {\bf B659}, 145 (2003) (hep-ph/0211467) 

\bibitem{colliders}
P.~Niezurawski, A.F.~Zarnecki, M.~Krawczyk, hep-ph/0403138\\
R.M.~Godbole, S.D.~Rindani, R.K.~Singh, Phys. Rev. {\bf D67}, 095009
(2003) (hep-ph/0211136) \\
A.~Dedes, S.~Moretti, Nucl. Phys. {\bf B576}, 29 (2000) (hep-ph/9909418) 
\\
A.~Dedes, S.~Moretti, Phys. Rev. Lett. {\bf 84}, 22 (2000) 
(hep-ph/9908516)\\
B.~Grzadkowski, J.F.~Gunion, J.~Kalinowski, Phys. Rev. {\bf D60}, 075011
(1999) (hep-ph/9902308) 

\bibitem{GGN} 
J.~Gunion, G.~Gamberini and S.~Novaes, Phys. Rev. {\bf D38},
3481 (1988)

\bibitem{thresh}
H.~Eberl, K.~Hidaka, S.~Kraml, W.~Majerotto, Y.~Yamada,
Phys. Rev. {\bf D62}, 055006 (2000) (hep-ph/9912463) 

\bibitem{factors}
J.~Ellis, M.K.~Gaillard, D.V.~Nanopoulos, Nucl. Phys. {\bf B106}, 292 
(1976) \\
A.I.~Vainshtein, M.B.~Voloshin, V.I.~Zakharov, M.A.~Shifman, 
Sov. J. Nucl. Phys., 30, 711 (1979) (Yad. Fiz. {\bf 30}, 1368 (1979))

\bibitem{SDGZ} M. Spira, A. Djouadi, D. Graudenz and P. M. Zerwas,
  Nucl. Phys. {\bf B453}, 17 (1995) (hep-ph/9504378) \\
M. Spira, Fortsch.\ Phys.\ {\bf 46}, 203 (1998) (hep-ph/9705337) 

\bibitem{comphep}
E. Boos, V. Bunichev, M. Dubinin, L. Dudko, V. Edneral, V. Ilyin, A. 
Kryukov, V. Savrin, A. Semenov, A. Sherstnev, CompHEP 4.4, 
hep-ph/0403113;
A. Pukhov et.al., CompHEP 3.3, hep-ph/9908288.

\bibitem{lanhep}
A.~Semenov, Nucl. Instr. and Meth., {\bf 389}, 293 (1997)
(hep-ph/9608488)

\end{thebibliography}
\end{document}